\documentclass[twocolumn]{aastex62}

\usepackage{savesym}
\savesymbol{tablenum}
\usepackage{siunitx}
\usepackage{mathrsfs}

\restoresymbol{SIX}{tablenum}
\usepackage{hyperref}

\usepackage{amsmath,bm}

\newcommand{ \angstrom}{\textup{\AA}}

\graphicspath{{./}{figures/}}
\usepackage[T1]{fontenc}
\usepackage[section]{placeins}
\usepackage{amssymb}
\usepackage{float}

\submitjournal{ApJ}

\title{Dynamical mass-to-light gradients}
\shortauthors{Mehrgan et al.}

\begin{document}
\title{Dynamical stellar mass-to-light ratio gradients:\\Evidence for  very centrally concentrated IMF variations in ETGs?}

\author{Kianusch Mehrgan}
\affil{Max-Planck-Institut f\"ur extraterrestrische Physik, Giessenbachstrasse, D-85748 Garching} \affil{Universit\"ats-Sternwarte M\"unchen, Scheinerstrasse 1, D-81679 M\"unchen, Germany}
\email{kmehrgan@mpe.mpg.de}
\author{Jens Thomas}
\affil{Max-Planck-Institut f\"ur extraterrestrische Physik, Giessenbachstrasse, D-85748 Garching} \affil{Universit\"ats-Sternwarte M\"unchen, Scheinerstrasse 1, D-81679 M\"unchen, Germany}

\author{Roberto Saglia}
\affil{Max-Planck-Institut f\"ur extraterrestrische Physik, Giessenbachstrasse, D-85748 Garching} \affil{Universit\"ats-Sternwarte M\"unchen, Scheinerstrasse 1, D-81679 M\"unchen, Germany}

\author{Taniya Parikh}
\affil{Max-Planck-Institut f\"ur extraterrestrische Physik, Giessenbachstrasse, D-85748 Garching} 
\author{Bianca Neureiter}
\affil{Max-Planck-Institut f\"ur extraterrestrische Physik, Giessenbachstrasse, D-85748 Garching} \affil{Universit\"ats-Sternwarte M\"unchen, Scheinerstrasse 1, D-81679 M\"unchen, Germany}

\author{Peter Erwin}
\affil{Max-Planck-Institut f\"ur extraterrestrische Physik, Giessenbachstrasse, D-85748 Garching}

\author{Ralf Bender }
\affil{Max-Planck-Institut f\"ur extraterrestrische Physik, Giessenbachstrasse, D-85748 Garching} \affil{Universit\"ats-Sternwarte M\"unchen, Scheinerstrasse 1, D-81679 M\"unchen, Germany}

\begin{abstract}
Evidence from different probes of the stellar initial mass function (IMF) of massive early-type galaxies
(ETGs) has repeatedly converged on IMFs more bottom-heavy than in the Milky Way (MW). This consensus
has come under scrutiny due to often contradictory results from different methods on the level
of individual galaxies. In particular, a number of strong lensing probes are ostensibly incompatible with a non-MW IMF. Radial gradients of the IMF -- related to gradients of the stellar mass-to-light ratio $\Upsilon$ -- can potentially resolve this
issue. We construct Schwarzschild models allowing for $\Upsilon$-gradients in seven massive ETGs with MUSE and SINFONI observations. We find dynamical evidence that $\Upsilon$ increases towards the center for all ETGs. The gradients are confined to sub-kpc scales. \textcolor{black}{Our results suggest that constant-$\Upsilon$ models may overestimate the stellar mass of galaxies by up to a factor $1.5$}. For all except one galaxy, we find a radius where the total dynamical mass has a minimum. This minimum places the strongest constraints on the IMF outside the center and appears at roughly $\SI{1}{kpc}$. We consider the IMF at this radius characteristic for the main body of each ETG. In terms of the IMF mass-normalization $\alpha$ relative to a Kroupa IMF, we find on average a MW-like IMF $<\alpha_{\mathrm{main}}> = 1.03 \pm 0.19$. In the centers, we find concentrated regions with increased mass normalizations that are less extreme than previous studies suggested, but still point to a Salpeter-like IMF,  $<\alpha_{\mathrm{cen}}> = 1.54 \pm 0.15$.

\end{abstract}


\section{Introduction}
The question of how much stars contribute to the total mass of distant galaxies remains one of the fundamental issues of extragalactic astronomy. The answer is critical for mass decompositions of these objects into stellar components, dark matter (DM) and supermassive black holes (SMBHs), as well as for our understanding of galaxy formation histories. The difficulty lies in the fact that the unresolved stellar populations of these galaxies contain both low-luminosity dwarf stars and stellar remnants -- both of which contribute to the galactic mass and follow the light of these galaxies, but contribute barely or not at all to the observed light.

The stellar initial mass function (IMF) describes the distribution function of stars as a function of stellar mass at the time of the star formation events in which the observed stellar populations of a galaxy were produced. It encompasses long-lived low-luminosity dwarf stars whose distribution essentially remains unchanged during galaxy evolution to the present epoch, and more massive stars which will have turned into remnants by the time of observation. Besides allowing an estimation of the total stellar mass, the IMF informs essentially every other part of galaxy evolution, such as star formation rates, stellar feedback, and heavy element production \citep[e.g.][]{Kennicutt1998, BasitanCoveyMeyer2010}.

Numerous studies have found that a Kroupa or Chabrier IMF can describe the IMF of the Milky Way (MW) across multiple different environments \citep[e.g.][]{Kroupa2001, Kroupa2002, Chabrier2003, BasitanCoveyMeyer2010}, as well as that of nearby spiral galaxies \citep[e.g.][]{Kassin2006, Brewer2012}. This prompts the question: is the IMF \textit{universal} to all galaxies? If so, the proposed IMF models could be used to a priori separate the baryonic, DM and SMBH content of distant galaxies in dynamical models, which would greatly improve the accuracy of SMBH and DM measurements.

Individual star counts, as performed for IMF probes of the MW, are infeasible in other galaxies, as the stellar populations are unresolved. Therefore, different methods have to be used to extract IMF information from the observed stellar light. There are two dominant techniques in use:

1) Fitting of IMF-sensitive stellar absorption features whose strength is regulated by the ratio of dwarf to giant stars with models based on single stellar population (SSPs) synthesis libraries. These models output a stellar mass-to-light ratio $\Upsilon^{\mathrm{SSP}}$, as well as an IMF model. However, in this manner we can only probe the low-mass end of the IMF of early-type galaxies (ETGs), as on the high-mass end (without replenishment from star formation) most stars have turned into remnants, which are invisible to SSP modeling.
2) Measurements of the galactic gravitational potential, via stellar dynamics and/or gravitational lensing. These do not directly distinguished between DM, stars and the central SMBH of the galaxy, but produce a total mass-to-light ratio $(M^{\mathrm{tot}}/L)^{\mathrm{dyn}}$. From this, a stellar mass-to-light ratio $\Upsilon^{\mathrm{dyn}}$ can be inferred relative to assumptions about the shape of the DM halo. $\Upsilon^{\mathrm{dyn}}$ can be driven up either by the mass contributions of dwarfs or remnants from the high-mass end of the IMF.

For either approach, it is convenient to characterize the IMF probe by a mass normalization factor $\alpha$ of the stellar mass-to-light ratio relative to a reference $\Upsilon^{\mathrm{SSP}}_{\mathrm{ref}}$ with a reference IMF, which in this study will be a Kroupa IMF.

Many of the earliest dynamical probes of the stellar mass content of ETGs did not directly attempt to separate DM from stellar masses. These, most notably the SAURON project \citep{deZeeuw+2002, Emsellem2004, Cappellari2006}, found that ETGs were fundamentally unlike spiral galaxies in their mass-light composition: Here, $(M^{\mathrm{tot}}/L)^{\mathrm{dyn}} > \Upsilon^{\mathrm{SSP}}_{Kroupa}$, with the ratio for some galaxies being large enough that the total mass budget could accommodate a Salpeter or super-Salpeter IMF. Such an IMF produces larger $\Upsilon$, due to a relative excess of low luminosity dwarf stars relative to a MW IMF, a phenomenon typically referred to as ``bottom-heaviness''. At this point, there was still no consensus on whether or not the mass excess relative to a MW IMF was due to unaccounted DM or an enhanced stellar contribution. However, even early (spherical) dynamical models with DM halo components found similar results for the remaining stellar contribution \citep{Gerhard2001}. Since then, a number of surveys and projects focused on dynamical and lensing models of ETGs have used a variety of DM models to produce measurements of the stellar mass-to-light $\Upsilon$. These included the work of the SLACS group, which analyzed 56 massive \textcolor{black}{lensing} galaxies combining strong-lensing with simple spherical Jeans models \citep{Treu2010, Auger2010}, and dynamical studies \textcolor{black}{of the ETGS of the} Coma cluster \citep{Thomas2007, Thomas2009, ThomasJens2011} and the cluster Abell~262 \citep{Wegner2012} using sophisticated axisymmetric Schwarzschild orbit superposition models \citep{Schwarzschild1979}. This was followed up by the ATLAS$^{3D}$ project \citep{Cappellari2012Nature, Cappellari2013a, Cappellari2013b}, which analyzed 260 ETGs using Jeans anisotropic modeling \citep[JAM; ][]{Cappellari2006, Cappellari2008}. These studies found galaxy-by-galaxy variation of the mass normalization $\alpha$, which correlated with a number of galactic properties, particularly galactic velocity dispersion \textcolor{black}{\citep[e.g. equation 6 of][]{Posacki+15}}. Notably, for massive ETGs with $\sigma_e \gtrsim \SI{250}{km/s}$ these studies predict a mass normalization at least twice the MW-level.

Various lensing studies have been used to more thoroughly investigate the central DM profiles of these galaxies, but found complementary trends of $\alpha$, even where more concentrated DM profiles were used \citep[e.g.][]{Spiniello2011, Sonnenfeld+15, Oldham&Auger2018Lensing, Sonnenfeld+2019}. \citet{Napolitano2011, Napolitano2014} used observations of globular clusters and planetary nebulae to derive dynamical constraints on the DM halos of massive ETGs out to several times the effective radius. With these constraints they found that unless the centers of the DM halos had undergone adiabatic contraction from baryonic infall, these galaxies required a Salpeter-level $\alpha$.

At the same time as mass probes converged on a comprehensive picture of a variation in $\alpha$, SSP modeling probes of the centers of ETGs, often from the same samples, supported the claim that the established trends of $\alpha$ indeed arise from variations of the IMF \citep{vanDokkum&ConroyNature2010, vanDokkum2011, vanDokkum2012, SmithLuceyCarter2012, Conroy&vanDokkum2012, Conroy&vanDokkum2014, Tortora2013, Ferreras2013, LaBarbera2013}. 

Since then, claims in favour of IMF variation among ETGs with mass and other properties, such as metallicity and [Mg/Fe] enrichment, have been accumulating \citep{Martin-Navarro2015b,Lyubenova+2016, Li2017, vanDokkum2017, Parikh+18, Poci2022, Bernadi2019}.

However, a number of problems remain with this framework, which have yet to be resolved before the IMF can conclusively be determined to be non-universal. While the overall \textit{trends} of the IMF found by dynamical/lensing and SSP measurements appear to be in agreement, on the level of individual galaxies, the measurements of $\alpha$ from the two methods often do not agree or not even correlate \citep{Smith2014, McDermid2014}. Furthermore, recent lensing measurements from the SNELLS and MNELLS surveys \citep{Smith+15, Newman+17, Collier2018, CollierSmith&Lucey2020}, as well as a survey of 23 lensed ETGs by \citet{Sonnenfeld+2019}, and individual dynamical measurements \citep{Rusli2013a,Thomas2016} have ruled out a mass normalisation $\alpha$ above the MW value for a number of very massive galaxies with $\sigma_e > \SI{250}{kms}$. 

Work by the CALIFA survey \textcolor{black}{\citep{Lyubenova+2016}} spanning all three methods suggested that the tension between different IMF probes can be partially alleviated by correcting for aperture effects. Consideration of aperture differences become crucial if ETGs posses intrinsic radial IMF gradients. \citet{Bernardi+18} and \citet{DominiqueSanchez2019} suggested that if such gradients exist, they could bridge the difference between galaxy-gravitational and stellar population probes of the IMF. Radial gradients for massive ETGs would not be unexpected in a two-phase formation scenario where the central stars are mostly formed in-situ at high redshift while most of the outer material is accreted later on from smaller sub units with potentially different star-formation conditions.

A number of stellar population modeling studies have already claimed internal IMF gradients confined to small spatial scales on the order of a few kpc \citep{Martin-Navarro2015a,vanDokkum2017, Parikh+18, LaBarbera2019, DominiqueSanchez2019}. 
There exist only a few dynamical and lensing studies related to IMF gradients and these found similar results for the massive ETG M87 \citep{Oldham&AugerM872018, Li_M872020}, the lensing galaxy ESO 325-G004 \citep{Collett2018}, as well as for several lensing galaxies from the samples of \citet{Oldham&Auger2018Lensing} and \citet{NewmanEllisTreu2015}.

Our goal in this study is to systematically investigate for the first time the possible existence of IMF-gradients with dynamical models. To this end we use our state-of-the-art orbit-based Schwarzschild dynamical modelling code which originally goes back to the code of \citet{RichstoneTremaine1988,Gebhardt2003,Thomas2004,Siopis2009}. This code has been advanced since then in many respects, most notably it accounts for the overfitting problem and respective biases by using a generalised model selection technique \citep{Lipka2021,ThomasLipka2022}. 

Central gradients in the stellar mass-to-light ratio $\Upsilon$ can only be reliably determined if SMBHs are taken into account. For this reason, we are here studying a sample of seven massive ETGs with a combination of two sets of previously published non-parametric 2D stellar kinematics from a) the Multi-Unit Spectroscopic Explorer (MUSE), and b) the spectrograph Integral Field Observations in the Near Infrared (SINFONI). While the wide-field MUSE data have a high SNR \citep{Mehrgan2023}, the SINFONI data, which are concentrated on the central regions of the galaxies, are adaptive optics (AO) supported and resolve the sphere of influence (SOI) of the SMBHs \citep{Rusli2011, Rusli2013a, Rusli2013b, Erwin2018}. 

While our sample is relatively small, we combine several crucial advancements compared to previous studies: (i) we systematically probe for dynamical gradients in ETGs combining spectroscopic data which allows us to simultaneously constrain the wide-field mass distribution as well as central SMBHs; (ii) we use Schwarzschild models that do not require any a priori assumption on the anisotropy of the stellar orbits; (iii) we use a new generalised model selection technique that overcomes known \textcolor{black}{limitations} in Schwarzschild fits and allows for mass measurements with very high precision; (iv) we consistently use non-parametric LOSVDs both in the center and for the wide-field data. Points (ii) to (iv) have been demonstrated to be sufficient to break known degeneracies and avoid biases in dynamical models even for (more complex) triaxial galaxies and to allow for dynamical mass determinations with a precision at the 10\%-level \citep{Lipka2021,deNicola2022,Neureiter2023a}.

This study is structured as follows: in Section 2, we present our MUSE and SINFONI kinematics for the seven ETGs, as well as our Schwarzschild modeling approach. in Section 3, we present the derived gradients of $\Upsilon$. Afterwards, in Section 4, we discuss them in terms of evidence for IMF gradients. Finally, we conclude our study in Section 6 by summing up our results and discussing their implications for future investigations of IMF variations in and between ETGs.

\section{Orbital dynamical modeling: technique and data}

We list the seven ETGs which we dynamically modeled for their $\Upsilon$ gradients in Table \ref{tab:generalTab}, together with some of their morphological properties and general information about the MUSE and SINFONI data which we used in this study. This sample is a sub-sample of the nine ETGs analysed in \citet{Mehrgan2023}. We have singled out the remaining two galaxies from that previous study, NGC~5419 and NGC~6861, for separate analysis elsewhere. NGC~5419 was modelled using our new triaxial Schwarzschild dynamical modeling code SMART in \citet{Neureiter2023}. NGC~6861 will be presented in Thomas et al. in prep.

All seven galaxies under study here were modelled previously but using other data, mostly long-slit, for the outer parts rather than the new MUSE data (\citealt{Rusli2011, Rusli2013a, Erwin2018}, which we will refer to as R+11, R+13 and E+18).

Using the sequencing of ETGs first introduced by \citet{KormendyBender1997} and \citet{Faber1997} 
into luminous ETGs with shallow central surface brightness cores and less luminous ETGs with steep power-law surface brightness profiles \cite[e.g.][]{NietoBenderSurma1991, Crane1993, Kormendy1994, Ferrarese1994, Lauer1995, Gebhardt1996, Faber1997,Kormendy1999,  Lauer2007,Kormendy2009}, our sample can be partitioned into four cored ETGs and three power-law ETGs  (R+11,13; E+18). We also classified these galaxies in our previous publication, \citet{Mehrgan2023} in accordance with the angular momentum classification scheme of \citet{Emsellem2007, Emsellem2011}. As is typical for the core/power-law dichotomy \citep[for review, see][]{Lauer2012}, the three power-law ETGs are fast rotating and have either disc components or disc-like components, while the cored ETGs have no disc components and have less rotation. Two of the cored ETGs are typical slow rotators, while two have an angular momentum that could be considered ``intermediate''

\begin{table*}
\centering
 \begin{tabular}{l l  c c c l}
 \hline
 \hline
Galaxy               &  Morphology (M+23)    & MUSE-PSF [$\arcsec$])                                                     & SINFONI SNR$_{\mathrm{min}}/\angstrom$  & study for SINFONI\\
\hline
NGC~307                      &  power-law/fast   & $2.10$                     & $30$                 & E+18 \\
NGC~1332                   &   power-law/fast   & $2.12$              & $83$                 & R+11 \\
NGC~1407                      &  core/interm.  & $1.93$                         & $30$                 & R+13 \\   
NGC~4751                      &  power-law/fast       & $1.59$                 & $30$                 & R+13 \\  
NGC~5328                      &  core/slow & $1.28$           & $30$                 & R+13 \\   
NGC~5516                     &  core/slow    & $2.00$                          & $30$                 & R+13 \\  
NGC~7619                     &  core/interm.  & $2.00$                         & $30$                 & R+13 \\
\hline
\end{tabular}
\caption{Selected properties of the observations and kinematic analysis of the sample galaxies with MUSE and SINFONI. The full width at half maximum (FWHM) of the point spread function (PSF) the MUSE observations are listed here as in Mehrgan et al 2023 (M+23) for the sake of convenience. The SINFONI observations were adaptive optic based and have a FWHM of the PSF of roughly $\SI{0.15}{\arcsec}$. We also list morphological classifications of the galaxies from M+23 according to their central regions and their angular momentum into fast, slow and intermediate rotating galaxies. In this study, we supplement our MUSE kinematics from M+23 with SINFONI kinematics from studies which are listed in the last column: \citet{Rusli2011} (R+11) \citet{Rusli2013a, Rusli2013b} (R+13), and \citet{Erwin2018} (M+18). 
We also list the minimum SNR of the SINFONI data. For all MUSE data SNR$_{\mathrm{min}}/\angstrom \sim 100$ (as described in M+23).}
\label{tab:generalTab}
\end{table*}

Below, in Section \ref{sec:modelsetup}, we describe our implementation of the axisymmetric Schwarzschild dynamical models which we used on our sample. 
As inputs, these models use 3D deprojections of (2D) imaging data along the line-of-sight, which we describe in Section \ref{subsec:lightDens}, and --  importantly --  stellar kinematics in the form of non-parametric line-of-sight velocity distributions (LOSVDs) derived from MUSE and SINFONI spectroscopy. These kinematics are described in Section  \ref{subsec:kinematics}.

\subsection{Axisymmetric Schwarzschild modeling}
\label{sec:modelsetup}
\subsubsection{Implementation of models with radial mass-to-light ratio gradients}
We dynamically model the sample galaxies under the assumption that they are axisymmetric. We discuss this assumption later on in Section \ref{subsec:axisymm}.

The dynamical models in this study consist of an advanced implementation of the axisymmetric Schwarzschild orbit superposition code of \citet{Thomas2004}. It allows for radial gradients of the stellar mass to light ratio, $\Upsilon(r)$.
We here only briefly summarize the key features of this implementation and highlight new additions and those parts of our approach which are specific to the present study. 

Following the Jeans theorem, in a stationary system, the phase-space density is constant along trajectories which typically obey three integrals of motion: $E$, $L_{z}$ and the non-classical $I_{3}$ (for axisymmetric systems).
Hence, we can think of stationary galaxies as the superposition of orbits which represent the system's phase-space \citep{Schwarzschild1979} and constitute all possible solutions to the collisionless Boltzmann equation.
A representative sampling of the integrals of motion $E$, $L_{z}$ and $I_3$ in a model gravitational potential $\Phi$ enables us to construct any allowed configuration of orbits and match all kinds of observed galaxy shapes and kinematics.
By linking $\Phi$ to different model mass (density) distributions via Poisson`s equation, we can thus optimize the mass model to best reproduce the observed stellar kinematics and imaging data of galaxies. 

Here, we use the following parameterization for the mass composition $\rho(r, \theta)$:

\begin{equation}
\rho(r, \theta) = \rho_{\star}(r, \theta) + M_{\mathrm{BH}}\delta(r) + \rho_{DM}(r),
\label{eq:fullrho}
\centering
\end{equation}

where $\theta$ is the polar angle, $M_{\mathrm{BH}}$ the mass of the central SMBH and $\rho_{DM}$ the DM halo.
For $\rho_{DM}$ we initially chose to adopt the generalised NFW-halo derived from cosmological N-body simulations by
\citet{Navarro1996,Zhao1996}, which is defined by three parameters,
$\rho_{10}$, the DM density at $\SI{10}{kpc}$, $r_{s}$, the
scale radius of the halo and $\gamma$, the inner slope of the DM
density profile. After extensive preliminary testing we found that for our sample galaxies the dynamical models always converged on cored DM-profiles, $\gamma = 0$ while $r_{s}$ was always on similar scales $\sim \SI{100}{kpc}$. We will discuss our DM halos and the implications of these findings in a different study. In the interest of avoiding parameter degeneracies with $\Upsilon(r)$ and saving computational time, we set $\gamma$ to zero and $r_{s}$ to a large value outside the spatial coverage of our kinematic data (in this case $\sim \SI{90}{kpc}$, the average best-fit $r_s$ of our preliminary models). Therefore, we only model one parameter for the DM halo, $\rho_{10}$.

The stellar mass-density distribution is tied to the three dimensional deprojection $\nu(r, \theta)$ of photometric imaging, as detailed in Section \ref{subsec:lightDens}, 
via $\Upsilon(r)$, 
\begin{equation}
    \rho_{\star}(r, \theta) = \Upsilon(r) \cdot \nu(r, \theta),
\end{equation}
 with $\nu(r, \theta)$, the 3D light density distribution, which is not a model parameter, but a constraint -- it is fixed to the profiles derived from imaging data. Furthermore, our implementation allows for the modeling of multiple morphological components with separate $\Upsilon(r)$ (e.g. \citealt{Nowak2010}, E+18). Therefore, for the fast rotating power-law galaxies, NGC~307, NGC~1332, and NGC~4751 we use a photometric decomposition to distinguish a bulge and a disc component.
 These are deprojected separately and have their own separate $\Upsilon_{\mathrm{bulge}}$ and $\Upsilon_{\mathrm{disc}}$. Since the disc components fade into DM-dominated regions at larger radii and are outshone by the bulge components in the center, they are locally less well constrained and we decided to fit the disc components without gradients $\Upsilon_{\mathrm{disk}}(r) \xrightarrow{} \Upsilon_{\mathrm{disk}}$. We fit the bulge components with gradients as with the cored ETGs, $\Upsilon_{\mathrm{bulge}}(r)$. 
 \begin{figure}
\includegraphics[width = 0.9\columnwidth]{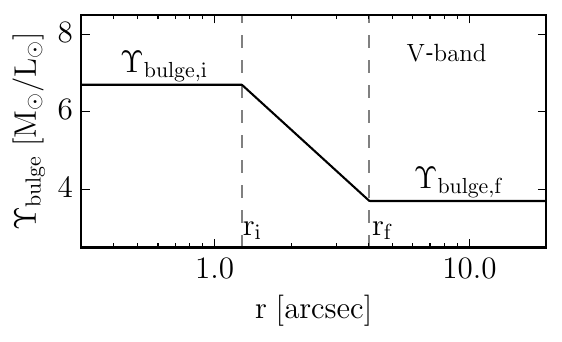}
 \caption{Example of our implementation of $\Upsilon_{\mathrm{bulge}}$ gradients from the dynamical modeling of NGC~5328. We show best-fit values for one spatial quadrant of the galaxy (see Section \ref{subsec:kinematics}).
 $r_i$ is fixed to the MUSE PSF $= \SI{1.28}{\arcsec}$, whereas $r_f$, $\Upsilon_{\mathrm{bulge, i}}$, $\Upsilon_{\mathrm{bulge, f}}$ are fit parameters. 
 }
   \label{fig:exampleGradient}
\end{figure}
 
Our implementation of mass-to-light ratio radial profiles operates by two values $\Upsilon_{i,f}=\Upsilon(r_{i,f})$ at different distances from the center of the galaxy, $r_{i,f}$.
We show an example of this implementation in Figure \ref{fig:exampleGradient}. Between $r_i$ and $r_f$, $\Upsilon_{\mathrm{bulge}}(r)$ is linearly interpolated over $\log(r)$. Outside $r_{i}$ and $r_f$, $\Upsilon_{\mathrm{bulge}}(r) = \Upsilon_{\mathrm{bulge}, i}$ for $r < r_i$, and  $\Upsilon_{\mathrm{bulge}}(r) = \Upsilon_{\mathrm{bulge}, f}$ for $r > r_f$.

Here, however, we face two challenges in particular: At both small and large radii, mass-contributions from the stars become much more difficult to differentiate from those of the ``dark'' components, i.e. the central SMBH and DM halo. Per definition, within the SOI of the central SMBH, the enclosed stellar mass is less than $M_{\mathrm{BH}}$. Towards the center then, $\Upsilon(r)$ becomes overshadowed by $M_{\mathrm{BH}}$ in terms of its impact on the observed stellar kinematics. In the opposite direction, with increasing distance from the galactic center, as the luminous component of the galaxy becomes ever fainter and the DM halo more dominant, it becomes more difficult to determine $\Upsilon$ locally.

Therefore, after trying a number of different approaches for the galaxies, we settled on the following setup: We defined the inner value $\Upsilon_{\mathrm{bulge}, i}$ at $r_i$ equalling one full width at half maximum (FWHM) of the point spread function (PSF) of the MUSE stellar kinematics (see the second column of Table \ref{tab:generalTab}) and the outer value, $\Upsilon_{\mathrm{bulge}, f}$ at a radius $r_{f}$, which in the fit is restricted to an interval between two times the FWHM of the PSF and two thirds of the MUSE FOV, i.e. up to $r = \SI{20}{\arcsec}$. Beyond this radius, the above mentioned \textcolor{black}{problem} with differentiating between DM and stellar mass contributions becomes too acute for a measurement.  We also do not add another $\Upsilon_{\mathrm{bulge}, j}$ inside the PSF, instead keeping $\Upsilon_{\mathrm{bulge}}$ constant,
 $\Upsilon_{\mathrm{bulge}}(r)= \Upsilon_{\mathrm{bulge}, i}$, for $r < r_i =$ PSF since the AO-supported SINFONI data which cover these spatial scales generally have a much lower SNR than our MUSE data (see Table \ref{tab:generalTab} and Section \ref{subsec:kinematics} below). 

For NGC~307, the spatial extent of the bulge component is too small, $r_{e, bulge} \sim \SI{2}{\arcsec}$, to warrant gradient models in our approach. Therefore, for this galaxy alone we set $\Upsilon_{\mathrm{bulge}, i} \equiv \Upsilon_{\mathrm{bulge}, f}$. 

Together with contributions from DM and the SMBH, and accounting for disk and bulge components where necessary, we fit a total of four to six parameters, depending on the galaxy: $M_{\mathrm{BH}}$,$\Upsilon_{\mathrm{bulge}, i}$,$\Upsilon_{\mathrm{bulge}, f}$, [$\Upsilon_{\mathrm{disk}}$,] $r_{f}$ and $\rho_{10}$.

\subsubsection{Model selection and Non-parametric LOSVD fits}
\label{subsec:modelselection}
Our modeling optimization entails sifting through different sets of ($M_{\mathrm{BH}}$,$\Upsilon_{\mathrm{bulge}, i}$,$\Upsilon_{\mathrm{bulge}, f}$, [$\Upsilon_{\mathrm{disk}}$,] $r_{f}$, $\rho_{10}$) with the optimization software NOMAD \citep{Audet2006, Ledigabel2011, Audet2017}  and
computing orbit libraries in the associated gravitational potentials $\Phi$($M_{\mathrm{BH}}$,$\Upsilon_{\mathrm{bulge}, i}$,$\Upsilon_{\mathrm{bulge}, f}$, [$\Upsilon_{\mathrm{disk}}$,] $r_{f}$, $\rho_{10}$). For each $\Phi$, tens of thousands of orbits, which are assigned individual weights, are generated from different ($E$, $L_{z}$, $I_{3}$). The Schwarzschild modeling code then optimizes these weights by maximizing 

\begin{equation}
\hat{S} = S - \hat{\alpha} \cdot \chi^2,
\label{eq:entropy}
\centering
\end{equation}
where $\chi^2$ is calculated from the model fit to the observed non-parametric LOSVDs, and $S$ is the Boltzmann entropy \citep{Thomas2004}.
The deprojected light-distributions are used as a constraint.

The parameter $\hat{\alpha}$ constitutes the smoothing of the models. \citet{Lipka2021} have shown that an optimal determination of $\hat{\alpha}$ is required for an unbiased dynamical recovery of the internal mass parameters. This can be achieved by taking the so-called effective degrees of freedom, $m_{\mathrm{eff}}$, a generalised measure of the degrees of freedom in a penalized system, into account. 
To that end, we minimize the generalized Akaike Information Criterion $AIC_p = \chi^2 + 2 \times m_{\mathrm{eff}}$ for penalized likelihood models \citep{ThomasLipka2022} over a grid of $\hat{\alpha}$ values. 

After determination of the optimal $\hat{\alpha}$-value for the current $\Phi$ the associated minimum AIC$_p$ value is passed to NOMAD. NOMAD minimizes the AIC$_p$ until the optimal ($M_{\mathrm{BH}}$,$\Upsilon_{\mathrm{bulge}, i}$,$\Upsilon_{\mathrm{bulge}, f}$, [$\Upsilon_{\mathrm{disk}}$,] $r_{f}$, $\rho_{10}$) to fit the LOSVDs is found. This approach not only optimizes the smoothing in each trial potential but also takes into account that the mass optimisation in Schwarzschild models is actually a model selection problem rather than a simple parameter estimation \citep{Lipka2021}. The model selection allows for very accurate and unbiased mass- and anisotropy-recoveries \citep{Lipka2021, Neureiter2023a, deNicola2022}.

\subsection{Galaxy light density profiles}
\label{subsec:lightDens}
The 3D light distribution in our dynamical models, $\nu(r,\theta)$, is constrained by -- or rather fixed to -- deprojections of 2D imaging data of the galaxies along the line-of-sight. We here re-use the imaging data, bulge/disc decompositions (where applicable) and deprojections from the studies which are listed in the last column of Table \ref{tab:generalTab}, with one exception, NGC~4751. 

\textcolor{black}{For the power-law galaxies, the inclination $i$ was assumed from the flattening of their discs at large radii (for an assumed intrinsic flattening $q=0.2$): $i = 75$ for NGC~307 (E+18) and $i = 90$ for both NGC~1332 and NGC~4751. For the four disc-less cored galaxies, we assumed $i = 90$. Axisymmetric Schwarzschild models of realistic triaxial N-body simulations of core galaxies suggest that even using the AIC$_p$ optimization technique, the models often fit the galaxies best at $i=90$. These tests further suggest that the bias of the mass-to-light ratio that can arise from the assumption of axial symmetry (and $i=90$) is on the order of $15\%$ (Lipka et al. in prep).}

All galaxies, including NGC~4751, have been assumed to be close-to or directly edge-on for the deprojections,based on their flattening at large radii

For NGC~4751, we performed a new disc/bulge decomposition (as none has been performed in R+13) based on the same HST NICMOS2 images we used in R+13, combined with K-Band observations with VIRCAM \citep{Vista1, Vista2}. We followed the same steps and approach as for the other galaxies to produce the disc/bulge decomposition and separate deprojections for both components. This is outlined in Appendix \ref{ap:ngc4751phot}.

\subsection{Non-parametric stellar kinematics}
\label{subsec:kinematics}
{\textbf{MUSE data}}:
The MUSE stellar kinematics of our sample were the result of the first systematic study of the detailed non-parametric shapes of the LOSVD of massive ETGs, which we published in \citet{Mehrgan2023}, from here on M+23. They were derived using the new non-parametric spectral fitting code WINGFIT (Thomas et al. in prep.), which also uses the data-driven AIC$_p$-optimisation technique of \citet{ThomasLipka2022}.
The details of the observations, derivation of the kinematics from them, as well as the resulting kinematics are presented in M+23.

The MUSE non-parametric LOSVDs are the main input for our orbital dynamical models: They cover a large $\SI{1}{\arcmin} \times \SI{1}{\arcmin}$ field of view (FOV), encompassing half to a full effective radius $r_{e}$ for each galaxy in our sample. 
Furthermore, \textcolor{black}{the data were Voronoi binned using the Voronoi tessellation method of \citet{Voronoi2003} for a very high SNR/$ \angstrom > 100$ (as described in M+23)}.

\textcolor{black}{For the dynamical models, we split the MUSE FOV into quadrants along the major and minor axes of each galaxy to ensure that we can 
provide a robust estimation of the error bars  of the best-fit model parameters from the scatter between the quadrants.} This resulted in roughly $15 - 100$ spatial bins per quadrant per galaxy, each with its own non-parametric LOSVD.

We sampled the LOSVDs either with $N_{vel} = 15$ velocity bins out to $\SI{1500}{km/s}$, or $N_{vel} = 17$ out to $\SI{1700}{km/s}$, depending on where the LOSVDs of each galaxy terminate \footnote{The sole exception here being NGC~307, the least massive ETG in our sample. Here, the LOSVDs terminate at $\sim \pm \SI{1000}{km/s}$, and we used 21 velocity bins, to properly sample its much narrower distribution function}. Therefore, all in all, we end up with roughly 225 to 1500 kinematic MUSE-data points per galaxy per quadrant for our dynamical models.

{\textbf{SINFONI data}}:
For the central regions of the galaxies, we also supply our dynamical models with non-parametric SINFONI stellar kinematics. These kinematics were derived earlier using the maximum pealized likelihood method (MP) from \citet{Gebhardt2000}.

\textcolor{black}{The SINFONI data was in binned into radial and angular segments as in \citet{Rusli2013a}. In Table \ref{tab:generalTab}, we list the SNR achieved with this binning. For the details surrounding the observations, binning, and kinematics, we refer to the studies listed in the last column of Table \ref{tab:generalTab}.}

Though covering a much smaller FOV, $\SI{3}{\arcsec} \times \SI{3}{\arcsec}$, corresponding to the $\SI{100}{mas}$-mode of SINFONI, these LOSVDs, which are adaptive-optics based, and thus not seeing limited, supply our models with vital constraints on the central mass-light profile of the galaxies as they can resolve the gravitational SOI of their central SMBHs (on a scale of $\lesssim \SI{1}{\arcsec}$). \textcolor{black}{For these data we supply the PSF in the form of 2D images to the dynamical models. The images typically have a FWHM around $\sim \SI{0.15}{\arcsec}$}.

We sampled the LOSVDs in the same way as the MUSE LOSVDs, resulting in $\sim 300 - 500$ kinematic data points per galaxy per quadrant for our dynamical models ($\sim 1000$ in the case of NGC~1332).

\begin{figure}[!t]
\centering
 \includegraphics[width=0.7\columnwidth]{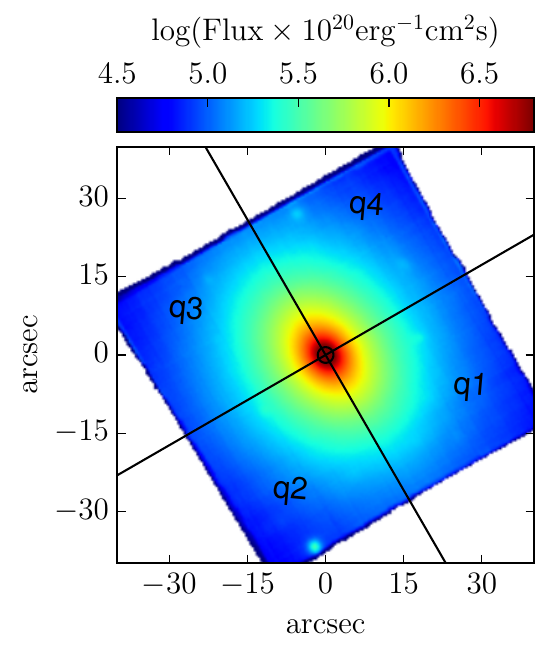}
 \includegraphics[width=1.\columnwidth]{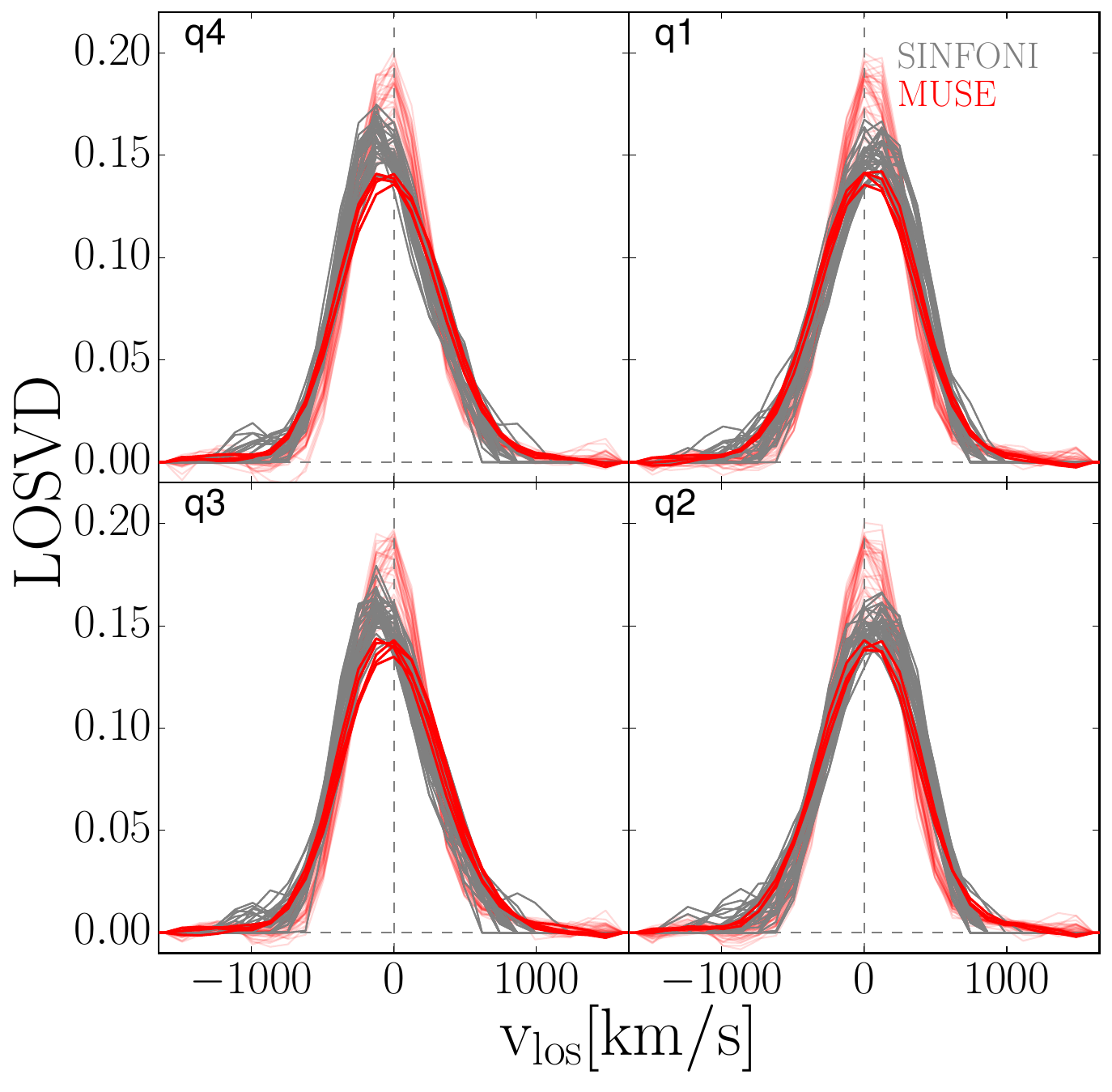}
 \caption{Example of the kinematic data used in this study. Top: Flux map of the MUSE data (north is up, east is left). Black lines show the major axis (position angle $ = 30^{\circ}$) and minor axis. The small black circle in the center ($r = \SI{1.5}{\arcsec}$) indicates the SINFONI FOV. Bottom: Non-parametric LOSVDs from MUSE (red) and SINFONI (grey) for NGC~7619, separated into the spatial quadrants indicated in the top panel. MUSE-LOSVDs from bins that spatially overlap with the SINFONI FOV are shown in solid red, whereas all other LOSVDs are shown in a fainter red. The stellar rotation of NGC~7619 increases towards the center (Figure~15 of M+23). This trend continues into the spatial regions resolved by SINFONI (i.e. the SINFONI LOSVDs are more strongly shifted in $\pm v_{\mathrm{los}}$.)}
   \label{fig:exampleLOSVDs}
\end{figure}

{\textbf{Combining the kinematic data}
In Figure \ref{fig:exampleLOSVDs} we show, as an example, all the LOSVDs of NGC~7619, including both MUSE and SINFONI LOSVDs, divided into quadrants. For the dynamical models we also include LOSVDs from MUSE which spatially overlap with those from SINFONI.}

\subsection{Approach to deriving results}
\label{subsec:approach}

We compute at least 2500 models per quadrant. The best-fit model parameters in terms of AIC$_p$, as well as the associated mass profiles, including $\Upsilon(r)$, are averaged over all quadrants to produce one final set of model parameters and mass distribution per galaxy.

For NGC~1332, an independent black hole mass measurement was available from direct observation of the circumnuclear disk in the central $\SI{200}{pc}$ of the galaxy \citet{Barth2015}, $M_{\mathrm{BH}} = 6.64(-0.63,+0.65) \times 10^8 M_{\odot}$.
We had previously dynamically determined a larger $M_{\mathrm{BH}}$ using Schwarzschild models in R+13. However the measurement from \citet{Barth2015} have a much higher spatial resolution of $\SI{0.044}{\arcsec}$ (versus $\sim \SI{0.15}{\arcsec}$) and are derived from the kinematics of a cold disk within the SOI of the central SMBH -- a simpler dynamical problem than our own models. 
Therefore we fixed $M_{\mathrm{BH}}$ for this galaxy to the measured value from \citet{Barth2015} and only varied the other model parameters to get better constraints on the central $\Upsilon (r)$.

For both NGC~1332, and NGC~1407 we had an especially large number of spatial bins available, with well over a 120 MUSE+SINFONI LOSVDs per Quadrant. The same assumption of axisymmetry that allowed us to split our dynamical models into quadrants and model those quadrants as ``separate'' galaxies, over which we average for the final results, allow us to sort all spatial bins in a quadrant according to radius and then group together every second spatial bin as a sub-quadrant to be modeled independently. Hence for these two galaxies, we model and average over eight instead of four dynamical best-fit models (for each sub-quadrant we also run at least 2500 models), which allows us to better sample the statistical uncertainties.

 We here treat the values of $\Upsilon_{\mathrm{bulge}, i,f}$ listed in Table \ref{tab:resultsTab} as \textit{nuisance} parameters and not as the primary measures of the gradients which we detect: \textcolor{black}{Firstly}, if two photometric components are present, as is the case for NGC~307, NGC~1332 and NGC~4751, the final gradient $\Upsilon(r)$ emerges from the superposition of the light profiles of the bulge- and disk-components times their respective $\Upsilon$-profiles, divided by the total light. In the case of NGC~1332 and NGC~4751, this produces a much more complex $\Upsilon (r)$ profile than for the bulge-component alone (for NGC~307, the gradient only emerges from the superposition of two constant-$\Upsilon$ components). Second, we take our $\Upsilon$ profiles as the average over the individual (sub-)quadrants of each galaxy at each radius. The resulting average profiles can be more complex than the parametric profiles of the individual quadrants. 
 
 Furthermore, for better comparison with stellar population models we project $\Upsilon$ along the line-of-sight. However, $\Upsilon$ as an intended purely {\it stellar} mass component, depends on assumptions in the mass decomposition. This is not so much of a concern in regions in the center that are at the same time still outside the SOI. Here $\Upsilon$ is essentially identical to the total inner dynamical mass-to-light ratio, $(M^{\mathrm{tot}}/L)(r)$, as the local mass-contribution of the DM-component is essentially drowned out by the stellar component. For all galaxies in our sample, except one (NGC~1407, see Section \ref{subsec:axisymm}), the SOI is very small compared to the innermost radius of our gradient-models, $r_i/SOI \gtrsim 3$.
 
 However, on scales of $0.5 - \SI{1}{kpc}$ from the center, $(M^{\mathrm{tot}}/L)(r)$ starts to diverge from $\Upsilon (r)$ because DM begins to assert more influence on the dynamics of the stars and $(M^{\mathrm{tot}}/L)(r)$ rises relative to $\Upsilon (r)$. At this point, disentangling DM from stars becomes more and more difficult and the derived $\Upsilon(r)$ will depend on the assumptions about DM (and vice versa).

 In order to overcome the difficulty related to the mass decomposition in the outer parts, we try to determine the stellar $\Upsilon(r)$ focusing entirely on spatial scales where $\Upsilon \sim M^{\mathrm{tot}}/L $, i.e. where the {\it stellar} $\Upsilon$ is least dependent on any assumption upon the mass decomposition. It turns out that this is possible, because the stellar dynamical gradients all fall very quickly with galactocentric radius (see next Section) and at larger radius the DM halo ``takes over''. As a consequence, the $(M^{\mathrm{tot}}/L)(r)$ profiles are effectively valley-shaped (see next Section and Figure~\ref{fig:MLgradsA}), with a global minimum in between the two regimes. This minimum is not only a characteristic property related to the central gradients but it is also key to determine the stellar mass-to-light ratio in the main body of the galaxies in a way that depends only little on the assumed DM profile: under the only assumption that the stellar mass-to-light ratio does not increase towards the outer parts, the minimum in the total $(M^{\mathrm{tot}}/L)(r)$ is the point of strongest constraint for the stellar mass-to-light ratio in the main body of the galaxy. More specifically, it sets an upper limit for this ratio. We therefore treat the stellar mass-to-light ratio $\Upsilon_{\mathrm{main}} = \Upsilon(r_{\mathrm{main}})$ associated to the radius $r_{\mathrm{main}}$ where the minimum in $(M^{\mathrm{tot}}/L)(r)$ occurs as the mass-to-light ratio of the galaxy main body. For the central stellar mass-to-light ratio, we define $\Upsilon_{\mathrm{cen}}$ simply as $\Upsilon(r)$ within the MUSE PSF ($r_{\mathrm{cen}} = r_i = PSF$). 

\section{Results}
\label{sec:results}
\begin{table*}[!ht]
\centering
 \begin{tabular}{l c c c c c c c c c c}
 \hline
 \hline
Galaxy            & Band & $\Upsilon_{\mathrm{bulge}, i}$ & $\Upsilon_{\mathrm{bulge}, f}$ & $r_f$ & $\Upsilon_{\mathrm{disk}}$ & $M_{\mathrm{BH}}$ & $\rho_{10}$ & $<(\chi^2+m_{\mathrm{eff}})/N>$\\
& &$\mathrm{[M_{\odot}/L_{\odot}]}$ &$\mathrm{[M_{\odot}/L_{\odot}]}$ & [$kpc$]& $\mathrm{[M_{\odot}/L_{\odot}]}$& [$\mathrm{10^9 \ M_{\odot}}$] & [$\mathrm{10^8 \ M_{\odot}/kpc^3}$] & \\
\hline
NGC~307   & K &   $1.13 \pm 0.04$ &  $1.13 \pm 0.04$  & $-$          & $0.63 \pm 0.27$ & $0.22 \pm 0.04$               & $1.9 \pm 0.4$ & $0.72$ \\
NGC~1332   & R &   $8.50 \pm 0.83$ &  $1.38 \pm 0.89$  & $1.0 \pm 0.4$             & $2.31 \pm 1.62$ & $0.66^{*}$        & $5.0 \pm 0.9$ & $0.76$\\
NGC~1407   & B &   $11.10 \pm 2.86$ &  $1.04 \pm 1.08$ & $1.2 \pm 0.5$            & $-$             & $5.50 \pm 1.58$    & $2.7 \pm 0.4$ & $0.78$\\
NGC~4751   & K &   $2.17 \pm 0.24$ &  $0.83 \pm 0.47$  & $1.3 \pm 0.6$             & $0.58 \pm 0.13$ & $1.75 \pm 0.34$   & $5.5 \pm 0.6$ & $1.39$\\
NGC~5328   & V &   $6.57 \pm 0.51$ &  $4.28 \pm 0.43$  & $1.9 \pm 1.4$            & $-$             & $1.63 \pm 0.89$     & $1.1 \pm 0.2$ & $0.99$\\
NGC~5516   & R &   $6.16 \pm 0.60$ &  $2.83 \pm 1.91$  & $2.9 \pm 1.8$              & $-$             & $2.50 \pm 0.53$     & $0.8 \pm 0.1$ & $0.95$\\
NGC~7619   & I &   $4.00 \pm 0.79$ &  $2.00 \pm 1.28$  & $2.0 \pm 0.6$            & $-$             & $3.25 \pm 1.40$     & $0.8 \pm 0.2$ & $0.62$\\

\hline
\end{tabular}
\caption{Results of Schwarzschild dynamical modeling using mass-to-light gradient models. Photometric bands, as well as extinction corrections for $\Upsilon$-values for all galaxies were taken over from R+11, R+13, and E+18, according to Table \ref{tab:generalTab}, except for NGC~4751. Model parameters are averages with standard deviations over all quadrants or sub-quadrants. $\Upsilon_{\mathrm{bulge}, i}$ was fitted at a set radius $r_i = $ PSF, whereas $r_f$, the radius of $\Upsilon_{\mathrm{bulge}, f}$ was a free parameter in the fit. We also list (sub-)quadrant averages of ($\chi^2+m_{\mathrm{eff}})/N$ of the fits to the non-parametric LOSVDs. (*) For NGC~1332, we used the $M_{\mathrm{BH}}$-value measured by \citet{Barth2015} as a fixed parameter.}

\label{tab:resultsTab}
\end{table*}

\begin{table*}
\centering
 \begin{tabular}{l c  c c c c c c}
 \hline
 \hline
Galaxy     &  $\Upsilon_{\mathrm{cen}}$    &  $\Upsilon_{\mathrm{main}}$  & $(M^{\mathrm{tot}}/L)_{\mathrm{main}}$  & $r_{\mathrm{main}}$ & $\alpha_{\mathrm{cen}}$ & $\alpha_{\mathrm{main}}$ & $\alpha^{\mathrm{tot}}_{\mathrm{main}}$ \\

     &  $\mathrm{[M_{\odot}/L_{\odot}]}$   &  $\mathrm{[M_{\odot}/L_{\odot}]}$ & $\mathrm{[M_{\odot}/L_{\odot}]}$ & [kpc] & & &  \\

\hline
NGC~307   & $4.65 \pm 0.04$ &  $4.12 \pm 0.09$ & $4.64 \pm 0.04$ & $1.0$ & $1.23 \pm 0.06$ & $1.23 \pm 0.06$ & $1.23 \pm 0.06$ \\
NGC~1332   & $9.69 \pm 0.46$ &  $2.54 \pm 0.34$ & $3.72 \pm 0.3$ & $0.9$ & $2.20 \pm 0.26$ & $0.59 \pm 0.20$ & $0.86 \pm 0.14$ \\
NGC~1407   & $7.69 \pm 1.54$ &  $1.29 \pm 0.71$ & $2.00 \pm 0.61$ & $1.1$ & $1.76 \pm 0.45$ & $0.30 \pm 0.18$ & $0.47 \pm 0.17$ \\
NGC~4751   & $9.54 \pm 0.25$ &  $4.12 \pm 0.22$ & $5.40  \pm 0.25$ & $0.8$ & $2.35 \pm 0.26$ & $1.246 \pm 0.28$ & $1.63 \pm 0.32$ \\
NGC~5328   & $6.56 \pm 0.11$ &  $4.54 \pm 0.13$ & $4.88 \pm 0.12$ & $1.3$ & $1.62 \pm 0.13$ & $1.12 \pm 0.15$ & $1.21 \pm 0.14$ \\
NGC~5516   & $7.83 \pm 0.16$ &  $4.66 \pm 0.43$ & $5.00 \pm 0.42$ & $1.3$ & $1.90 \pm 0.19$ & $1.22 \pm 0.51$ & $1.29 \pm 0.50$  \\
NGC~7619   & $5.81 \pm 0.29$ &  $3.96 \pm 0.42$ & $4.12 \pm 0.42$  & $1.1$ & $1.55 \pm 0.31$ & $0.92 \pm 0.37$ & $0.97 \pm 0.36$\\
\hline
\end{tabular}
\caption{V-band stellar and total mass-to-light ratios, $\Upsilon$, $M^{\mathrm{tot}}/L$ measured from our best-fit dynamical models. These values are projected along the line-of-sight for later comparison with SSP models. Therefore, these values should not be confused with the modeling-parameters in Table \ref{tab:resultsTab}. Inner $\Upsilon_{\mathrm{cen}}$-values are essentially identical to inner $(M^{\mathrm{tot}}/L)_{\mathrm{cen}}$. The stellar mass-to-light ratios of the main body of the galaxy, $\Upsilon_{\mathrm{main}}$ are defined at the global minimum of $(M^{\mathrm{tot}}/L)(r)$, $M^{\mathrm{tot}}/L(r_{\mathrm{main}}) = min(M^{\mathrm{tot}}/L) = (M^{\mathrm{tot}}/L)_{\mathrm{main}}$, for all galaxies except NGC~307, where we manually set $r_{\mathrm{main}} = \SI{1}{kpc}$. Finally we also show these mass-light values relative to the stellar mass-to-light ratio assuming a Kroupa IMF for these galaxies from the SSP analysis of Parikh et al. submitted to MNRAS, in the form of the excess parameter $\alpha$. }
\label{tab:gradientsTab}
\end{table*}

\begin{figure*}
\centering
\includegraphics[width = 1.7\columnwidth]{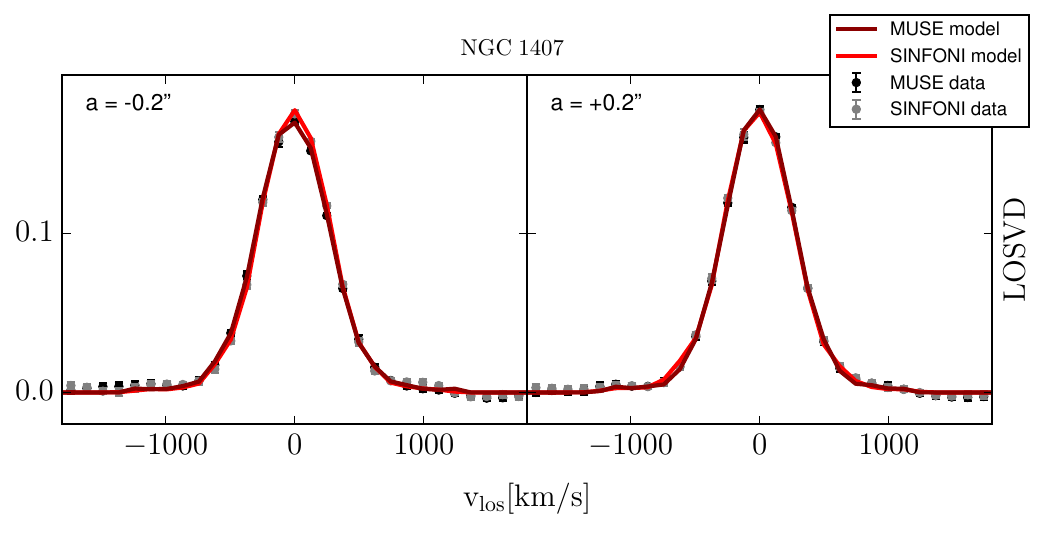}
 \caption{Dynamical fits to the non-parametric LOSVDs from the center of NGC~1407. We show fits at two positions near the major-axis $a$ on opposite sites of the minor axis of the galaxy. MUSE and SINFONI data spatially overlap at these positions. The MUSE data LOSVDs with statistical uncertainties are shown as black points with error bars, whereas the respective best-fit dynamical model is shown as a solid dark red line. Analogously, we show the SINFONI data and model LOSVDs in grey and light red. The best-fit LOSVDs are here shown at the full MUSE velocity resolution.}
   \label{fig:Losvdfitn1407}
\end{figure*}

The best-fit model parameters for all galaxies are listed in Table \ref{tab:resultsTab}.  The best-fit models have on average $\chi^2/N \sim 0.6$ over all (sub-)quadrants. Such low $\chi^2/N$ values for best-fit models have long been typical for Schwarzschild models, due to the large number of the degrees of freedom involved. Taking the effective degrees of freedom, $m_{\mathrm{eff}}$ into account, $(\chi^2 + m_{\mathrm{eff}})/N \sim 0.9$ (see last column of Table \ref{tab:resultsTab}). The remaining difference between $(\chi^2 + m_{\mathrm{eff}})$ and $N$ likely originates from covariances between the individual velocity bins of the LOSVDs. 

For all intents and purposes our $(\chi^2 + m_{\mathrm{eff}})/N$ values demonstrate that our dynamical models produced good fits to the kinematic data -- At least for all galaxies except NGC~4751. Here $(\chi^2 + m_{\mathrm{eff}})/N \sim 1.4$ was larger than for the other galaxies, due to the presence of dust-lanes covering almost the entirety of the major axis within $r_e$ (see Appendix \ref{ap:ngc4751phot}) We also had to exclude one quadrant entirely for this galaxy as we could not find a good fit to the data $(\chi^2 + m_{\mathrm{eff}})/N \sim 3$. We treat the results for this galaxy with some added caution. This is discussed later in Section \ref{subsec:kinproblems}.

We show one example-fit to central LOSVDs of NGC~1407 in Figure \ref{fig:Losvdfitn1407}. LOSVD- and radial kinematic fits for all galaxies are included in Appendix \ref{ap:kinfits}.

We show AIC$_p$ model selection curves converging on the best-fit parameters of the (sub-)quadrants of the galaxies in Figure \ref{fig:AICEnvelopes}. 

\begin{figure*}
\centering
\hspace{9mm}\includegraphics[width=\columnwidth]{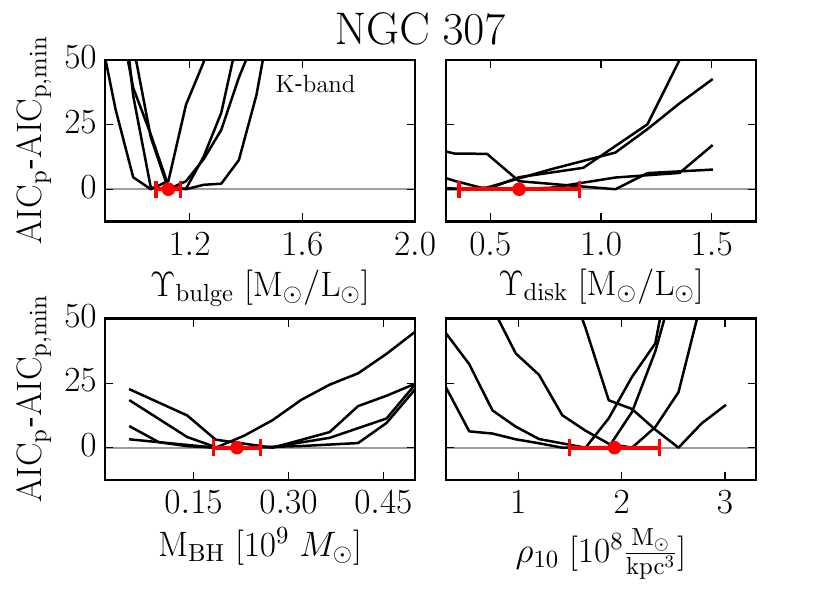}
\hspace{-9mm}\includegraphics[width=\columnwidth]{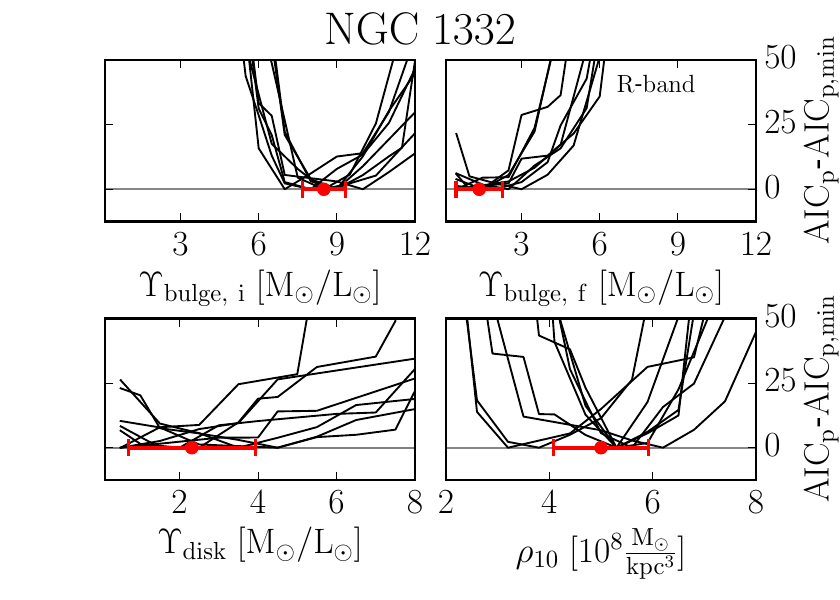} \\
\hspace{9mm}\includegraphics[width=\columnwidth]{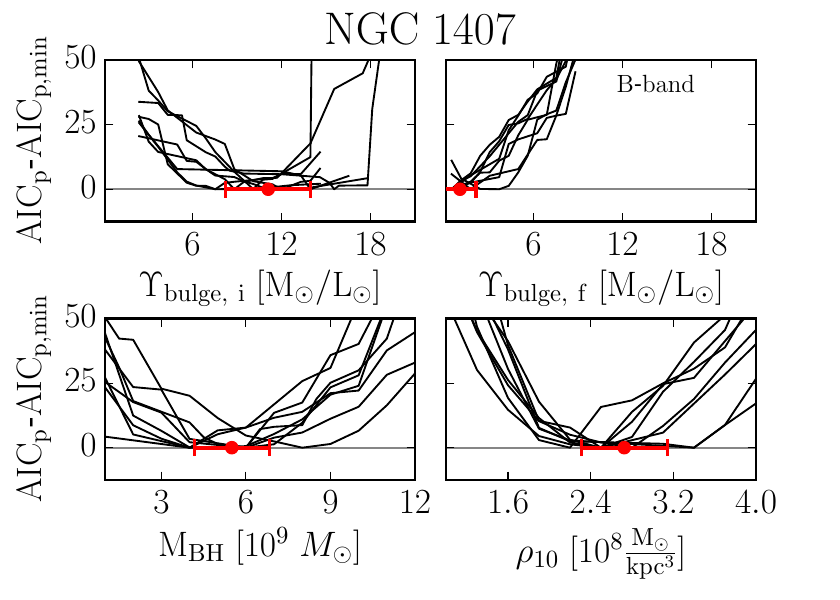}
\hspace{-9mm}\includegraphics[width=\columnwidth]{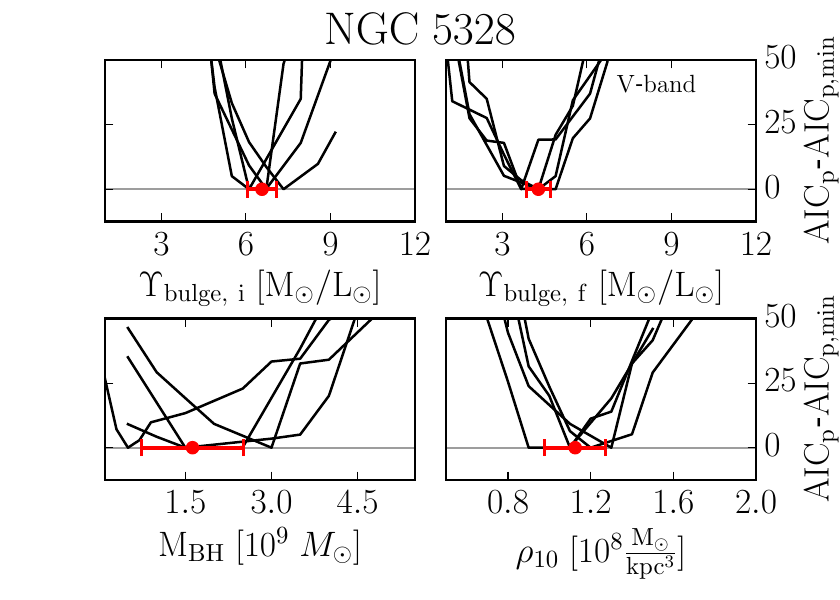} \\
\hspace{9mm}\includegraphics[width=\columnwidth]{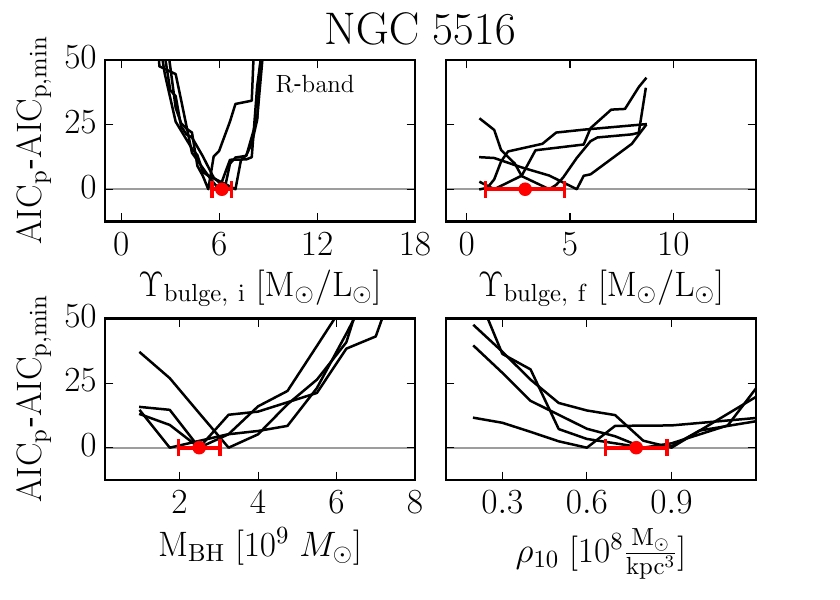}
\hspace{-9mm}\includegraphics[width=\columnwidth]{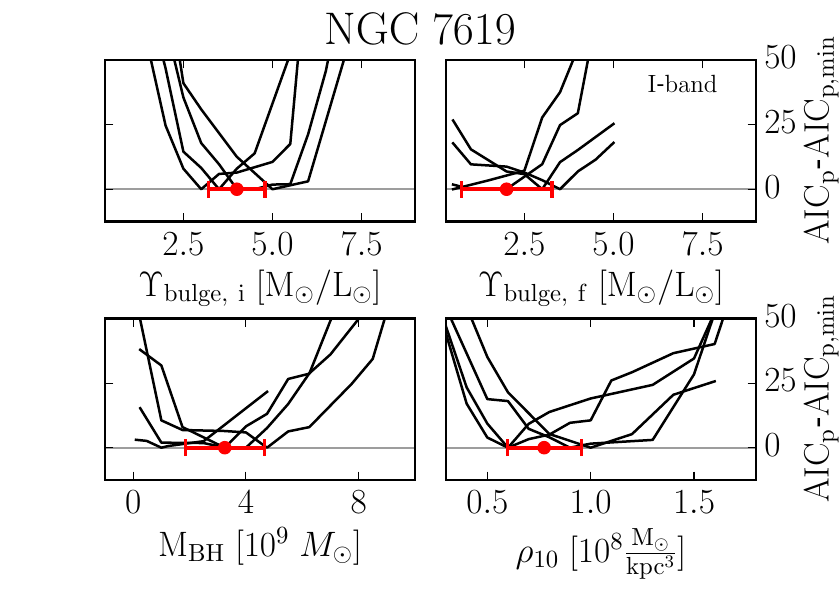}

 \caption{AIC$_p$ model selection curves, the equivalent of the classical $\chi^2$ curves for Schwarzschild models. Each curve represents the (independent) modelling result of one quadrant/sub-quadrant and is derived from the lowest AIC$_p$ value of each sampling point of the corresponding modeling parameter. The best-fit values (red points with errorbars) are determined at the minima of the AIC$_p$ and the variation between the AIC$_p$ minima represents the statistical uncertainties of each measurement for each galaxy.}
   \label{fig:AICEnvelopes}
\end{figure*}
\addtocounter{figure}{-1}
\begin{figure*}[!htbp]
\centering
\includegraphics[width=1.5\columnwidth]{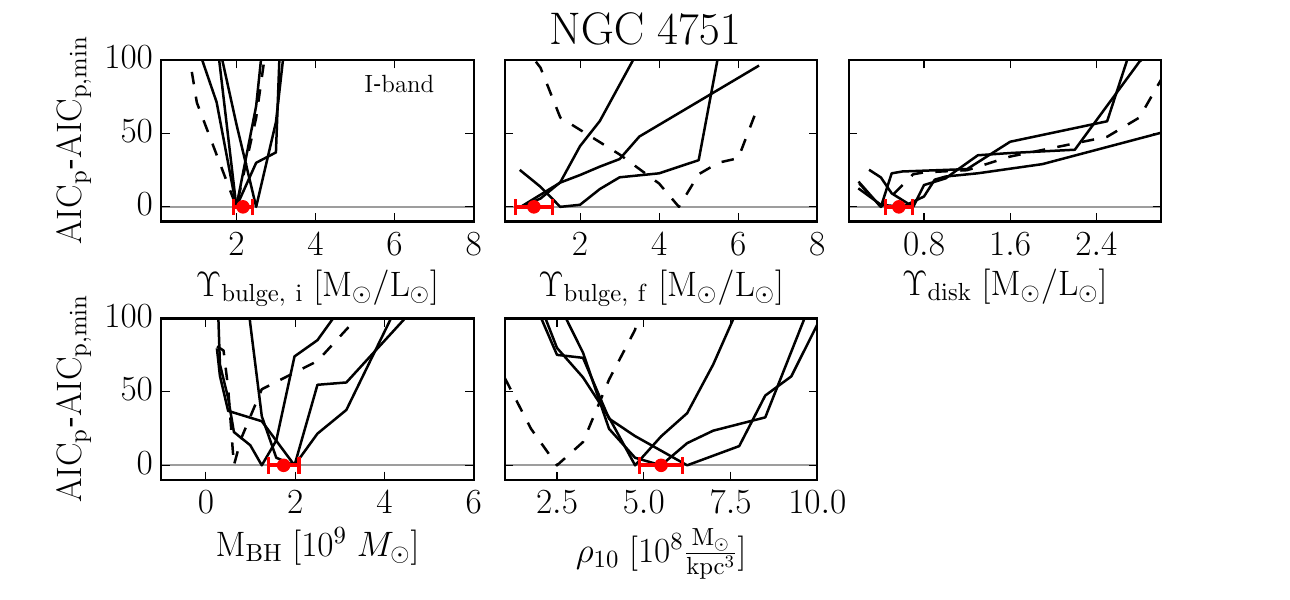}
 \caption{(continued) For NGC~4751 one of the quadrants (dashed curve), was not used for the calculation of the best-fit parameters or any other modeling results.}
\end{figure*}

In the following we \textcolor{black}{examine} the mass-to-light ratio gradients $\Upsilon(r)$ and discuss the effect of gradients $M_{\mathrm{BH}}$ measurements.

\subsection{Mass-to-light ratio gradients}

The main result of our study is that we have found stellar dynamical evidence in favour of radial gradients of the stellar mass-to-light ratio, $\Upsilon(r)$ for all galaxies in our sample. These gradients are confined to the very centers of the galaxies and occur on spatial scales of $r \sim \SI{1}{kpc}$. For all galaxies, $\Upsilon$ becomes larger towards the center of the galaxy (Figure~\ref{fig:MLgradsA}). 

Moreover, in all our galaxies a well-defined global minimum of the total dynamical mass-to-light ratio $(M^{\mathrm{tot}}/L)(r)$ occurs. We call the radius where this minimum occurs $r_{\mathrm{main}}$. As explained above, the mass-to-light ratio at this radius poses strong constraints on the mass-to-light ratio of the stars in the main body of the galaxy largely independent of the detailed assumptions upon the mass decomposition.

NGC~307 is an exception since the galaxy does not show a minimum in $(M^{\mathrm{tot}}/L)$. Here we set $r_{\mathrm{main}} \sim \SI{1}{kpc}$, which  coincides roughly with the point where $(M^{\mathrm{tot}}/L)(r)$ begins to rise from the center.

For a few individual (sub-)quadrants of the galaxies the AIC$_p$-curves of the outer $\Upsilon_{\mathrm{bulge}, f}$, (and/or $\Upsilon_{\mathrm{disk}}$) did not converge to a minimum, but instead hit the lower boundary of our sampling range. This amounts to the mass contribution of the DM component displacing mass contribution of the stellar component, and $\Upsilon_{\mathrm{bulge}, f}$ getting as close to zero as our models allow. As explained, this does not concern us since $r_{\mathrm{main}} < r_{f}$ (cf. Tables \ref{tab:resultsTab} and \ref{tab:gradientsTab}) for all galaxies, and in our approach we focus on the parts of the galaxies least affected by DM, while treating the mass decomposition past $r_{\mathrm{main}}$ as a curtain we do not look behind -- the dynamical mass of our models can reproduce the kinematics in this region without us knowing the details of the mass decomposition.

For the gradient-plots in Figure \ref{fig:MLgradsA}, we normalized all gradients relative to $\Upsilon_{\mathrm{main}}$ to illustrate by how much the stellar mass-to-light ratio appears to increase in the centers of the individual galaxies.

For the four core galaxies in our sample we supplement our gradient models with models that assume a spatially constant stellar $\Upsilon$ both as a consistency check and for better comparison with previous measurements (Appendix~\ref{ap:constant}). These models without gradients were worse fits to the kinematic data for all (sub-)quadrants and galaxies. Compared to their counterparts with gradients the $\Delta$AIC$_p \sim 10 - 20$ is significant. In general, the best-fit $\Upsilon$ derived from models without a gradient lie between $\Upsilon_{\mathrm{cen}}$ and $\Upsilon_{\mathrm{main}}$. Note that because the actual gradients occur on very small spatial scales, this means that the models without gradients tend to overestimate the stellar mass in the main body of the galaxy by a factor of $1.5$ on average. This effect of overestimating $\Upsilon$ when such gradients remain unaccounted for had also previously been suggested by \citet{Bernardi+18, DominiqueSanchez2019}.

In Table \ref{tab:gradientsTab}, we list the characteristic inner and main-body mass-to-light ratios of our models in the V-band, as well as the IMF normalization $\alpha$ relative to a Kroupa IMF for these values. We discuss the mass normalisation in Section~\ref{sec:discussion}.

\begin{figure*}
\centering
\includegraphics[width=\columnwidth]{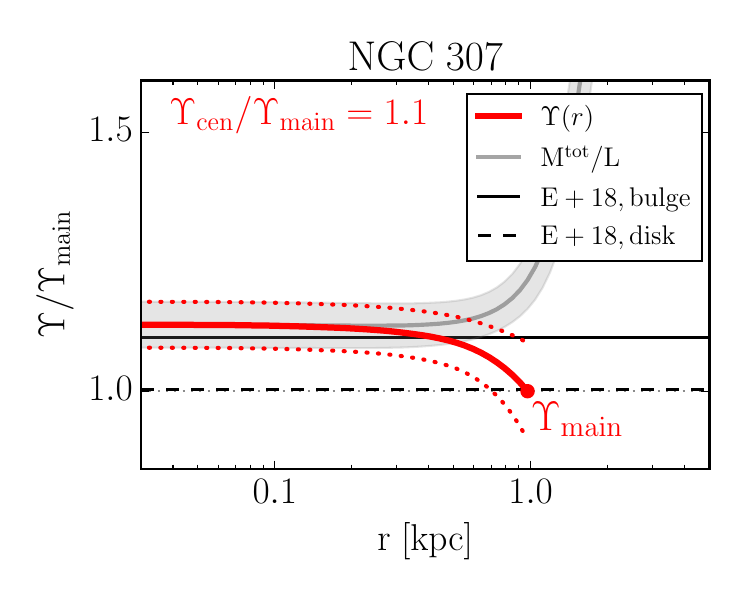}
\includegraphics[width=\columnwidth]{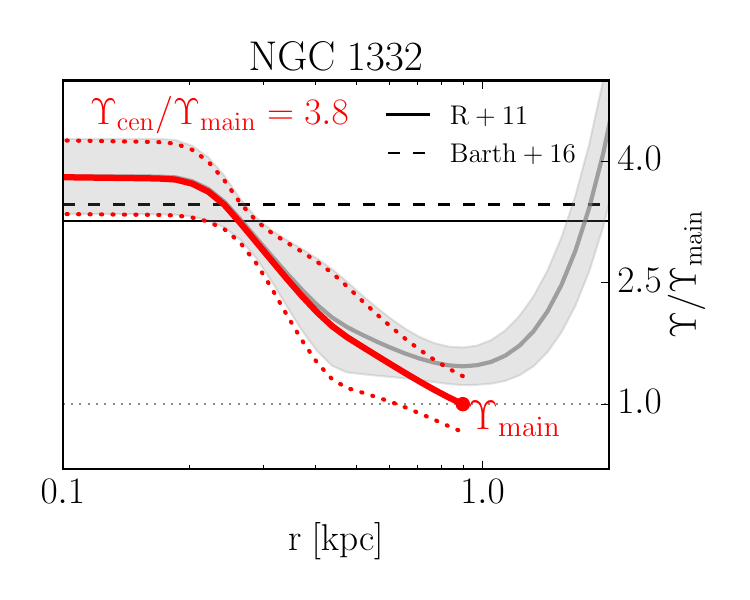}

 \caption{Mass-to-light ratio profiles of the sample galaxies. The best-fit stellar $\Upsilon(r)$ (solid red lines, uncertainties indicated by dotted red lines) of the galaxies are shown relative to the values of the galaxy main body, $\Upsilon_{\mathrm{main}} = \Upsilon(r_{\mathrm{main}})$. The radius $r_{\mathrm{main}}$ is defined at the minimum of the {\it total} mass-to-light ratio $(M^{\mathrm{tot}}/L)(r)$, for all galaxies except NGC~307 (see text). While the {\it stellar} $\Upsilon(r)$ depends on assumptions upon the mass decomposition, $(M^{\mathrm{tot}}/L)(r)$ is directly derived from the observations. 
 We indicate $(M^{\mathrm{tot}}/L)(r)$ and its uncertainties by solid grey lines and grey shaded areas, respectively.
 Note that $\Upsilon(r)$ profiles are projected along the line-of-sight (differences to the non-projected profiles are small). The figure includes comparisons with previous dynamical models without gradients (R+11,13; E+18 and \citealt{Barth2015}). For the four core galaxies we run comparison models without gradients as well (denoted as ``$\Upsilon$ = const.''). Models without gradients are worse fits to the kinematic data and tend to overestimate the mass in the main body of the galaxy.}
   \label{fig:MLgradsA}
\end{figure*}

\addtocounter{figure}{-1}
\begin{figure*}[!htbp]
\begin{flushleft}
\includegraphics[width=\columnwidth]{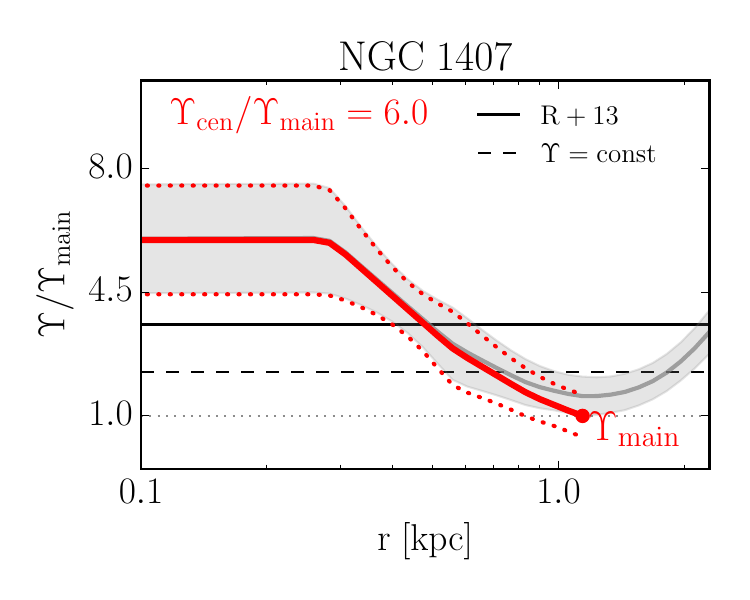}
\includegraphics[width=\columnwidth]{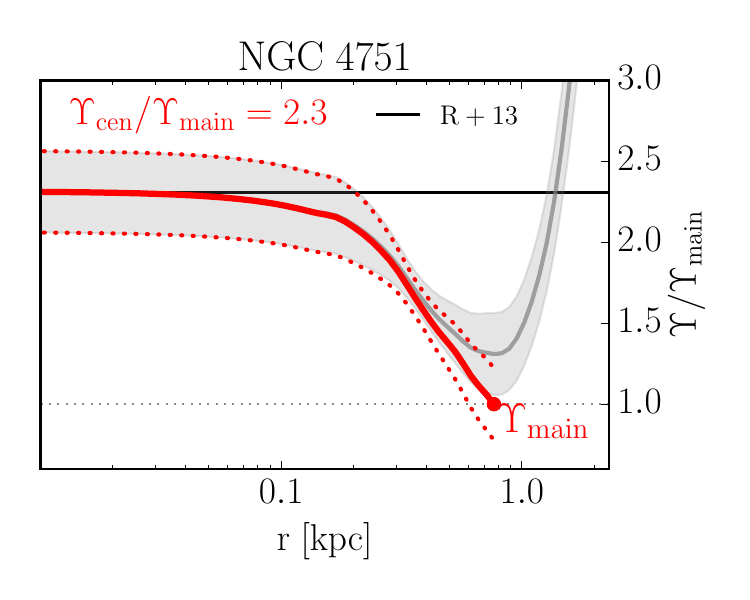}
\includegraphics[width=\columnwidth]{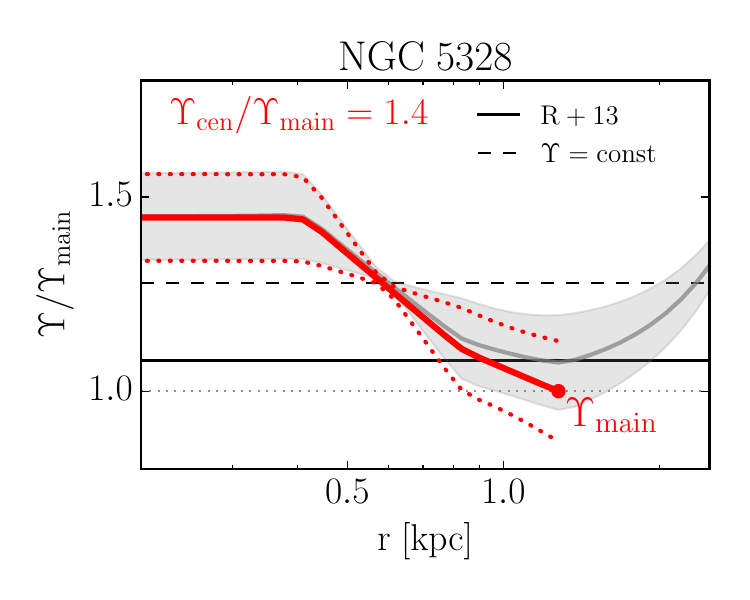}
\includegraphics[width=\columnwidth]{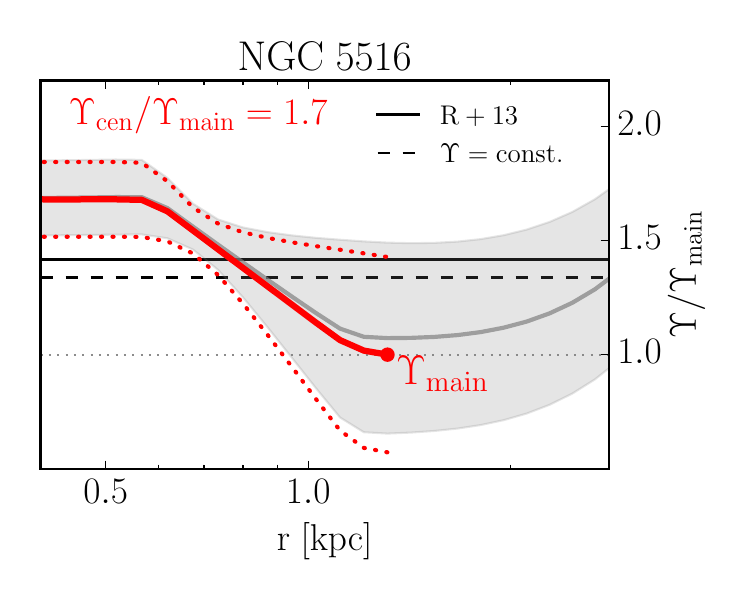}
\includegraphics[width=\columnwidth]{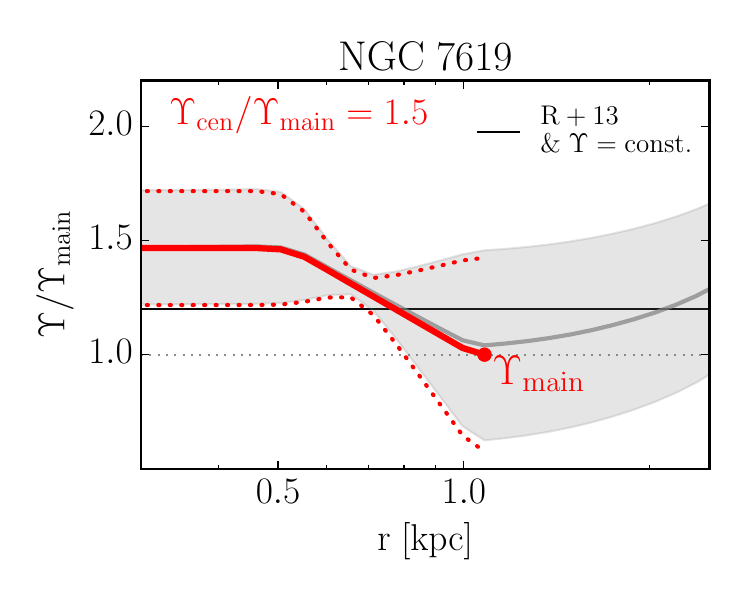}
\end{flushleft}
 \caption{(continued)}
\end{figure*}

We briefly \textcolor{black}{describe} the $\Upsilon$-gradients of the galaxies.

\textbf{NGC~307}: As stated above, the bulge of this galaxy was too small to warrant the implementation of gradients. There is, however, a weak composite $\Upsilon(r)$ gradient from the superposition of the two constant $\Upsilon_{\mathrm{bulge}}$, $\Upsilon_{\mathrm{disk}}$. The increase of our composite $\Upsilon(r)$ within $\SI{1}{kpc}$ is consistent with the one found by E+18, $\Upsilon_{\mathrm{bulge}}/\Upsilon_{\mathrm{disk}} = 1.1$ (their values). Considering our $\Upsilon_{\mathrm{bulge}}$, $\Upsilon_{\mathrm{disk}}$ best-fit model parameters, our $\Upsilon_{\mathrm{bulge}}$-value is identical to the one from E+18. For the disk component, our value is overall lower, but still roughly consistent with theirs within the uncertainties: $\Upsilon_{\mathrm{disk}} \sim 0.63 \pm 0.27$ versus $1.0 \pm 0.1$ in E+18 (I-band).

\textbf{NGC~1332}: We find a significant, almost factor-of-four increase towards the center of this galaxy from the superposition of the disk and bulge components. The central parts of this gradient ($r \lesssim \SI{0.3}{kpc}$) have a slightly larger $\Upsilon$ than our constant-$\Upsilon$ models from R+11. Over most of the galaxy's spatial extent, however, our new models produce significantly lower $\Upsilon$. Our central $\Upsilon$ is furthermore in agreement with the models by \citet{Barth2015} for the central $\SI{0.2}{kpc}$. 

\textbf{NGC~1407}: This galaxy has by far the most notable $\Upsilon$-gradient in our sample, with a factor six increase towards the center. This is the only galaxy in our sample for which the SOI of the central SMBH, $r_{\mathrm{SOI}} = (0.34 \pm 0.076) \ kpc$, extends to scales larger than the inner part of the $\Upsilon$-gradient, $r_{\mathrm{cen}} \sim \SI{0.3}{kpc}$. Furthermore, the outer mass-to-light ratio is surprisingly low, $\Upsilon^{\prime}_{f}= 1.29 \pm 0.71$ in V-band. Even accounting for uncertainties in the mass decomposition, the total $(M^{\mathrm{tot}}/L)_{\mathrm{main}} \sim 2$ is by far the lowest in our sample. However, the comparison models without gradient yield $\Upsilon = 3.0 \pm 0.20$ closer to our outer mass-light profile and lower than measured by R+13 ($\sim 4.6$ in V-band). The latter appears consistent with a radial average of our $\Upsilon$-gradient, roughly bisecting our mass-light profile
in the middle in Figure \ref{fig:MLgradsA}.

\textbf{NGC~4751}: As with NGC~1332, we find a $\Upsilon$-gradient within the bulge component, which in superposition with the constant-$\Upsilon$ disc component produces an effective total $\Upsilon$-gradient of slightly more than a factor two. The maximum of the gradient, within $r < \SI{0.1}{kpc}$, matches our previously published constant-$\Upsilon$ value from R+13.

\textbf{NGC~5328}: For this galaxy, the constant-$\Upsilon$ measurement is roughly an average over radius of our gradient model $\Upsilon(r)$. At the point where our gradient intersects with the constant $\Upsilon$-model ($\Upsilon \sim 5.8$ in V-band, $r \sim \SI{0.6}{kpc}$), it is also the most well defined with respect to the uncertainties. Our previously published $\Upsilon$-measurement from R+13 appears to be consistent with our $\Upsilon_{\mathrm{main}}$, but a factor $\sim 1.3$ smaller than $\Upsilon_{\mathrm{cen}}$.        

\textbf{NGC~5516 \& 7619}: For both of these galaxies we find gradients of a similar magnitude as for NGC~5328, and for which both our new constant-$\Upsilon$ models and previous measurements from R+13, are roughly averages over radius.

\subsection{SMBH measurements}

Unless one has kinematic data that resolve the SOI of a central SMBH very well, there is always some covariance between dynamically determined stellar mass-to-light ratios and the respective black hole mass, $M_{\mathrm{BH}}$ (e.g. \citealt{Rusli2013a}). Our previous SMBH mass measurements for the galaxies studied here were based on models without gradients hence we expect that after allowing for gradients the SMBH masses will change to some extent. However, a direct comparison is difficult, since the previous measurements used older (mostly long-slit) kinematic data outside the central regions. If we directly compare the SMBH masses from the old (gradient-free) and the new (gradient) models then we find two galaxies where $M_{\mathrm{BH}}$ goes up and two where it goes down\footnote{We restrict the discussion to the four core galaxies where we ran comparison models without gradients. For NGC~1332 we took $M_{\mathrm{BH}}$ from \citet{Barth2015} and for NGC~307 and NGC~4751 the new and old $M_{\mathrm{BH}}$ are almost identical.}. The difference can be up to 50\%. This is surprising, since our new, central stellar mass-to-light ratios $\Upsilon_{\mathrm{cen}}$ are always larger than the previous $\Upsilon$ from the gradient-free models. However, if we take our new comparison models without gradients as reference (which are based on the same data and modelled with the same advanced Schwarzschild code) then we find that in the gradient models, $M_{\mathrm{BH}}$ is \textit{always} smaller than in the gradient-free models -- as expected. The average decrease is 25\%. 

The remaining scatter when comparing the old (gradient-free) models with the new (gradient) models stems from the fact that the new MUSE data and advancements in the dynamical modelling have a non-negligible effect on our SMBH mass measurements. 
Still, our new values of $M_{\mathrm{BH}}$ are consistent with those found in R+13 and E+18 within the uncertainties for all galaxies except NGC~5328 (see Section~\ref{subsec:kinproblems} and also the discussion in Appendix C of M+23).

In Figure \ref{fig:Msig} we compare our new dynamical models to established trends between $M_{\mathrm{BH}}$ and galaxy velocity dispersion $\sigma$. We take the data for galaxies from \citet{Saglia2016} but use the updated values for the seven galaxies of this study.
We also added a number of the most recent Schwarzschild-based measurements from the literature:
From our own work we include axisymmetric Schwarzschild modeling results for the massive ETGs NGC~1600 \citet{Thomas2016} and Holm~15A \citet{Mehrgan2019}, which were both noted for their particularly massive SMBHs, as well as NGC~5419, which was modeled with our new triaxial modeling code SMART \citep{Neureiter2021}. 
Moreover we add results from triaxial Schwarzschild modeling of NGC~1453 rom \citet{Quenneville2022}, using the $\sigma_e$ value from \citet{Veale2018}, as well as triaxial models for M87 from \citet{Liepold2023}. Finally, we add seven more axisymmetric Schwarzschild measurements for low-mass fast-rotating ETGs from \citet{Thater2019} and \citet{Thater2022}.

With all of these new measurements added, we find the following relation:
\begin{equation}
    \log(\frac{M_{\mathrm{BH}}}{M_{\odot}}) = (5.05 \pm 0.41) \cdot \log(\frac{\sigma}{\SI{200}{km/s}}) + (8.46 \pm 0.06)
\end{equation}

This updated $M_{\mathrm{BH}}$ - $\sigma$ relation for ETGs is consistent with the relation of \citet{Saglia2016} (``CorePowerE'' in Table 11 of that study) within the uncertainties, though slightly steeper.

\begin{figure}
\centering
\includegraphics[width = \columnwidth]{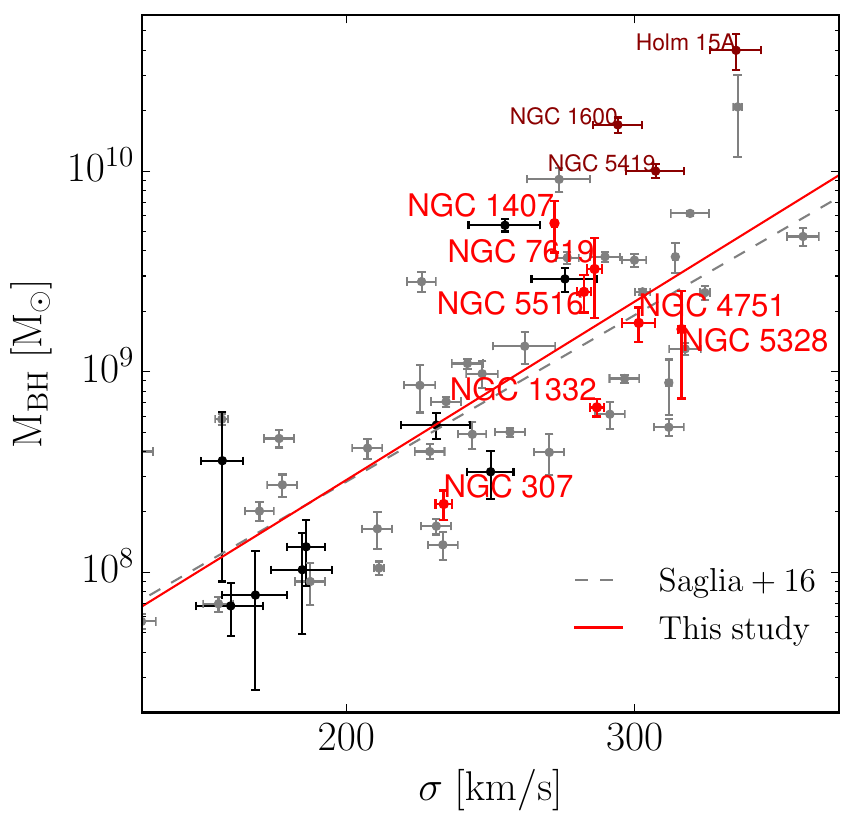}
 \caption{Our new dynamical $M_{\mathrm{BH}}$ measurements from the Schwarzschild models with gradients of this study (red) compared to the ETGs in \citet{Saglia2016} (grey) and new SMBH measurements from \citet{Thomas2016, Mehrgan2019} and \citet{Neureiter2023} (all in dark red) and \citet{Thater2019, Thater2022, Quenneville2022, Liepold2023} (black). The $M_{\mathrm{BH}} - \sigma$ relation of \citet{Saglia2016} is shown as a dashed grey line. We re-fitted the relation using all of the above measurements (solid red). For the same galaxy, using the same data and modelling codes we find that $M_{\mathrm{BH}}$ from models that allow for gradients are 25\% smaller than in models without gradients. }
   \label{fig:Msig}
\end{figure}

\section{Discussion}
\label{sec:discussion}
\subsection{On the stellar IMF}
\label{subsec: IMF}
In this section we evaluate our measured radial mass-light gradients in the context of a potential IMF variation within galaxies. To this end we calculate the mass normalization of our $\Upsilon (r)$ and $M^{\mathrm{tot}}/L (r)$ profiles relative to SSP-based measurements assuming a Kroupa IMF, $\Upsilon^{\mathrm{SSP}}_{Kroupa}$ (Parikh et al. submitted to MNRAS). While this is not a direct measurement of the shape of the IMF itself, it allows us to explore what level of bottom-heaviness is compatible with the dynamics of the galaxies, since the presence of low-luminosity dwarf stars is expected to be the main driver of IMF variation in ETGs \citep{vanDokkum&ConroyNature2010}.

\subsubsection{Radial IMF gradients}
The radius $r_{\mathrm{main}}$ is particularly relevant for our IMF-probes. The total-mass profiles from our dynamics effectively serve as upper-limits for the bottom-heaviness of the IMF: Formally all IMF models which produce $\Upsilon_{\mathrm{IMF}}(r)$ below our derived $(M^{\mathrm{tot}}/L)^{\mathrm{dyn}}(r)$ are consistent with our analysis -- we only need to account for the difference between $\Upsilon_{\mathrm{IMF}}(r)$ and $(M^{\mathrm{tot}}/L)^{\mathrm{dyn}}(r)$ by local mass-density corrections to our DM-halo models. Thus, at $r_{\mathrm{main}}$, the radial position of the global minimum of $(M^{\mathrm{tot}}/L)(r)$, the constraints on the maximum bottom-heaviness of the IMF are strongest. We here formulate the IMF mass normalization for a Kroupa IMF for both our $\Upsilon(r)$ and $(M^{\mathrm{tot}}/L)(r)$, and refer to them as $\alpha(r)$ and $\alpha^{\mathrm{tot}}(r)$, respectively. As explained above $\Upsilon$ depends on the mass decomposition but is projected along the line-of-sight (as the SSP measurements are). The directly measured quantity $(M^{\mathrm{tot}}/L)(r)$ is independent of any mass decomposition but its projection is useless as it carries all the DM in the outskirts of the galaxy/model with it. 

\begin{figure*}
\centering
\includegraphics[width=1.5\columnwidth]{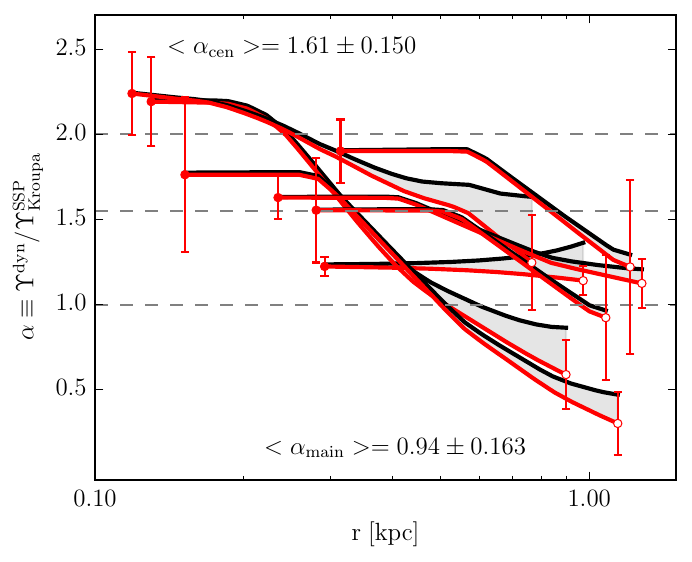}
 \caption{Profiles of the mass normalization $\alpha(r)$ of our dynamical mass-to-light ratio profiles relative to a Kroupa IMF. Normalization profiles of the stellar mass component $\Upsilon$ are shown in red and those relative to the entire dynamical $M^{\mathrm{tot}}/L$ in black. Stellar and total $\alpha$ profiles for the same galaxy are connected by grey shaded areas.  Grey horizontal dashed lines indicate Kroupa, Salpeter and ``heavyweight'' IMFs ($\alpha = 1$, $1.55$ and $2$, respectively). We highlight the two ends of the gradients, 
 $\alpha_{\mathrm{cen}}$ and $\alpha_{\mathrm{main}}$ (Table~\ref{tab:gradientsTab}), with (red) filled and open symbols, respectively. Uncertainties of the total mass profiles are comparable to those of the stellar component. The gradients are spatially very concentrated and confined to the central kpc of the galaxies. The gradients are plotted up to the point where the total dynamical $M^{\mathrm{tot}}/L$ has its minimum. This point constraints the $\alpha_{\mathrm{main}}$ of the main body of the galaxy strongest (see text for details).}
   \label{fig:IMFgrads}
\end{figure*}

Values of the main body of each ETG, $\alpha_{\mathrm{main}}$, $\alpha^{\mathrm{tot}}_{\mathrm{main}}$, as well as of the inner regions, $\alpha_{\mathrm{cen}}$ are listed in Table \ref{tab:gradientsTab}. As stated before, towards the center, the $\Upsilon$-gradients become essentially identical to $(M^{\mathrm{tot}}/L)(r)$, and this carries over to $\alpha$. We show the full $\alpha(r)$ profiles up to $r_{\mathrm{main}}$ for all galaxies in Figure \ref{fig:IMFgrads}.

At roughly $\SI{1}{kpc}$ our dynamical models are on average consistent with the $\Upsilon$ from a Kroupa or Chabrier IMF,
$<\alpha_{\mathrm{main}}> = 0.94 \pm 0.16$ ($\Upsilon_{\mathrm{Chabrier}} = 0.9 \times \Upsilon_{\mathrm{Kroupa}}$). Considering our total mass profiles, at $\SI{1}{kpc}$, a local IMF with a Salpeter level bottom-heaviness is inconsistent with the fits at a level between one and two sigma for all galaxies except NGC~5516 and NGC~4751. We find $<\alpha^{\mathrm{tot}}_{\mathrm{main}}> = 1.16 \pm 0.14$.

Interior to $\SI{0.3}{kpc}$ our dynamical models are on average consistent with the $\Upsilon$ of a Salpeter IMF, $<\alpha_{\mathrm{cen}}> = 1.61 \pm 0.15$ ($\Upsilon_{\mathrm{Salpeter}} \sim 1.55 \times \Upsilon_{\mathrm{Kroupa}}$). 
A Salpeter-level bottom-heaviness is consistent with our dynamical models for all but one galaxy, the least massive galaxy in our sample, NGC~307. For more than half of the sample levels of bottom-heaviness up to a ``heavyweight'' $\alpha = 2$, are consistent with the fits at a one sigma level.

\subsubsection{IMF variation with galaxy $\sigma$}
Many previous studies of the IMF using various methods found a trend between $\alpha$ and galaxy velocity dispersion $\sigma$ which suggests that galaxies with higher $\sigma$ have higher $\alpha$. The majority of existing $\alpha$ determinations are based on models without gradients. Different measurements are also derived over different spatial scales. SSP probes typically focus on the very center of a galaxy, i.e. within $r_e/8$. Dynamical probes, by tendency try to capture as much of the galaxy as possible within $r_e$. Apertures of gravitational lensing probes are identical to the observed Einstein rings, $\theta_{Ein}$, and lie usually in between SSP and dynamics measurements in terms of spatial coverage. \citet{Lyubenova+2016} found that part of the tension between different IMF probes could be alleviated by matching apertures. Here we address the question what trends with $\sigma$ our $\alpha$-gradient models produce for different apertures.

To this end we compare light-weighted averages of our $\alpha$ profiles \footnote{We here assume that $\alpha(r) = \alpha_{\mathrm{main}}$ for $r > r_{\mathrm{main}}$} and $\sigma$ to different IMF probes from the literature while adapting our aperture sizes to the respective comparison sample.

\begin{figure*}
\centering
\includegraphics[width=0.9\columnwidth]{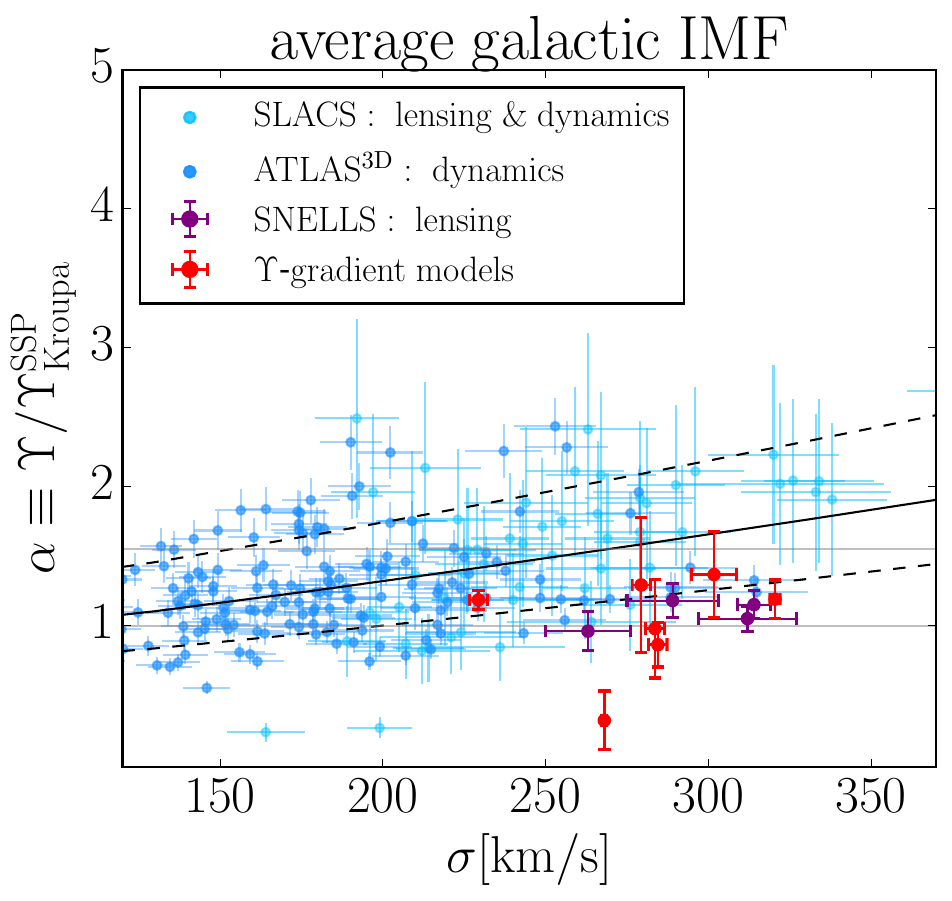}
\includegraphics[width=0.9\columnwidth]{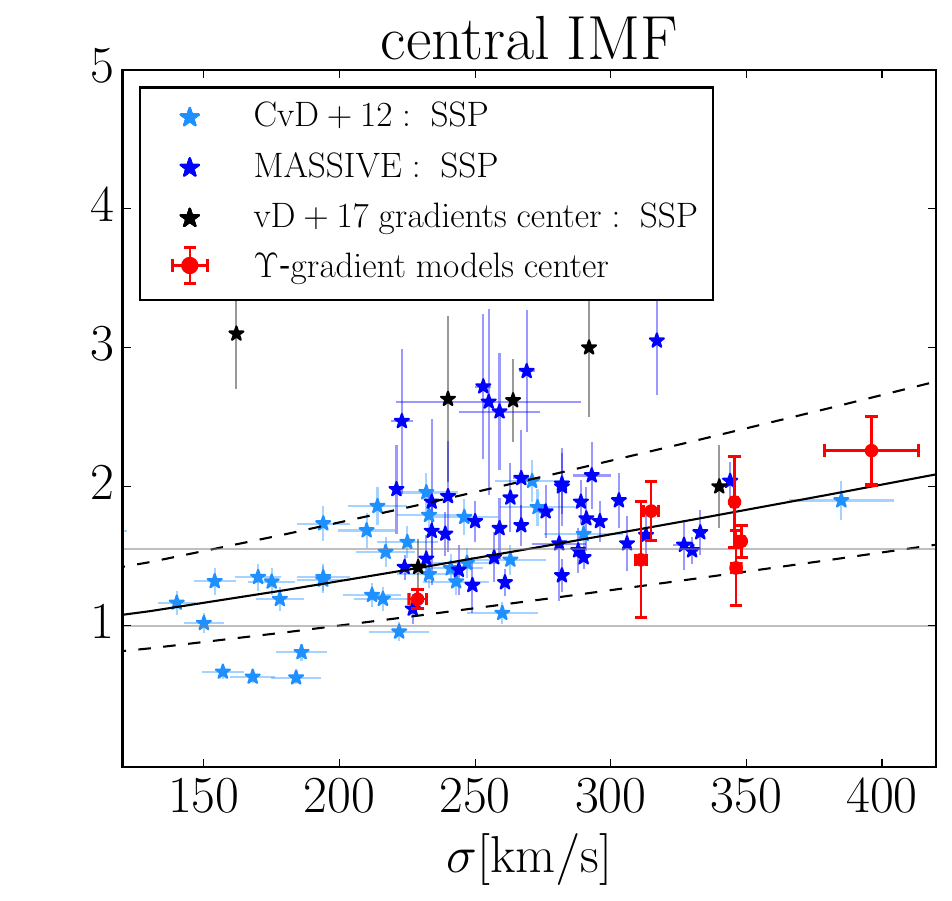}

 \caption{Comparison of our dynamical $\alpha(r)$ and velocity dispersion $\sigma$ within different apertures (red), compared to different IMF probes. Left: Light-weighted averages of our $\alpha(r)$ and $\sigma$ for an aperture of $\SI{2}{kpc}$ compared to $\alpha$ measurements from the ATLAS$^{3D}$ \citep{Cappellari2013b}, SLACS \citep{Treu2010, Auger2010, Posacki+15}, and 
 SNELLS \citep{Smith+15, Newman+17} surveys. Right: the innermost $\alpha$ and $\sigma$ of our models compared to the SSP measurements of \citep{Conroy&vanDokkum2012} and from the MASSIVE survey \citep{Gu2022}. We also show central $\alpha$-measurements from the $\alpha$-gradient models of \citet{vanDokkum2017}.
 In both panels the solid/dashed black lines shows the quadratic $\alpha-\sigma$ relation from \citet{Posacki+15} and its scatter. Horizontal light-grey lines indicate Kroupa- and Salpeter-levels of bottom-heaviness.
 Both panels taken together illustrate how spatially concentrated the detected gradients are: already over scales of only $\SI{2}{kpc}$ (left panel) the central high $\alpha$ seen in the right panel are washed out and the IMF becomes Kroupa-like, consistent with nearby strong lenses.}
   \label{fig:IMFsigma}
\end{figure*}

First, in Figure \ref{fig:IMFsigma} we compare our $\alpha$ measurements on both small and large spatial scales. In the left panel of the Figure we consider the ``overall'' IMF of the galaxy. By this we mean the light-weighted average $\alpha$ within an isophote with a circularized radius $r_{ap} > r_{\mathrm{main}}$ (see below). We compare this $\alpha$ to stellar dynamical $\alpha$-measurements from ATLAS$^{3D}$ \citep{Cappellari2013b} \textcolor{black}{and dynamics+lensing measurements from SLACS, as well as lensing measurements} from the SNELLS lensing survey \citep{Smith+15, Newman+17}. For the SLACS sample we use the updated values from \citet{Posacki+15}. We also show the quadratic $\alpha-\sigma$ relation from \citet{Posacki+15} which simultaneously fits the  ATLAS$^{3D}$ and updated SLACS measurements. 

The ATLAS$^{3D}$ values were determined for an aperture of $r_{ap} = r_e$. The SLACS values are a combination of stellar dynamics and strong lensing constraints and the average $\theta_{Ein}$ is roughly $r_e/2$. Thus, they still probe  similar spatial scales. 
The SNELLS lens-measurements on the other hand probe more confined absolute scales, $\theta_{Ein} \sim \SI{2}{kpc}$, which translates into $\sim 20-70\%$ of $r_e$ depending on the galaxy's distance.

For the comparison with our measurements, these varying spatial scales are not a problem, however. The gradients which we found are so spatially concentrated, that between $r_{ap} = \SI{1}{kpc}$ and $r_{ap} = r_e$, the integrated $\alpha$ changes on average by less than $4 \%$ for all galaxies in our sample (we find similarly small changes with aperture past $\SI{1}{kpc}$ for $\sigma$). Since
$\alpha(r)$ seems to correlate well with physical radius we here use $r_{ap} = \SI{2}{kpc}$ (average extent of the SNELLS lenses).

On the $\alpha-\sigma$ diagram for the overall galaxy-wide IMF, our gradient models appear to follow a different, much less bottom-heavy trend than the ATLAS$^{3D}$ and SLACs galaxies. Six out of seven of our sample galaxies are more massive than $\sigma = \SI{250}{km/s}$, yet our sample scatters around a MW IMF normalization $\alpha = 1.03 \pm 0.33$ (or $\alpha = 1.15 \pm 0.17$ if we do not count the outlier NGC~1407), whereas the relation of \citet{Posacki+15} predicts a Salpeter or above-Salpeter level bottom-heaviness, $\alpha \gtrsim 1.55$ for $\sigma > \SI{250}{km/s}$. However, our gradient models agree well with the SNELLS lensing results, which find a MW-level normalization even for ETGs with $\sigma > \SI{250}{km/s}$. 

In the right panel of Figure \ref{fig:IMFsigma}, we compare the bottom-heavy centers $r_{ap} = r_{\mathrm{cen}}$ of our models to \textit{SSP} IMF probes from the MASSIVE survey \citep{Gu2022}, as well as from \citet{Conroy&vanDokkum2012}, since their probes are also focused on the centers of the ETGs. \citet{vanDokkum2017} also measured radial IMF gradients for a set of six ETGs using SSP models. We here add the centermost $\alpha$ values from these gradients to the diagram. 

For the most part, within the uncertainties our central $\alpha$-values seem to be consistent with the SSP trends of the MASSIVE, \citet{Conroy&vanDokkum2012} and \citet{vanDokkum2017} samples, which also agree with the dynamics-based trend of \citet{Posacki+15} (despite the latter originating from measurements from much larger apertures).

There is, however, a distinct band of galaxies with extremely bottom-heavy SSP measurements $\alpha \gtrsim 2.5$. Among these galaxies is also NGC 1407 whose SSP-measured $\alpha = 3$ is much larger than our central $\alpha = 1.76 \pm 0.516$. Since on the relevant spatial scales uncertainties in the mass decomposition are insignificant, the dynamical and SSP measurements are hard to reconcile. This is indicative of a still unresolved broader \textcolor{black}{problem} of matching SSP and dynamical measurements of $\Upsilon$ on the level of individual galaxies \citep{Smith2014, McDermid2014}.

\begin{figure*}
\centering
\includegraphics[width=0.9\columnwidth]{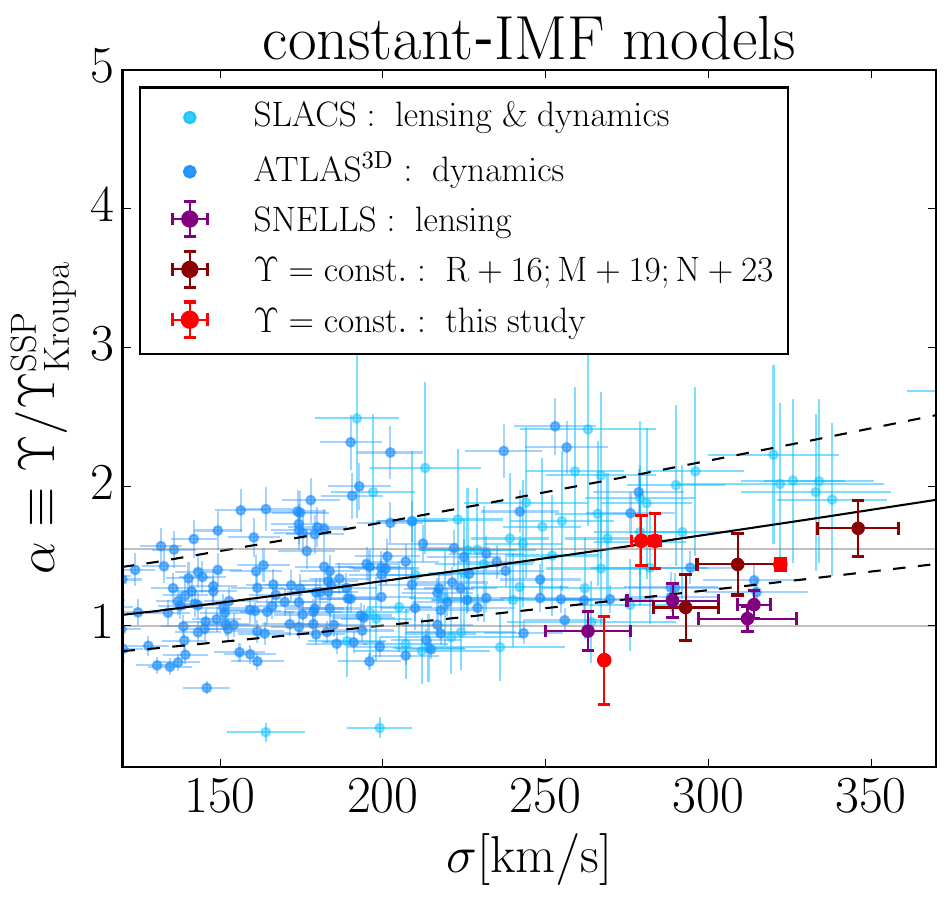}
\includegraphics[width=0.9\columnwidth]{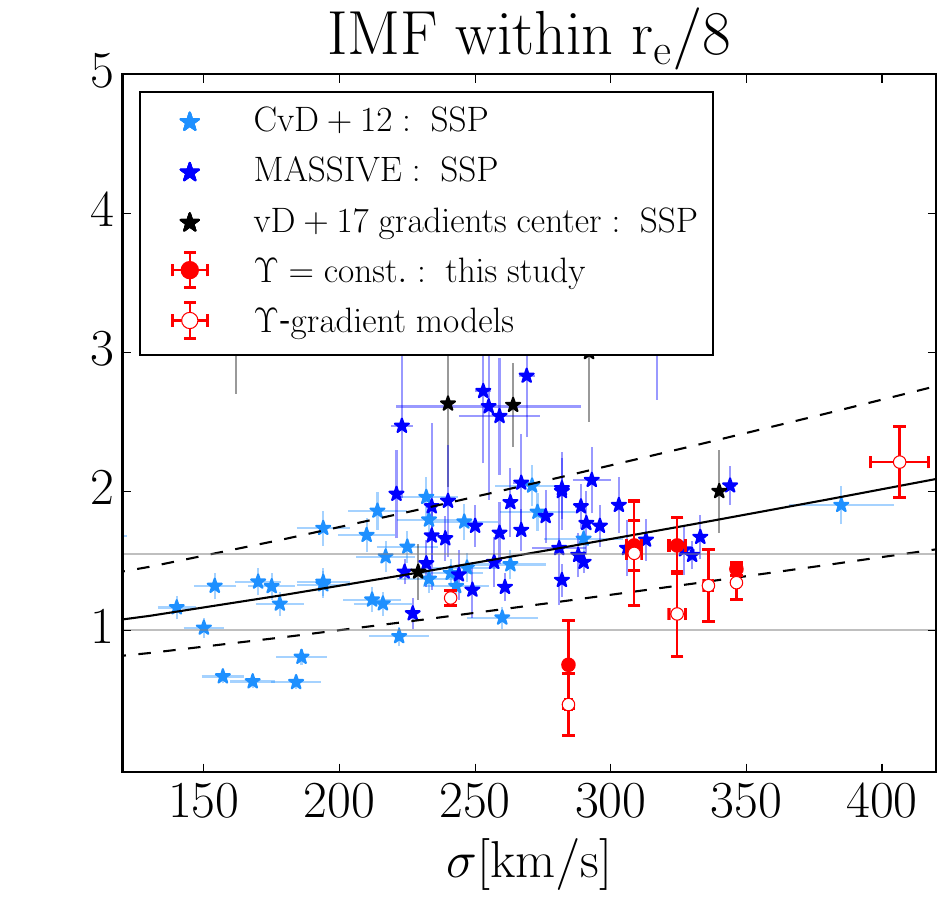}
 \caption{Same as Figure \ref{fig:IMFsigma}, but for comparison models without gradients (left, only for the four galaxies for which we generated constant-$\Upsilon$ models), to which we add recent constant-$\Upsilon$ Schwarzschild models of \citet{Thomas2016, Mehrgan2019}, and \citet{Neureiter2023} (here T+16, M+19, and N+23). For NGC~5328 from our new models (at $\sigma \sim \SI{325}{km/s}$), the errobars are of similar size as the scatter point. For the right-hand panel we consider an aperture of $r_e/8$, instead of $r_{\mathrm{cen}}$ for $\alpha$ and $\sigma$ for our models (the literature data remains unchanged from the right panel of Figure \ref{fig:IMFsigma}). Models without gradients not only fit the data worse but lead to an overestimation of the stellar mass. Still, even without gradients our stellar mass normalisation $\alpha$ derived with state-of-the-art Schwarzschild models is on the low side of previously found distributions. Within $r_e/8$ the spatially very confined dynamical gradients are already partly washed out and the mass normalisation is on average lower than in Figure~\ref{fig:IMFsigma} (right).}
   \label{fig:IMFpart2}
\end{figure*}

Nonetheless, considering the overall trends, the two panels of Figure \ref{fig:IMFsigma} could be seen to imply that our models are in agreement with SNELLS lensing results (at large scales) and SSP modeling results (at small scales) and at tension with dynamical measurements from ATLAS$^{3D}$ and SLACS. However, there are unaccounted differences between the measurements which we discuss in Figure~\ref{fig:IMFpart2}.

As we stated in the previous section (and as also discussed by \citealt{Bernardi+18, DominiqueSanchez2019}),  if $\Upsilon(r)$ intrinsically rises towards the center, this biases $\alpha$ high for models without gradients. With the exception of the SSP measurements by \citet{vanDokkum2017}, all of the literature measurements we showed here were based on the assumption of gradient-free $\Upsilon$.

Hence, for a more consistent comparison, the left-hand panel of Figure~\ref{fig:IMFpart2} compares the dynamical measurements from ATLAS$^{3D}$, SLACS, and SNELLS to 
our own gradient-free models. As stated above, these models provide worse fits to the kinematics than models with gradients and are here used merely to understand where the differences between the various IMF determinations could arise from. We also add recent Schwarzschild-based constant-$\Upsilon$ measurements of the ETGs NGC~1600, Holm~15A and NGC~5419 \citep{Thomas2016, Mehrgan2019, Neureiter2023}. The figure confirms that models with a spatially constant $\Upsilon$ lead to higher $\alpha$. Thus they are more consistent with the measurements from ATLAS$^{3D}$ and SLACS, as expected. However, our measurements are still on the lower side of those distributions. This may be an artefact of our small sample size. It may also be due to the differences in the modelling approach. The ATLAS$^{3D}$ and SLACS measurements were determined using Jeans an-isotropic modeling \citep[JAM; ][]{Cappellari2006, Cappellari2008} while we use Schwarzschild models. Schwarzschild models provide the most general solutions to the Collisionless Boltzmann Equation which governs the dynamics of stars in galaxies. We have shown that using adaptive regularisation, our generalised model selection and non-parametric LOSVDs, Schwarzschild models allow for very accurate mass reconstructions \citep{Lipka2021,ThomasLipka2022,Neureiter2023a,deNicola2022}. 

Considering the central regions of our models, in the right-hand panel of Figure~\ref{fig:IMFpart2} we repeat the same diagram as in the right panel of Figure~\ref{fig:IMFsigma} but take the {\it exact} aperture of the SSP measurements, $r_{ap} = r_e/8$. Over this aperture, our gradient models for all galaxies except NGC~4751 are similarly offset with respect to the SSP measurements from MASSIVE and \citet{Conroy&vanDokkum2012} as they are for a $\SI{2}{kpc}$ aperture to the dynamical measurements from SLACS and ATLAS$^{3D}$ (cf. left panel of Figure \ref{fig:IMFsigma}). This demonstrates again how concentrated our gradients are. Adding once again the \textit{actual} constant-$\Upsilon$ models for our galaxies to the diagram, we find the same results as for the dynamical, galaxy-wide comparison: Broadly consistent with previous trends within the uncertainties, but with $\alpha$ that tend to be lower overall.

We might summarize the contents of Figures \ref{fig:IMFsigma} and \ref{fig:IMFpart2} as follows: In the centers of the galaxies, our  Schwarzschild dynamical $\Upsilon$-measurements reveal increased 
levels of stellar mass that confirm and agree with previously suggested mass normalization factors larger than that of a Kroupa IMF in ETGs. Most likely, this mass excess points to a bottom heavy IMF in the centers (but see Section~\ref{subsec:topheavy}).
The gradients are so centrally concentrated, however, that already for apertures of only $r_{ap} = \SI{2}{kpc}$ the mass enhancement disappears and the IMF converges to a Kroupa level, consistent with measurements in nearby lenses. This largely alleviates the differences between previous studies. Not accounting for existing centrally rising gradients of $\Upsilon$ biases $\alpha$ high -- for some galaxies high enough to ostensibly yield a Salpeter-level $\alpha$. However, there remain some inconsistencies. Even when compared on equivalent spatial scales and when matching the use of constant-$\Upsilon$ models, for both small and large apertures, our $\alpha$ values are overall less extreme than previous probes.

\subsubsection{Comparison with SSP-based gradients}

\begin{figure}
\centering
\includegraphics[width=1.\columnwidth]{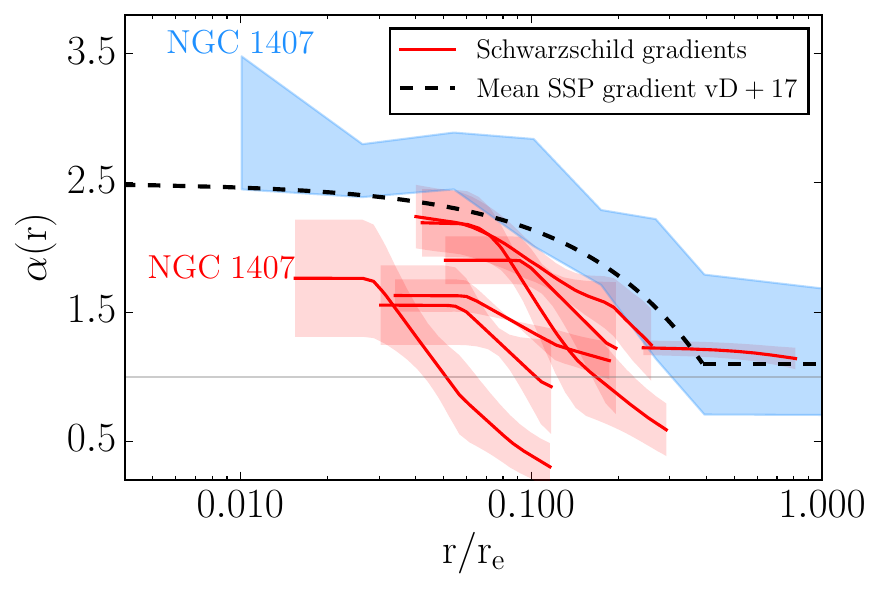}
 \caption{Comparison of the IMF normalization $\alpha$ of our Schwarzschild dynamical $\Upsilon$-gradients (red lines, with red shaded areas indicating uncertainties) to the mean SSP-based gradient of \citet[][black dashed curve]{vanDokkum2017}. We also show the SSP-based gradient with uncertainties for NGC~1407 (light blue area), which is also in our sample. While the resolved dynamical and SSP gradients both indicate an increase in $\alpha$ towards the center, the dynamical gradients are steeper, more centrally concentrated, and have a lower mass normalisation. For NGC~1407 the two approaches yield inconsistent results.}
   \label{fig:IMFvD}
\end{figure}

After having compared the central values of $\alpha$ from our dynamical $\Upsilon$-gradient models to the central values of the SSP-based $\Upsilon$-gradient models from \citet{vanDokkum2017}, we will now compare the full radial $\alpha$-gradients with each other. In Figure \ref{fig:IMFvD}, we show all seven models from our study together with the average $\alpha$ gradient determined by \citet{vanDokkum2017} over the six ETGs of their sample. One galaxy, NGC~1407, is mutual to both studies.

The figure confirms many of the trends we have found in the previous subsections. Both our dynamical models and the SSP models show radial profiles that at large radii converge on a MW-like IMF normalization on average. Both approaches yield an increased mass normalisation near the center around the Salpeter level. However, the dynamical masses are about $1.6$ times smaller than the SSP models of \citet{vanDokkum2017} imply. This difference can not be explained by uncertainties in the dynamical mass decomposition, as $\alpha \sim \alpha^{\mathrm{tot}}$ in the center. 

For NGC~1407 the discrepancy is even larger: At no radius is the dynamical profile consistent with the extremely bottom-heavy $\alpha$ profile measured by \citet{vanDokkum2017}. At the radius where the total dynamical mass-to-light ratio reaches its minimum, the dynamical models yield a very low stellar mass normalisation $\alpha_{\mathrm{main}} = 0.30 \pm 0.19$ whereas the SSP models produce a ``heavyweight'' normalisation of $\alpha \sim 2.5$. Even considering the total dynamical mass, this value remains surprisingly high compared to the dynamical $\alpha^{\mathrm{tot}}_{\mathrm{main}} = 0.48 \pm 0.18$. This does not appear to be a problem originating from our gradient models per-se, as even our dynamical models without gradient result in a low $\alpha = 0.66 \pm 0.044$ consistent with the gradient models within the uncertainties. \textcolor{black}{In principle, the lower dynamical $\Upsilon$ could be matched with the very bottom-heavy IMF of \citet{vanDokkum2017} by increasing the low-mass cut-off of the IMF. However, the central IMF of the galaxy was also studied with non-parametric IMF-models in a companion SSP analysis \citep{Conroy2017}. This study suggests that the low-mass IMF slope remains very steep down to $\SI{0.1}{M_{\odot}}$($dN/dM_{\star} \propto M_{\star}^{-2.7}$).}

In Section \ref{subsec:problems} we suggest that our dynamical models of NGC~1407 could be partly biased by the galaxy being triaxial. On the other hand, however, we already noted that NGC~1407 is among the handful of galaxies for which the SSP analysis results in distinctly high mass normalizations (Figure~\ref{fig:IMFsigma}). Even if triaxiality might bias the dynamical analysis by up to factor of 2 in extreme cases \citep{Thomas2007Triax} it seems unlikely that this can explain the entire difference between our dynamical models and the SSP analysis (which amounts to a factor of $\sim 5$).

A similar case is the massive ETG NGC~1600: the Schwarzschild models of \citet{Thomas2016} produce a MW-like $\alpha = 1.1 \pm 0.24$ which is consistent with our results for similar core galaxies presented here (though the models of \citealt{Thomas2016} are without gradients). However, this low mass normalisation is at tension with the gradient SSP-models of \citet{vanDokkum2017} that point to a Salpeter-level or higher bottom-heaviness at most radii and with the gradient-free models of \citet{Gu2022} (who found a super-Salpeter normalization $\alpha = 1.67 \pm 0.16$).

\subsection{Evaluation of uncertainties}
\label{subsec:problems}
Our new state-of-the-art dynamical models yield very spatially concentrated gradients together with an almost Kroupa-like mass normalisation for the galaxies outside the center.
We have seen that taking into account aperture effects and gradients can bring different IMF probes closer together that at first glance seem to yield inconsistent results. In this section we discuss some of the possible systematics which could contribute to the remaining inconsistencies between methods.

Generally, there is the potential of a bias towards high $\alpha$ 
in some of the SSP models to which we have compared our dynamical results here. Such a bias could arise from incomplete stellar libraries. If for instance, elemental abundances associated with certain IMF-sensitive features such as Na I were underrepresented in the stellar modeling libraries of low-mass dwarf stars, more of them would be needed to reproduce this feature in observed spectra, driving up the measured bottom-heaviness.  A more detailed discussion of stellar population \textcolor{black}{uncertainties} will be given in the companion paper by Parikh et al. (submitted to MNRAS). We here focus only on our own dynamical models, though a complete evaluation of the discrepancies of different IMF probes among each other has to take into account the combined effects of biases of all methods.

\subsubsection{Input stellar kinematics}
\label{subsec:kinproblems}

As discussed in Section \ref{sec:modelsetup}, MUSE and SINFONI LOSVDs are generally consistent with each other within the uncertainties. Differences still arise due to spatial, spectral, and seeing differences, particularly as the SINFONI kinematics are supported by adaptive optics, while the MUSE kinematics are limited by natural seeing. We expect the stellar dynamical models to be able to fit both sets equally well as they take the above mentioned differences into account. 

Overall, our models were successful in fitting both sets of non-parametric LOSVDs for six out of the seven galaxies, as the values of $(\chi^2 + m_{\mathrm{eff}})/N$ for the fits in Table \ref{tab:resultsTab} show (with NGC~4751 being the exception). In particular, the models were generally able to fit MUSE and SINFONI kinematics simultaneously in areas where they spatially overlap, $r \leq \SI{1.5}{\arcsec}$. We show individual LOSVD fits for all galaxies in such overlapping regions in Appendix \ref{ap:kinfits}. In our kinematics paper, M+23, we had noted forms of ``hidden template-mismatch'' which can not be unambiguously diagnosed from the spectral analysis alone. Since our models were mostly able to reproduce both (independent) LOSVD sets simultaneously at the same spatial locations, it seems that the hidden template-mismatch in our data was low. In M+23 we had taken deliberate steps to render this outcome more likely, as is detailed in that study. Nonetheless we faced some \textcolor{black}{problems} for a few galaxies, which we briefly describe here:

{\textbf{NGC~4751}}: This was the only galaxy in our sample for which $(\chi^2 + m_{\mathrm{eff}})/N > 1$. Moreover, one of the four quadrants even produced a $(\chi^2 + m_{\mathrm{eff}})/N > 3$. This anomalous quadrant was also an outlier in terms of the best-fit model parameters (see the last plot of Figure \ref{fig:MLgradsA}). We therefore excluded this quadrant from our analysis entirely. However, the large $(\chi^2 + m_{\mathrm{eff}})/N$ value in Table \ref{tab:resultsTab} was already derived without this quadrant. The main \textcolor{black}{limitation} here appears to be dust contamination of the LOSVD signal. As described in Section \ref{subsec:lightDens} and Appendix \ref{ap:ngc4751phot}, most of the major axis of the galaxy is covered with dust all the way to the effective radius on both sides of the center of the galaxy. Our imaging data was derived in the K-band and the most severely contaminated regions were masked before the photometric decomposition. The SINFONI LOSVDs were derived in the infrared. The MUSE kinematics by contrast were measured in the optical and therefore potentially more affected by dust. In general, the presence of dust in a galaxy should not affect the symmetry of the LOSVDs, only emphasise the LOSVD signal from some part of the galaxy more than others -- those parts of the LOSVD which originate from behind the dust along the line-of-sight being dampened. This is consistent with both asymmetric spatial variation and biases of even order Hermite moments if the LOSVDs are parameterized with Gauss-Hermite polynomials. In Appendix \ref{ap:kinfits}, we discuss to to what extent the LOSVDs are likely distorted by the dust in terms of $h_4$.

To what extent our dynamical models of NGC~4751 might be biased by dust cannot be evaluated easily. To be conservative, we quote our sample-averaged IMF normalization measurements \textit{without} NGC~4751: $<\alpha_{\mathrm{cen}}> = 1.54 \pm 0.15$, $<\alpha_{\mathrm{main}}> = 0.91 \pm 0.172$, and $<\alpha^{\mathrm{tot}}_{\mathrm{main}}> = 1.13 \pm 0.142$. However, our previous conclusions on IMF gradients remain essentially the same even without NGC~4751.

{\textbf{NGC 7619 (and NGC~5516)}}: While we can successfully fit the MUSE and SINFONI kinematics for NGC~7619 for the majority of our spatial coverage, there are some small \textcolor{black}{problems} at the largest and smallest radii of the MUSE data (the SINFONI data is reproduced well over the full SINFONI coverage, see Figure \ref{fig:radKin}). At large radii ($r > \SI{20}{\arcsec}$) the $h_4$ of our models rises towards the edges of the MUSE FOV, whereas the MUSE data appears to follow the opposite trend. Within $\SI{2}{\arcsec}$, our models underpredict the dispersion of the MUSE data (while reproducing all of the SINFONI data correctly). This could be indicative of a bias in the MUSE LOSVDs arising from the aforementioned hidden template mismatch. Whatever the cause of these differences between the model and the MUSE data, they are comparatively small as evidenced by non-parametric LOSVDs themselves, as seen in Figure \ref{fig:Losvdfit} (which, after all are the target and deciding factor of our dynamical models). Furthermore, the reduced $\chi^2$ for our dynamical fits are still favourable (see Table \ref{tab:resultsTab}). Similarly, but less significantly the dynamical models for NGC~5516 underpredict the centermost MUSE $\sigma$ value, and $h_4$ within $\SI{3}{arcsecond}$. However, once again, the difference in the nonparametric LOSVDs themselves is small.

{\textbf{NGC~5328}} is a galaxy where fitting both data sets, MUSE and SINFONI, simultaneously turned out to be particularly difficult. For this galaxy one of two CO band-heads -- the spectral features on which the SINFONI kinematics for all galaxies were based -- was obstructed by residual OH emission, limiting the accuracy of the SINFONI LOSVDs to an extent such that the central LOSVDs were assumed to have a Gaussian shape (R+13). We thus used the Gaussian fits from R+13 as the input SINFONI LOSVDs and not the original non-parametric LOSVDs. That the shape of these LOSVDs (Gaussian) is not consistent with the measured shape of the MUSE LOSVDs is not surprising. This could have biased our determination of $M_{\mathrm{BH}}$, but the inclusion of the SINFONI kinematics (basically the velocity dispersion scale) still provided vital constraints on the recovery of the $\Upsilon(r)$ profile (Appendix~\ref{ap:kinfits}). 

\subsubsection{Assumption of Axisymmetry}
\label{sec:Gradsim}
We have here dynamically modeled the sample galaxies under the assumption that they are axisymmetric systems. For galaxies with strongly ordered velocity fields like the fast rotating power-law galaxy NGC~307 or even the ``intermediate'' rotator NGC~7619, which has the most symmetric velocity field of all our cored ETGs, this assumption is generally justified. However, cored ETGs as a whole must have triaxial shapes in general \citep[e.g.][]{Bender1988, KormendyBender1997, Cappellar2007, Emsellem2007}.

The potentially negative effects of triaxiality on the accuracy of axisymmetric models are generally viewing angle and shape dependant \citep[e.g][]{Thomas2007Triax,Remko2010}. \citet{Thomas2007Triax} find that the mass-to-light ratio of triaxial galaxies can be underestimated in axisymmetric models by as much as a factor two. The effects of triaxiality in the case of mass-to-light ratio gradients have not been investigated yet. However, a factor of two bias holds only in extreme cases. For example, axisymmetric Schwarzschild models of the triaxial galaxy M87 from \citet{Gebhardt&Thomas2009}, using an earlier version of our modeling code, determined a SMBH mass of $M_{\mathrm{BH}} = (6.4 \pm 0.4) \times 10^9 M_{\odot}$, which was later confirmed by direct imaging of the shadow of the SMBH by the Event Horizon telescope ($M_{\mathrm{BH}} = (6.5 \pm 0.8) \times 10^9 M_{\odot}$, \citealt{Akiyama2019}). 

While M87 might be special (it appears nearly round in its central regions) such an accuracy is not entirely surprising. Numerical merger simulations suggest that core-formation, which involves the ejection of stars from the center of a forming core by binary SMBHs, preferentially ejects stars on box-orbits from the center of merger remnants, which essentially ``removes'' triaxiality from within the core break-radius $r_b$ \citep{Frigo2021}. This means that even in the centers of core galaxies there is no a priori reason to expect axisymmetric gradient models to be particularly biased.

In addition to triaxiality, allowing for $\Upsilon$-gradients poses new challenges. E.g. the extended parameter space and the larger freedom in the stellar mass distribution might cause degneracies or complications that were not yet encountered in models assuming only a single galaxy-wide $\Upsilon$ for the stars. 

In order to test for potential systematics in our fits we have fitted mock data based on a realistic numerical N-body simulation from \citet{Rantala2018}. Since this simulation was tuned to resemble NGC~1600, it represents quite realistically a massive triaxial elliptical galaxy with a DM halo and SMBH. Specifically, the simulation, as we have set it up here, is a cored ETG with a SMBH of $8.5 \times 10^9 M_{\odot}$ and a $\Upsilon(r)$ gradient that resembles the gradients of real galaxies. I.e. it consists of an increased $\Upsilon^{\mathrm{sim}}_{\mathrm{cen}} = 2$ inside $r \sim 1 - \SI{2}{kpc}$ that is two times larger than the main-body $\Upsilon^{\mathrm{sim}}_{\mathrm{main}} = 1$ (see Appendix~\ref{ap:stresstest} for details). We model this mock galaxy with exactly the same approach that we use for our observed galaxies.

We find that the input main-body $\Upsilon$ could be successfully recovered ($\Upsilon_{\mathrm{main}} = 0.93 \pm 0.11$ at $r_{\mathrm{main}}= \SI{1.9}{kpc}$). We have already argued above that we do not expect too strong biases of axisymmetric models around the core region of massive galaxies. This is supported by the result of these mock tests. In addition, the tests show that even when modelling a steep DM halo with a cored profile (i) the main-body mass-to-light ratio is highly robust and (ii) the spatial confinement of the gradient can be well recovered.

The central mass-to-light ratio of the simulation was overestimated by a factor of roughly $1.6$ ($\Upsilon_{\mathrm{cen}} = 3.16 \pm 1.13$). As argued above, triaxiality may not be the main driver behind this bias. There are several other reasons, why the central $\Upsilon_{\mathrm{cen}}$ is more difficult to measure than the mass in the main body. First, the central potential is dominated by the black hole (in this case $r_{\mathrm{SOI}} \sim \SI{0.5}{kpc} \sim r_{\mathrm{cen}}$) and the stars contribute less and less to the total mass. Second, the line-of-sight is more and more dominated by foreground and background light while the signal from the region physically close to the center is weak. Hence the increased uncertainty in the very central parts of the gradient is not entirely surprising. However, where the bias comes from is not clear yet. We note that the black hole is recovered within one sigma ($M_{\mathrm{BH}} = (7.4 \pm 2.7) \times 10^9 M_{\odot}$). Likewise, the central DM halo mass of the simulation is recovered within $10\%$. 

Overall this stress test leaves the possibility that the central Salpeter mass normalization which we inferred for our sample might actually be an upper limit. We plan fully triaxial gradient models for our galaxies as well as more extended tests with simulations to clarify this issue. 

Nonetheless, our finding that the IMF of the sample galaxies becomes MW-like at $\SI{1}{kpc}$ is a very robust result.
As we have seen in Section \ref{subsec: IMF}, this in itself is already an important step in potentially closing the gap between different IMF probes. 

\subsubsection{Uncertain cases: NGC~1407 and NGC~1332}

Two galaxies in our sample deserve deeper consideration. Firstly, while our axisymmetric dynamical models provided good fits to all available data for all galaxies, there was an \textcolor{black}{problem} with fitting our 2D kinematic data for NGC~1407, which we encountered for none of the other seven galaxies:
As shown in Figure~\ref{fig:1407map}, our dynamical models, while producing overall excellent fits to the kinematics (see also Table \ref{tab:resultsTab} and Figure \ref{fig:Losvdfitn1407}), were unable to reproduce the velocity signal $|v_{\mathrm{rot}}| > 0$ along the minor axis of the galaxy (the y-axis of the maps in the Figure). \textcolor{black}{As a counterexample, in Figure~\ref{fig:0307map}, we show kinematic maps of NGC~307, for which the full 2D rotation signal is captured by our dynamical models. The difference lies in the fact that} the velocity field of NGC~1407 is visibly distorted, the peaks of $v_{\mathrm{rot}}$ not being aligned with the major axis (M+23)
and the $v_{\mathrm{rot}} = 0$ line not being aligned with the minor axis but pointing along a diagonal direction outside the central few arcseconds. \textcolor{black}{The full extent of this kinematic pattern cannot be captured by axisymmetric models. Nonetheless, the kinematic signal in each quadrant can be \textit{individually} reproduced by the axisymmetric models. The only exception to this is the rotation directly on the minor axis, which cannot be reproduced with tube orbits. However, in NGC~1407 as well as in all other core galaxies in our sample, the velocity signal is overall very weak and thus carries little of the galaxy's energy.
Hence, mismatch in the rotation can be expected to result only in a small mass bias.}

The velocity pattern could be well caused by the galaxy being triaxial. However, the velocity signal is not very strong and we have seen above from the simulation test that triaxiality is not necessarily a driver for strong biases. In fact, our measured SMBH corresponds to a $r_{\mathrm{SOI}} = (2.41 \pm 0.546)\arcsec$ consistent with $r_b = \SI{2.01}{\arcsec}$ (R+13), as is expected for cored ETGs \citep{Thomas2016}. Furthermore, the core of this ETG (as well as that of the other cored ETGs in our sample), shows the characteristic orbit structure of a core, with the orbital anisotropy parameter $\beta$ transitioning from positive, i.e. radial anisotropy, $\beta \sim 0.55$ outside the core region to negative, i.e. tangential anisotropy, $\beta \sim -0.55$ within the core. This is predicted by numerical simulations of core formation \citep{Rantala2018}. We therefore consider the central $<\alpha_{\mathrm{cen}}>$ of NGC~1407 robust.

However, at large radii the IMF normalisation in NGC~1407 is worryingly low, even considering uncertainties in the mass decomposition, $\alpha^{\mathrm{tot}}_{\mathrm{main}} = 0.44 \pm 0.18$. ``Worrying'', because the outer parts of massive galaxies are thought to be assembled from material of less massive galaxies and satellites -- objects for which a MW-like IMF is strongly expected. Therefore, either we have accidentally detected a rare bottom-light IMF at $r_{\mathrm{main}}$ and $\Upsilon(r)$ rises again past $r_{\mathrm{main}}$ to $\alpha \sim 1$ (so that the $\Upsilon$ profile rises at both ends), or -- more likely -- our dynamical model is somewhat biased. Strong triaxiality (stronger than in the tested simulation) could in principle explain such a low mass normalisation. Another possibility might be that the distortions in the velocity field do not originate from triaxiality but instead the galaxy might be slightly out of equilibrium (e.g. due to a recent merger). In any case, we 
revise the sample-average of the outer IMF normalization from Section~\ref{subsec: IMF} by excluding this ETG from the calculation, $<\alpha_{\mathrm{main}}> = 1.05 \pm 0.18$ ($<\alpha^{\mathrm{tot}}_{\mathrm{main}}> = 1.25 \pm 0.15$). Excluding also NGC~4751, we find $<\alpha_{\mathrm{main}}> = 1.03 \pm 0.19$ ($<\alpha^{\mathrm{tot}}_{\mathrm{main}}> = 1.23 \pm 0.15$).
However, the conclusions of our study remain unchanged.

The second galaxy that deserves closer inspection is NGC~1332. While we have tested our setup on a (static) triaxial merger remnant, real galaxies can be even more complex and involve a rotating gravitational potential. 
Specifically for NGC~1332, we had inferred the possible presence of an end-on bar from a  comparison of the galaxy's 2D stellar kinematics with the kinematical signature of boxy/peanut bulges of simulated disk galaxies from \citet{Iannuzzi2015} (see Section 6.3 of M+23 for a detailed discussion). 

For the dynamical fits, we did not encounter any significant issues with reproducing both the MUSE and SINFONI LOSVDs for this galaxy (see Figs.~\ref{fig:Losvdfit} and \ref{fig:radKin}), which is also evidenced by the value of $<(\chi^2 + m_{\mathrm{eff}})/N> = 0.76$. 

While the main body $\alpha_{\mathrm{main}} \sim 0.6$ is also somewhat low for this galaxy, considering the total 
$\alpha^{\mathrm{tot}}_{\mathrm{main}} \sim 0.9$, the difference to a MW IMF can easily be attributed to the uncertainties of the dynamical mass decomposition. 

We have here used the $M_{\mathrm{BH}}$ of \citet{Barth2015} from the circumnuclear gas disc detected with ALMA, for which they measured $v_{\mathrm{disk}} \sim 450 - \SI{400}{km/s}$ at $r \sim \SI{1}{\arcsec}$. Not fixing the central black hole to the value of \citet{Barth2015} produces a $M_{\mathrm{BH}} = (1.58 \pm 0.43) \times 10^9 M_{\odot}$, which would be consistent with our results from R+13, but in excess of the ALMA $M_{\mathrm{BH}}$ by a factor of two. At $\SI{1}{\arcsec}$ this  higher-$M_{\mathrm{BH}}$ model
would imply a circular velocity $v_{\mathrm{circ}} \gtrsim \SI{500}{km/s}$ which is higher than the ALMA measurements and a central stellar mass normalisation that would be smaller by a factor $1.3$, though still above Salpeter, $\alpha_{\mathrm{cen}} = 1.65 \pm 0.49$. While it is possible that the ALMA measurement is biased low, the higher spatial resolution of the ALMA data makes it more plausible that the mismatch is due to an end-on bar which our current models do not account for.

However, for the models which we present here and use the ALMA $M_{\mathrm{BH}}$, the circular velocity  at $\SI{1}{\arcsec}$, $v_{\mathrm{circ}} = (459 \pm 43.3) \ km/s$
is consistent with the ALMA data. Moreover, the stellar $\Upsilon$ derived by \citet{Barth2015} is consistent with the central value of our gradient models (Figure~\ref{fig:MLgradsA}). 
For all these reasons from our dynamical point-of-view, we see little reason to discount our measurements of $\Upsilon_{\mathrm{cen}}$ at this stage.

\subsubsection{Can DM explain the Gradients?}
In our simulation tests in Section \ref{sec:Gradsim}, we have demonstrated that our assumption about the inner slope of the DM halo has no significant influence on the recovered stellar mass-to-light ratio. \citet{Cappellari2012Nature} also found that the dynamically inferred increased stellar mass normalizations of massive elliptical galaxies do not depend strongly on the assumed DM halo profile. 

Of course, under extreme assumptions this independence breaks down. In particular, if one considers a component of dark matter that follows the light and thus would become indistinguishable from stellar mass. Such a component could explain our central measured mass excess $<\alpha> \sim 1.5$ while the IMF would still be Kroupa in all galaxies at all radii. On average, the fraction of mass in our fitted DM components is about three percent at $r_{\mathrm{main}} \sim \SI{1}{kpc}$. 
Considering the values of $\alpha^{\mathrm{tot}}_{\mathrm{main}}$ at that radius (see Table \ref{tab:gradientsTab}), we can see that even if we assume that all the dynamical mass in excess of a Kroupa stellar mass would be dark matter, the DM fraction would still remain low. Hence, if we also assume that the IMF is Kroupa in the very centre, the DM fraction would have to rise from three to almost \textit{fifty} percent over a mere $\SI{1}{kpc}$ towards the galactic center. This would be difficult to explain. 

In summary, there is no reason to believe that our dynamical gradients are biased towards a centrally increasing stellar mass-to-light ratio due to our adopted DM halo profiles. In case of an exotic DM component that follows the light, a Kroupa IMF in all galaxies at all radii would still be consistent with the data though unlikely (but see Section~\ref{subsec:topheavy}).

\subsection{Origins of bottom-heavy galactic centers}
\label{subsec:axisymm}
In the following, we briefly speculate as to possible origins of the bottom-heavy IMF which we have potentially measured in the centers of the galaxies.

If the IMF is different in the centers of ETGs, necessarily, the conditions and/or mechanisms of the originating starbursts of the stellar populations had to be very different from those found in any environment in the MW.

Recent studies have proposed that the conditions in the centers of ETGs when they were first assembled, $z \gtrsim 2$ were unlike any environment found in the MW. In this picture, massive compact galaxies, which are up to 60 times denser than local ETGs and virtually absent from the local universe are the progenitors of the centers of massive ETGs. It is proposed that they have formed on very short times scales from the in-fall and compaction of cold gas triggering intense in-situ star-formation, followed by extreme quenching from stellar- and/or AGN feedback, turning them into ``red nuggets''. Around these nuggets stellar components accumulate via merger- and accretion-driven inside-out-growth, forming what will become local ETGs \citep{Bezanson+09, Oser2010, Barro2013, Nelson2014, vanDokkum2015, Zolotov2015, Barro2016}. It has been suggested that the intense nature of the starbursts which formed these red nuggets, meaning the exceptional intensity of the gravito-turbulent fragmentation of the in-falling gas, where radiation pressure is ramped up by the rate of star-formation, competing with gravitational collapse, could have created a relative excess of low-mass dwarf stars in the centers of ETGs \citep[e.g.][]{Laesker2013, ChabrierHennebelleCharlot2013, Belli2014, vandeSande2013}.

While this matter remains speculative, the fact that the correlation of $\alpha$ with [Mg/Fe] has been found to be tighter than with $\sigma$ \citep{Conroy&vanDokkum2012}, has been seen as indication that rapid starbursts are correlated with the excess production of dwarf stars, as the above scenario also suggests.
In our companion paper (Parikkh et al. submitted to MNRAS), however, 
we show that while all galaxies in our sample are strongly enriched in [Mg/Fe], [Mg/Fe] $\sim 0.3 - 0.4$, we do not find radial gradients for this abundance. The [Mg/Fe] - $\alpha$ correlation has also been called into question by other studies \citep{Smith2014, LaBarbera2015}.

On the other hand, if the above formation scenario for ETGs holds true, we would expect central gradients of the IMF to correlate more with physical radius than radius relative to $r_e$ (as the outer parts were assembled later on), which, as we have shown, is the case for our models. This had also previously been suggested by \citet{vanDokkum2017}.

The main conceptual \textcolor{black}{problem} with this framework is our understanding of the merger hierarchies of massive ETGs: High-mass ETGs are thought to have assembled from dry major mergers of less massive ETGs \citep[e.g.][]{NietoBender1987, KormendyBender1997, Hopkins2009, vanderWel2009, Lauer2012, KormendyBender2013}. Numerical merger simulations suggest that in dry major mergers the compact central regions of the progenitors sink to the center where a SMBH binary sling-shots stars to larger radii and forms a (cuspy) core \citet{Rantala2018, Rantala2019}. If the merger is wet, the new-born core is ``covered up'' by new star formation, which we expect to produce stars in line with a MW IMF (since the conditions around nugget-formation have past at this point). If the merger is dry, the diluted core remains as-is \citep[e.g.][]{KormendyBender1997, Kormendy1999, Kormendy2009}. Either way therefore, we expect that IMF gradients in massive galaxies become less steep the more they merge. We note that the two galaxies with the highest central mass normalizations in our sample, NGC~1332 and NGC~4751, are both power-law galaxies. On the other hand the least massive galaxy in our sample, NGC~307, has the smallest $\alpha_{\mathrm{cen}}$. It remains to be seen if larger samples of galaxies modelled with $\Upsilon$-gradients support the implied dichotomy between cored and power-law ETGs. 

Finally, the fact that our gradients seem to all have the same spatial scale of $\sim \SI{1}{kpc}$ could point to a characteristic size for the detectable remnants of red nuggets in the centers of ETGs. As of now, it is unclear what physical processes are the driver for the spatial size of our measured IMF gradients.

\subsection{On the possibility of top-heavy galactic centers}
\label{subsec:topheavy}

Similar to the ``DM following stars'' scenario, BHs could follow the luminous component and explain the high mass normalizations $\alpha_{\mathrm{cen}}$ which we found. The only difference here would be that the IMF would then no longer be MW-like, as the BHs would be the remnants of a population of giant stars which made up a much larger fraction of the IMF than in the MW, i.e. the IMF would be top-heavy. This scenario is rarely considered since SSP models cannot probe for top-heaviness, as once the massive stars become remnants they become invisible to spectral analysis. But not to dynamical modeling, which simply measures (enclosed) mass as a function of radius. As such, our results are fully consistent with a central top-heavy IMF -- any mass decomposition follows from other assumptions.

There is yet no consensus on the possible origins of this kind of IMF in the centers of ETGs. However, first-epoch JWST NIRCam imaging from the \textit{Cosmic Evolution Early Release Science} (CEERS) Survey has provided some insight into the possibility of an early-universe IMF evolution in this direction: For a sample of galaxies with $z \gtrsim 9$, \citet{Finkelstein2023} have found an excess of UV luminosity per unit halo mass at $z \sim 11$ relative to extrapolations of the UV luminosity function at lower redshifts. They argue that this excess could be accounted for if star formation in these galaxies was dominated by a top-heavy IMF. This, in principle, would be compatible with predictions of the fragmentation of metal-less gas into stars \citep{Bromm&Larson2004}, i.e. with predictions of the IMF in a very low-metallicty environment. Since these galaxies are very compact, $r_e \sim \SI{0.5}{kpc}$, some of the arguments that we have used for the possibility of bottom-heavy red nuggets ending up in the centers of massive ETGs would apply for top-heavy progenitors. But would these top-heavy populations remain intact in the centers of ETGs? As with the bottom-heavy centers some level of dilution of the IMF is expected. Particularly if core scouring events on similar spatial scales as these centers are sustained. It is also unclear why the excess of black holes from these populations would not be driven to the very center by dynamical friction and merge with the central SMBH. Nonetheless it will be interesting to see what further probes of the early-universe IMF from the JWST era will uncover on this matter.

\section{Summary and Conclusions}

We have constructed state-of-the-art axisymmetric Schwarzschild models to systematically probe for the existence of IMF variations within seven massive early-type galaxies. Our study utilises novel dynamical techniques to improve the accuarcy of the results: 
\begin{itemize}
    \item We consistently use non-parametric LOSVDs both in the center (from AO-based SINFONI data with high spatial resolution to resolve the central SMBHs) and for the galaxy main body (from high-SNR MUSE spectroscopy, \citealt{Mehrgan2023}).
    \item We use mass models that allow for radial gradients of the stellar mass-to-light ratio $\Upsilon(r)$.
    \item We use a generalized model selection technique to account for the varying model flexibility of Schwarzschild models \citet{Lipka2021,ThomasLipka2022}.
\end{itemize}
In previous papers we have shown that using non-parametric LOSVDs and the generalised model selection allows us to break known degeneracies and to avoid potential biases in dynamical models even in the more complex case of triaxial galaxies \citep{deNicola2022,Neureiter2023a}. We showed that with the above improvements dynamical mass determinations at the 10\% precision level are possible. 

Applying these models, we have found radial gradients of $\Upsilon$ in all seven galaxies, with $\Upsilon(r)$ always increasing towards the center of the galaxies. We have found the following results concerning these gradients:

\begin{itemize}
    
\item Gradients of $\Upsilon(r)$ are concentrated on very small spatial scales of less than $\sim \SI{1}{kpc}$.

\item The total dynamical mass-to-light of the galaxies has a minimum and this minimum occurs at roughly $r_{\mathrm{main}} \sim \SI{1}{kpc}$ from the center. Under the assumption that the stellar mass-to-light ratio does not increase with radius this point provides a strong constraint for $\Upsilon_{\mathrm{main}}$ in the main body of the galaxies.

\item Relative to the stellar mass-to-light ratio of the main body of the galaxy, $\Upsilon_{\mathrm{main}}$, the inner $\Upsilon_{\mathrm{cen}}$ increases on average by a factor $2.6$.

\item Models without gradients fit the data worse and yield $\Upsilon$-values between the $\Upsilon_{\mathrm{cen}}$ and $\Upsilon_{\mathrm{main}}$ of gradient models. Since gradients occur on small spatial scales, models without gradients can lead to an overestimation of the stellar mass content of a galaxy by up to a factor of $\sim 1.5$.

\item Models with gradients yielded $M_{\mathrm{BH}}$ that are on average $25 \%$ smaller than for constant-$\Upsilon$ models in our sample.
\end{itemize}

In order to probe for gradients of the IMF, we calculated radial profiles of the IMF mass normalization $\alpha$ relative to SSP measurements assuming a Kroupa IMF. Our probes revealed the following IMF trends:

\begin{itemize}

\item At $r_{\mathrm{main}} \sim \SI{1}{kpc}$ we find an IMF normalization which is on average Kroupa-like $<\alpha_{\mathrm{main}}> = 1.03 \pm 0.19$. Considering the total mass at this radius, which is independent of any assumption related to the mass decomposition, we find 
$<\alpha^{\mathrm{tot}}_{\mathrm{main}}> = 1.23 \pm 0.15$. A Salpeter-level bottom-heaviness is inconsistent with the dynamics for five out of seven galaxies in our sample at a one- to two-sigma level at this radius.

\item In the center of the galaxies we find concentrated regions of increased mass normalizations with $\Upsilon$-gradients rising to roughly a Salpeter-like normalization, $<\alpha_{\mathrm{cen}}> = 1.54 \pm 0.15$.

\item In the center, the DM contribution essentially vanishes. Therefore,
for many galaxies, there is a spatial interval that is still central enough for DM to be insignificant, but is at the same time outside the SOI of the central SMBH, so that $\alpha \sim \alpha^{\mathrm{tot}}$, i.e. $\alpha$ becomes independent of any assumption related to the mass decomposition. Considering this total dynamical mass, five out of seven galaxies in our sample are consistent with a Salpeter- or higher-level bottom-heaviness of the IMF in the very center.

\item Taking into account aperture effects and the difference between models with and without gradients our results produce similar, but overall less extreme levels of bottom-heaviness compared to many previous studies.

\item Not taking into account gradients biases $\alpha$ high.

\item The dynamically detected gradients are so spatially concentrated that even within central apertures as small as $r_e/8$ (typical for SSP measurements) aperture effects can affect the comparison.

\end{itemize}

Our study confirms previous claims in favor of the non-universality of the IMF. The main issue with this claim is that while the different SSP, dynamics and lensing studies all agree on the fact of non-universality, and sometimes the same IMF-trends, they often do not produce consistent results for individual galaxies. 
\citet{Bernardi+18} and \citet{Lyubenova+2016} already suggested that gradients put play a crucial role in matching different IMF probes. 
Our dynamical evidence for very concentrated $\Upsilon$-gradients makes the necessity of matching spatial apertures for comparisons between different works even more crucial. Moreover, the gradients that we find are so spatially concentrated that taking into account central SMBHs is important.

Modelling larger samples of galaxies with next-generation Schwarzschild models similar to the ones used here and direct comparisons with SSP models galaxy-by-galaxy will be important to constrain the IMF better. We plan to do this in a future paper, also combining gradient models with triaxial symmetry \citep{Neureiter2021, deNicola2022,Neureiter2023}.

\section*{Acknowledgement}
We acknowledge project/application support by the Max Planck Computing and Data Facility. All dynamic computations were performed on the HPC system Raven and Cobra at the Max Planck Computing and Data Facility.
\appendix

\section{Bulge/disc decomposition and deprojection of NGC~4751}
\label{ap:ngc4751phot}

While we used the same NICMOS2 high resolution imaging as in \citet{Rusli2013b}, we supplemented this with more recent large scale K-band imaging from the near infrared camera VIRCAM at the 4m VISTA telescope at La Silla \citep{Vista1, Vista2}. The imaging data consists of two 180 second exposures taken in the context of the VISTA hemisphere survey (Program ID 179.A-2010) and was taken from the ESO archive.

\begin{figure*}
\centering
\includegraphics[width = 0.4\columnwidth]{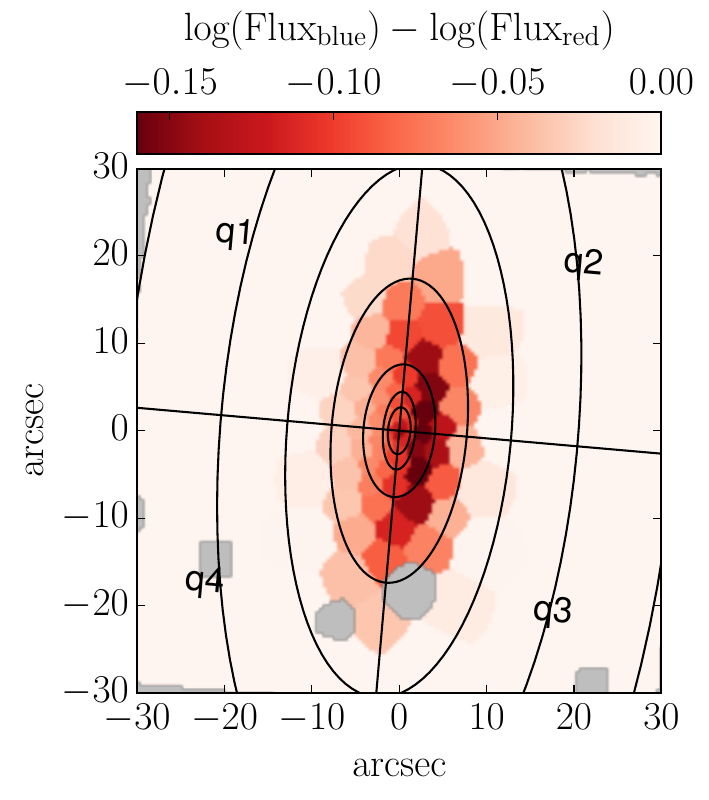}
 \caption{Dust map of NGC~4751, derived from the binned MUSE datacube (M+23) by computing the difference between the logarithms of the integrated flux between $7870 - \SI{8500}{\angstrom}$ and $4870 - \SI{5500}{\angstrom}$. Grey areas indicate regions spatially masked during the kinematic analysis (cf. Figure 13 in M+23). The line at $PA = 176^{\circ}$ indicates the major axis, the line orthogonal to this, the minor axis, while $q1-q4$ are labels for the quadrants used in the modeling. We also show exemplary isophotal ellipses. The figure shows that every spatial bin of our kinematic input data which lies on the major axis is contaminated by dust all the way to the effective radius $r_e = \SI{22.76}{\arcsec}$. The western (right) regions are affected more significantly than the eastern. The south-side of the major axis (bottom) appears to be slightly worse affected than the north-side major axis. This makes q3 the quadrant which is the most affected by dust.}
   \label{fig:dust4751}
\end{figure*}

The decomposition was derived from simultaneous fits to the VISTA and HST images using the “multimfit” extension of imfit \citep{Erwin2015} which allows us to fit the same model to multiple images. There was also very strong dust contamination in the nuclear region and along the major axis (see Figure \ref{fig:dust4751}), which we masked during the fit with imfit. The dust disproportionally affects one side from the major axis of the galaxy more than the other.
Due to the extent of the dusty regions, covering most of the galaxy's major axis within $r_e$, some of the LOSVDs from M+23 for this galaxy, which were derived in the MgB region, are likely affected by them. This is discussed in Section \ref{subsec:kinproblems}.

Our best fit was formally constructed from 4 components, which are listed in Table \ref{tab:4751decomp}. We decided to make component 3 the ``disc'', as it was the most flattened component, and we combined components 1, 2 and 4 into one ``bulge'' component.

During the dynamical  modeling-process, we sample $\Upsilon_{\mathrm{disc}}$ on the same grid as $\Upsilon_{\mathrm{bulge},i,f}$. Therefore, if our decomposition was in error, in the sense of there not being two distinct morphological components in the same way as there are in the other two power-law galaxies in our sample, the modeling can still find a solution which essentially amounts to just fitting one (bulge) component.

\begin{table*}
\centering
 \begin{tabular}{l c c c c c l}
 \hline
 \hline
Component            &  $\epsilon$  & PA [$^{\circ}$]& n & $r_e$ [$\arcsec$] & $\%$ of total light & description\\
\hline
1                    &  $0.21$   & $179$ & $3.2$ & $1.1$  & $22.6$      &  nuclear component \\
2                    & $0.44$    & $175$ & $0.8$ & $3.5$ & $15.9$       &  inner part of the main body \\
3                    & $0.62$    & $176$ & $1.7$ & $21.9$  & $48.3$     &  outer part of the main body \\
4                     & $0.23$   & $176$ & $0.8$ & $73.8$ & $13.1$        &  outer envelope 
\\
\hline
\end{tabular}
\caption{Photometric decomposition of NGC~4751 with imfit, showing the ellipticity, position angle, effective radius, and fraction of the total galaxy light for each component. The position angle of the fourth component was fixed to the one of the third.}
\label{tab:4751decomp}
\end{table*}
As with the other galaxies, we used the algorithm of \citet{Magorrian1999}, which utilizes a penalized log-likelihood function to produce 3D non-parametric axisymmetric luminosity density distributions $\nu_{depro}(\bm{r})$ which are consistent with the 2D input surface brightness profiles, under the assumed viewing angle $i$. We deprojected NGC~4751 which is close to edge on for $i = \ang{90}$, with the bulge and disc components treated separately.

\section{Kinematic fits}
\label{ap:kinfits}

\begin{figure*}
\centering
\includegraphics[width = 0.48\columnwidth]{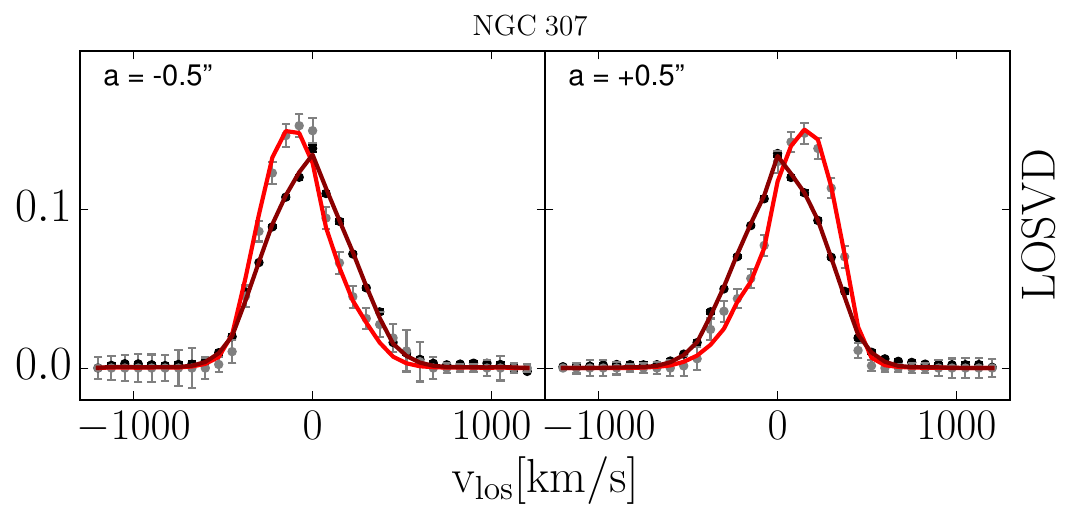}
\includegraphics[width = 0.48\columnwidth]{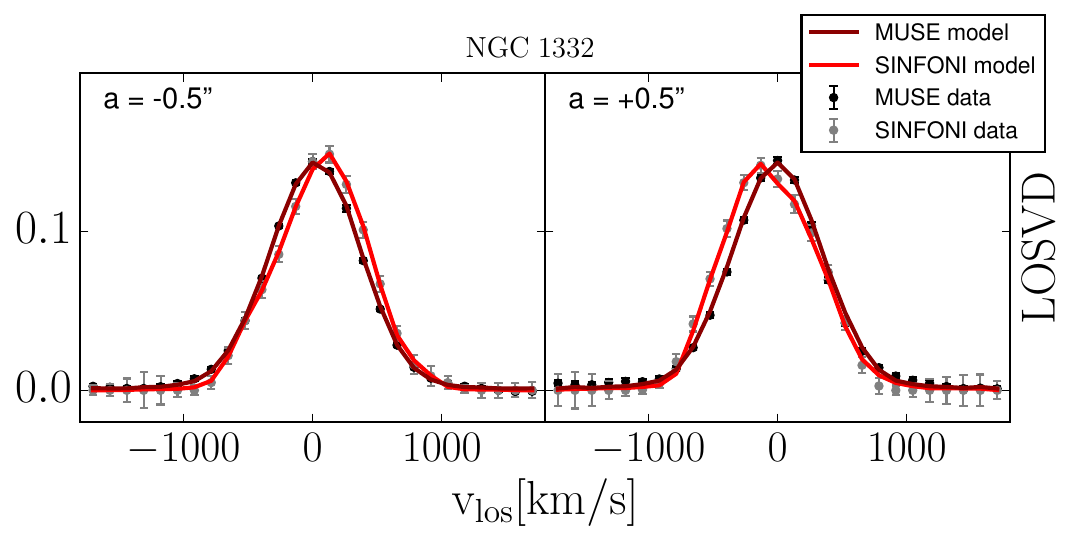}
\includegraphics[width = 0.48\columnwidth]{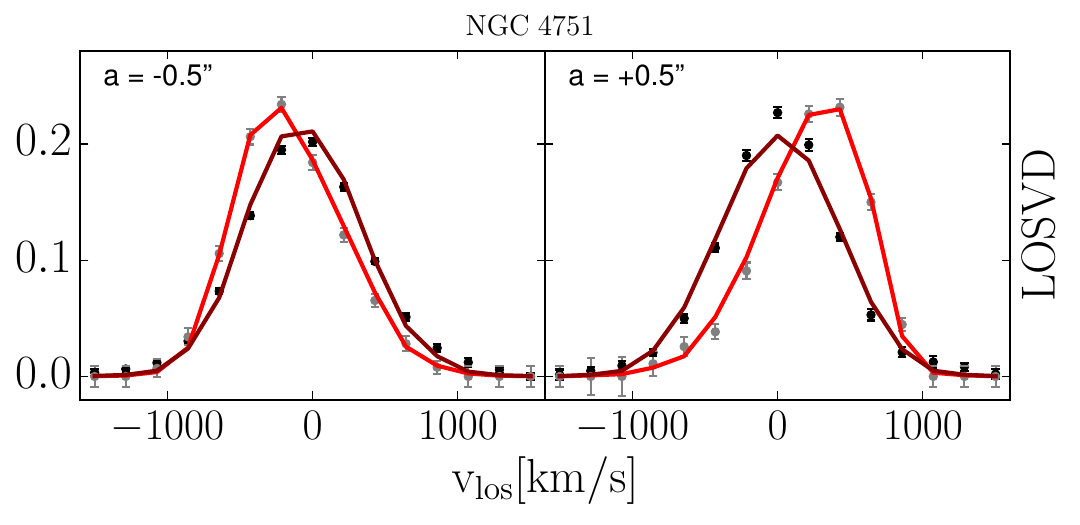}
\includegraphics[width = 0.48\columnwidth]{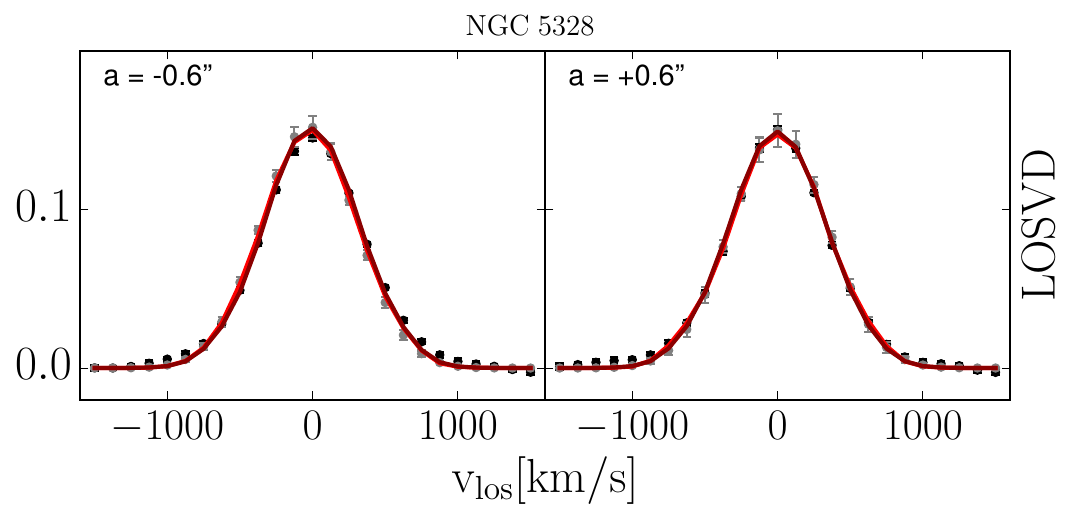}
\includegraphics[width = 0.48\columnwidth]{Figure12d.pdf}
\includegraphics[width = 0.48\columnwidth]{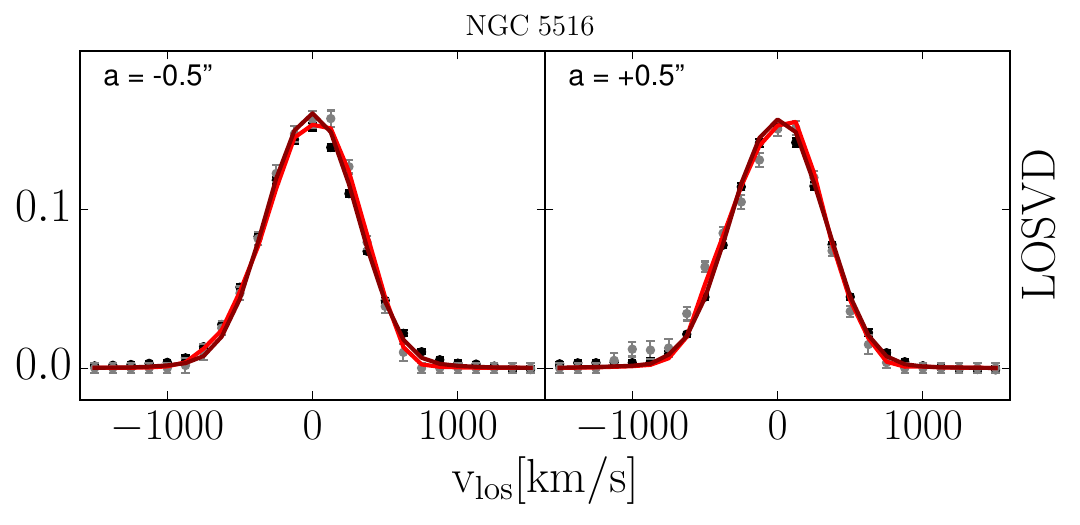}
 \caption{Non-parametric LOSVD fits from the centers of the sample galaxies except NGC~1407, which we present separately in Figure \ref{fig:Losvdfitn1407}. We show spatially overlapping fits to the MUSE and SINFONI LOSVDs along the major axis for a distance $a$ from the center. MUSE data- and model-LOSVDs are shown in black and dark red, respectively, SINFONI data- and model-LOSVDs in grey and light red. All data-LOSVDs are non-parametric, except the SINFONI data of NGC~5328, whose losvds are simple Gaussian LOSVDs (see Section \ref{subsec:kinproblems}). The uncertainties of these LOSVDs are adopted from the original non-parametric LOSVDs of the SINFONI data for this glaxy.}
   \label{fig:Losvdfit}
\end{figure*}

In Figure \ref{fig:Losvdfit} we present LOSVD fits to central MUSE and SINFONI LOSVDs in spatially overlapping regions for all galaxies, except NGC~1407 which we present separately in Figure \ref{fig:Losvdfitn1407}. As discussed in Section \ref{sec:modelsetup}, MUSE and SINFONI LOSVD-sets are generally consistent with each other within the uncertainties.   Differences in the shapes of the LOSVDs arise due to spatial, spectral, and seeing differences, particularly as the SINFONI kinematics are supported by adaptive optics. Baring fundamental kinematic inconsistencies with either set, we expect the stellar dynamical models to be able to fit both sets equally well at the same spatial location as the models take the above mentioned differences into account. Fortunately this is the case for our sample, and we produced good fits to both kinematic data sets, $(<\chi^{2} + m_{\mathrm{eff}})/N \sim 0.8$ (see Table \ref{tab:resultsTab}), which indicates a low amount of template-mismatch in the MUSE data from M+23, as we discuss in Section \ref{subsec:kinproblems}.

In  Figure \ref{fig:radKin} we show the full radial kinematic profiles of the MUSE, SINFONI, and dynamical model LOSVDs parameterized by fourth order Gauss-Hermite polynomials for all galaxies. We here add some special notes on the kinematics and kinematic fits of NGC~1332, 4751 and 5328:

{\textbf{NGC~1332}}: The radial kinematic profiles for NGC~1332 show that we can simultaneously reproduce both the MUSE and SINFONI kinematics over the full spatial coverage of our data, despite the bar-like kinematic signatures noted in M+23. There we had noted a particular $h_3$ butterfly-shape, which we can see in the radial profile as the crisscrossing of the $h_3$-model lines from two sides of the galaxy at around $r \sim \SI{6}{\arcsec}$ and $\SI{15}{\arcsec}$. The only outliers are within $\sim \SI{0.5}{\arcsec}$. Here the models slightly underpredict the $h_4$ of the MUSE data. While the difference appears significant in these figures, it is in fact minuscule when considering the underlying non-parametric LOSVDs (the actual concern of our dynamcal models). The LOSVDs belonging to NGC~1332 which we present in Figure \ref{fig:Losvdfit}, are from this problematic region.

{\textbf{NGC~5328}}: For this galaxy, radial kinematic profiles are also overall good, but within the SINFONI coverage, $r = \SI{1.5}{\arcsec}$, the $h_4$ of the MUSE data is significantly underpredicted by our models, much more so than for NGC~1332. This is due to the obstruction of one of the two CO band-heads from which the SINFONI kinematics were measured. This produced spurious $h_3$, indicating that there were either not enough constraints on the \textit{full} LOSVD-shape in the face of possible contamination from sky emission. Therefore R+13 corrected the LOSVDs such that they subtracted the higher order $h_3$ and $h_4$ signal, resulting in a suppression of light at higher velocities, which is very much present in our MUSE kinematics. These differences are shown for the non-parametric LOSVDs in Figure \ref{fig:Losvdfit}. There, these differences are also relatively small, but nonetheless show that the MUSE-model LOSVD-signal is suppressed around $v_{\mathrm{los}} \sim \pm \SI{1000}{km/s}$. This slightly biased the fit to MUSE LOSVDs in the center, as seen in Figure \ref{fig:radKin}, whereas the SINFONI LOSVDs were fit well (since there were more SINFONI LOSVDs within $r = \SI{1.5}{\arcsec}$ the latter dominated the fits in the central regions.): Within the SINFONI FOV our MUSE data has $h_4 \sim 0.03 \pm 0.01$. The models however, produce a $h_4$ that is roughly zero, which corresponds to the $h_4$ of the SINFONI data/models. As a consequence, the $<(\chi^2 +m_{\mathrm{eff}})/N> = 0.99$  while still good, is the largest in our sample. The SOI of the SMBH, $r_{\mathrm{SOI}} = (0.50 \pm 0.12)\arcsec$, is also the only one amongst our four cored galaxies which is inconsistent with the break-radius of the core, $r_b = (0.85 \pm 0.04)$. Typically in cored galaxies $r_b \sim r_{\mathrm{SOI}}$ \citep{Thomas2016}. For dynamical models without SINFONI LOSVDs, $<(\chi^2 +m_{\mathrm{eff}})/N> = 0.93$ becomes lower. However, this produces spurious results: The SMBH and SOI become even less consistent with $r_b$ as $M_{\mathrm{BH}}$ becomes significantly smaller, $M_{\mathrm{BH}} \sim 0.7 \times 10^9 M_{\odot}$. The $\Upsilon$-gradient, at the same time, becomes much steeper, $\Upsilon_{\mathrm{cen}} \sim 9$, $\Upsilon_{\mathrm{main}} \sim 0.6$ (V-Band). This essentially amounts to $\Upsilon(r)$ vanishing entirely into the DM. Put in terms of the IMF, this would mean a far below-MW bottom-light IMF normalization $\alpha_{\mathrm{main}} \sim 0.2$, compared of the perfectly MW-like IMF $\alpha_{\mathrm{main}} \sim 1$ which we found for our full models (see Table \ref{tab:gradientsTab}). As we argue for NGC~1407, such a bottom light outer IMF is extremely unlikely to be physical. We therefore suggest that the use of the AO-assisted SINFONI data might have biased our SMBH measurement, but still provided \textit{necessary} constraints on the larger shape of the $\Upsilon$-profile, via constraints on the central orbital anisotropy and SMBH. Finally, in M+23, we had noted a small counter-rotating region in the central few arcseconds of our MUSE FOV. Closer inspection of  Figure \ref{fig:radKin} shows that the lines tracking our model-$v_{\mathrm{rot}}$ for the MUSE kinematics from two sides of the galaxy cross and switch signs at around $r \sim \SI{3}{\arcsec}$ to fit this counter rotating region correctly.

{\textbf{NGC~4751}}: Considering the distribution of the dust in NGC~4751 (see Figure \ref{fig:dust4751}), the dust appears to be somewhat evenly distributed within $r_e$.
However, the distribution of dust is slightly more extended on the south and west side from the center. The quadrant which we had to exclude, q3, is the south western quadrant of the galaxy. The effects of the dust on the kinematics could potentially explain why the $(\chi^2 + m_{\mathrm{eff}})/N$ of our fits was higher in this galaxy. Considering the radial profiles of the dynamic fits parameterized by Gauss-Hermite polynomials (see Figure \ref{fig:radKin}), the main problem with the fits appears to be an elevated $h_4$ signal within the central $\SI{4}{\arcsec}$ for the MUSE data which the models cannot reproduce. Considering the non-parametric LOSVDs from this region (see Figure \ref{fig:Losvdfit}), we can see that while the fit to the SINFONI LOSVDs is quite good, the models have problems reproducing the LOSVD signal of the peak of the MUSE LOSVDs (roughly between $\pm \SI{250}{km/s}$). This problem appears to be worse on the side where $v_{\mathrm{rot}} < 0$ (right side), which corresponds to the southern, dustier side of the galaxy. At large radii ($r \sim 20 - \SI{30}{\arcsec}$ in Figure \ref{fig:radKin}), there also appears to be some bias in $h_3$ -- a telltale sign of template-mismatch. At the same time $h_4$ at radii large than $\SI{10}{\arcsec}$ are biased somewhat low. In the kinematic maps shown in Figure 13 of M+23 it can be seen that this bias towards low $h_4$ originates from one side of the galaxy, where $h_4$ becomes overall negative, the south side, whereas the north side has overall positive $h_4$. This again makes dust the likely candidate. The large radius template mismatch could also be associated with this, as the template selection was performed in the same spectral region as the main kinematic fits (M+23).

\begin{figure*}
\centering
\includegraphics[width = 0.49\columnwidth]{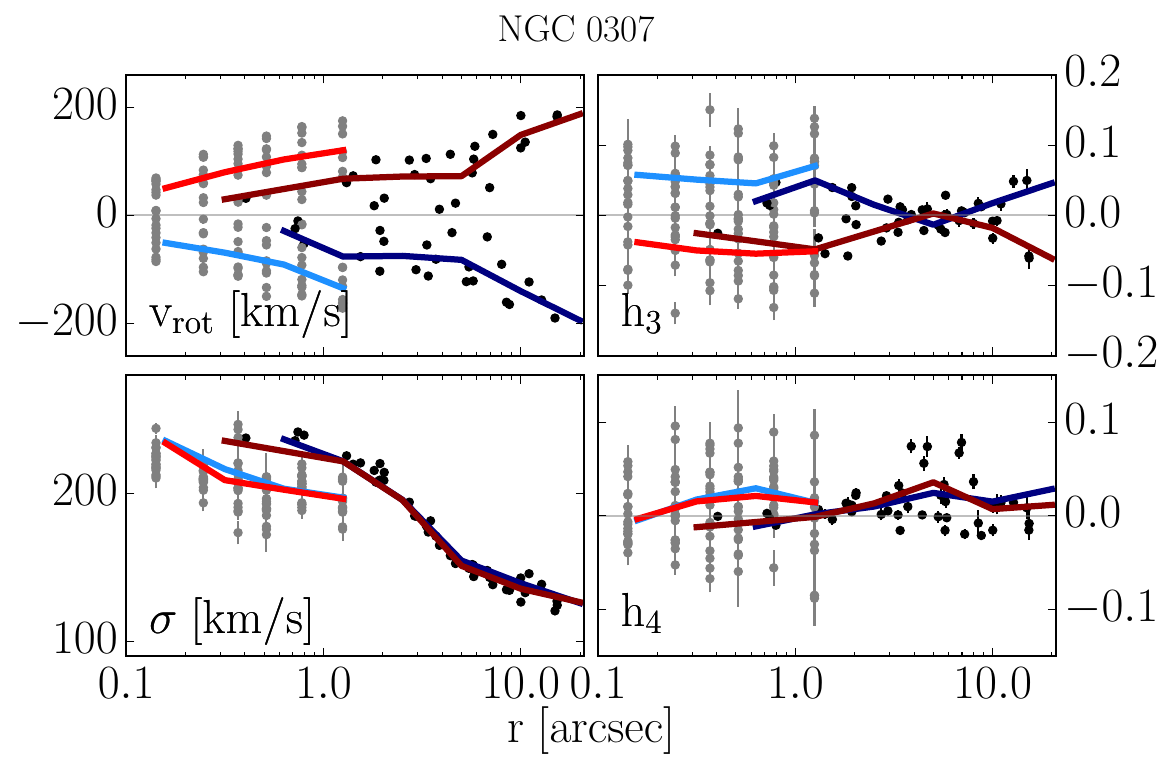}
\includegraphics[width = 0.49\columnwidth]{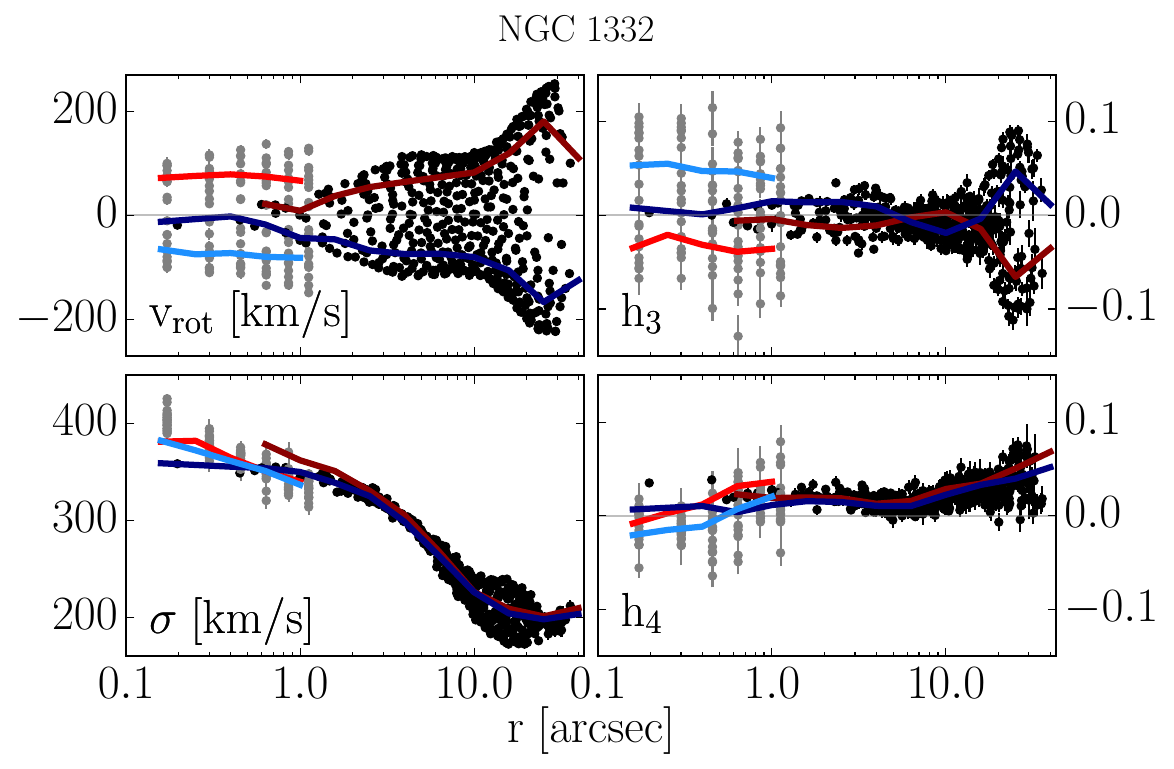}
\includegraphics[width = 0.49\columnwidth]{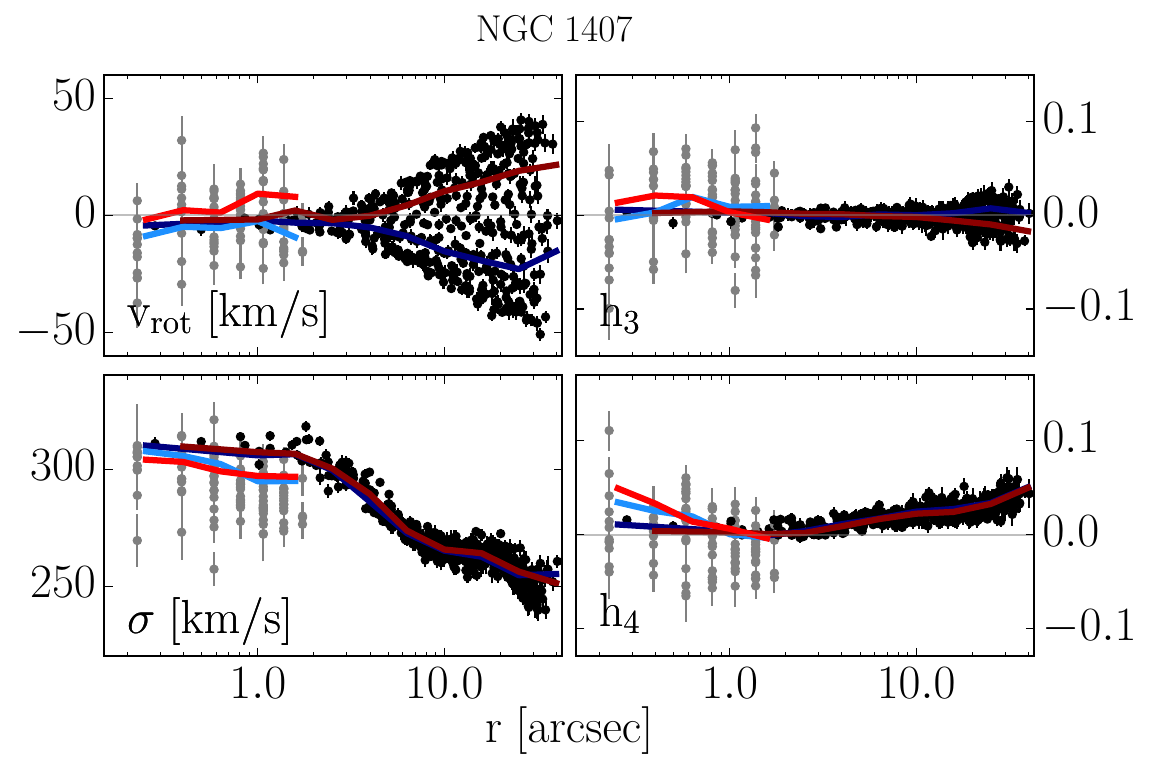}
\includegraphics[width = 0.49\columnwidth]{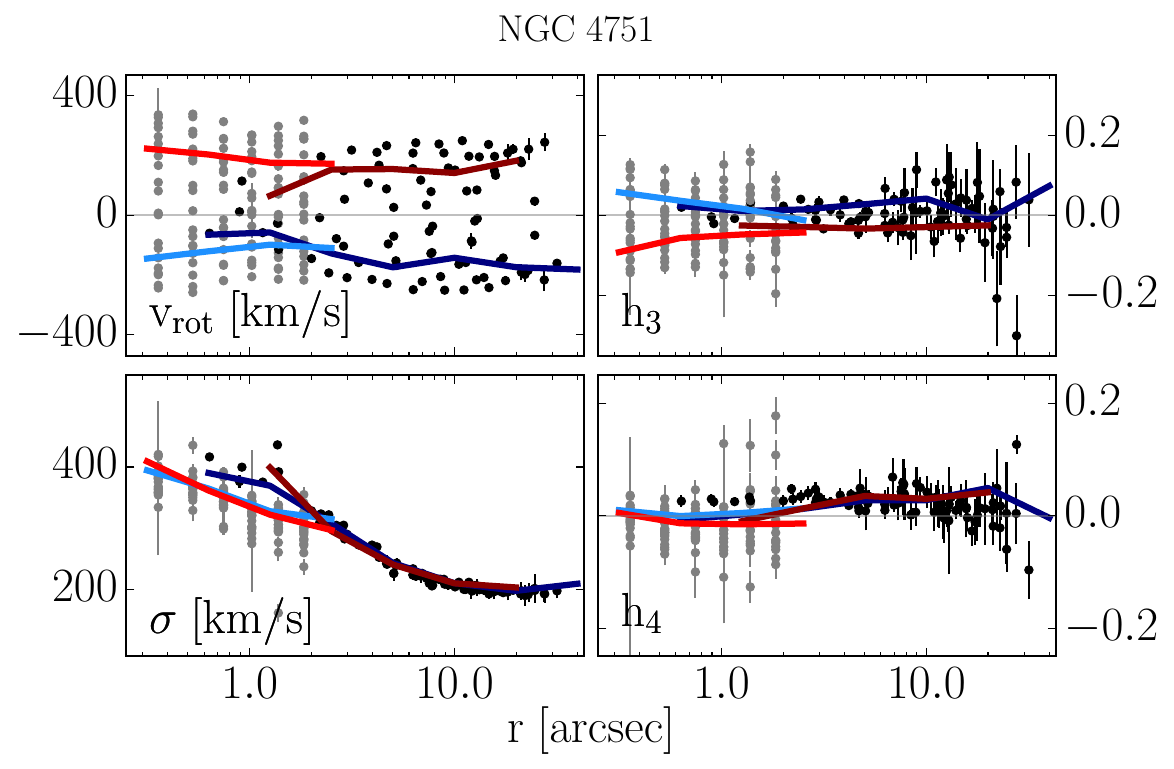}
\includegraphics[width = 0.49\columnwidth]{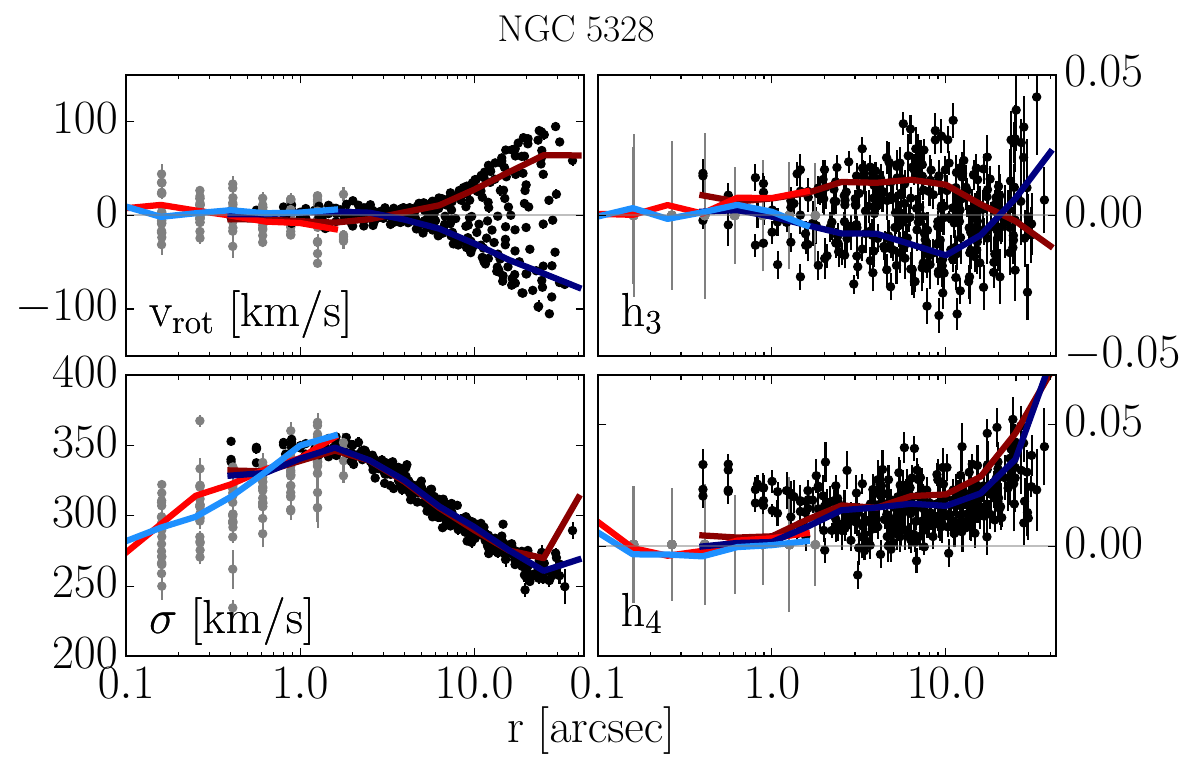}
\includegraphics[width = 0.49\columnwidth]{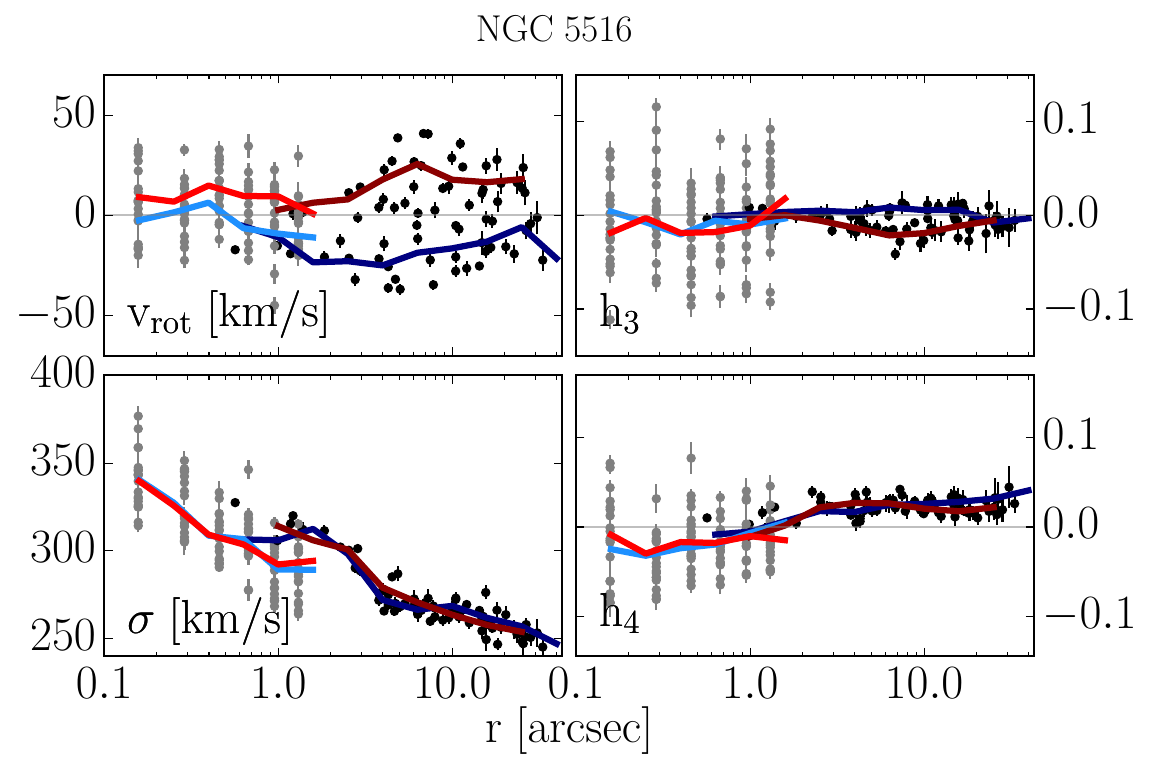}
 \caption{Dynamic fits to MUSE and SINFONI kinematics shown as radial profiles of Gauss-Hermite parameters. Gauss-Hermite parameters were derived from 8th order Gauss-Hermite polynomial fits to the non-parametric data and model LOSVDs -- though we only show the first four orders here. The MUSE data kinematics are shown as black points with error bars. Since our dynamical models fit all LOSVDs there should be one model LOSVD point per data LOSVD point, but for the sake of visibility we show our models as radial averages split into two for the two sides of rotation of each galaxy (red and blue lines). The dynamical fits to the MUSE data are shown as solid dark red and dark blue lines. Analogously, we show the SINFONI data in grey and the SINFONI model LOSVDs in light blue and light red. Points without visible errorbars have statistical uncertainties smaller than the symbols. For NGC~5328, the SINFONI kinematic data points all have $h_3 = h_4 = 0$, since only for this galaxy we used simple Gaussian LOSVDs which were derived from fits to the non-parametric LOSVDs (see Section \ref{subsec:kinproblems}). The errorbars of the data points were derived from fitting noisy realisations of these LOSVDs based on the noise of the original non-parametric LOSVDs.}
   \label{fig:radKin}
\end{figure*}
\addtocounter{figure}{-1}
\begin{figure*}[!htbp]
\centering
\includegraphics[width = 0.49\columnwidth]{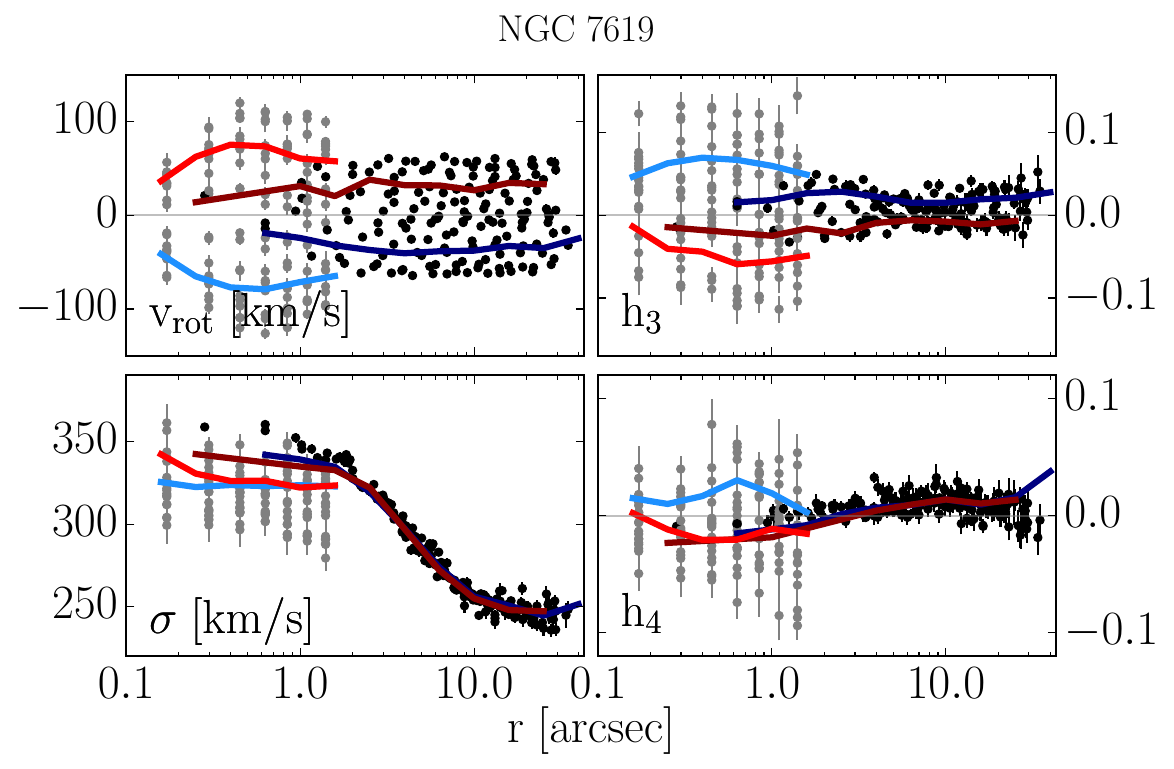}
 \caption{(continued)}
\end{figure*}
\begin{figure*}
\centering
\includegraphics[width = 0.49\columnwidth]{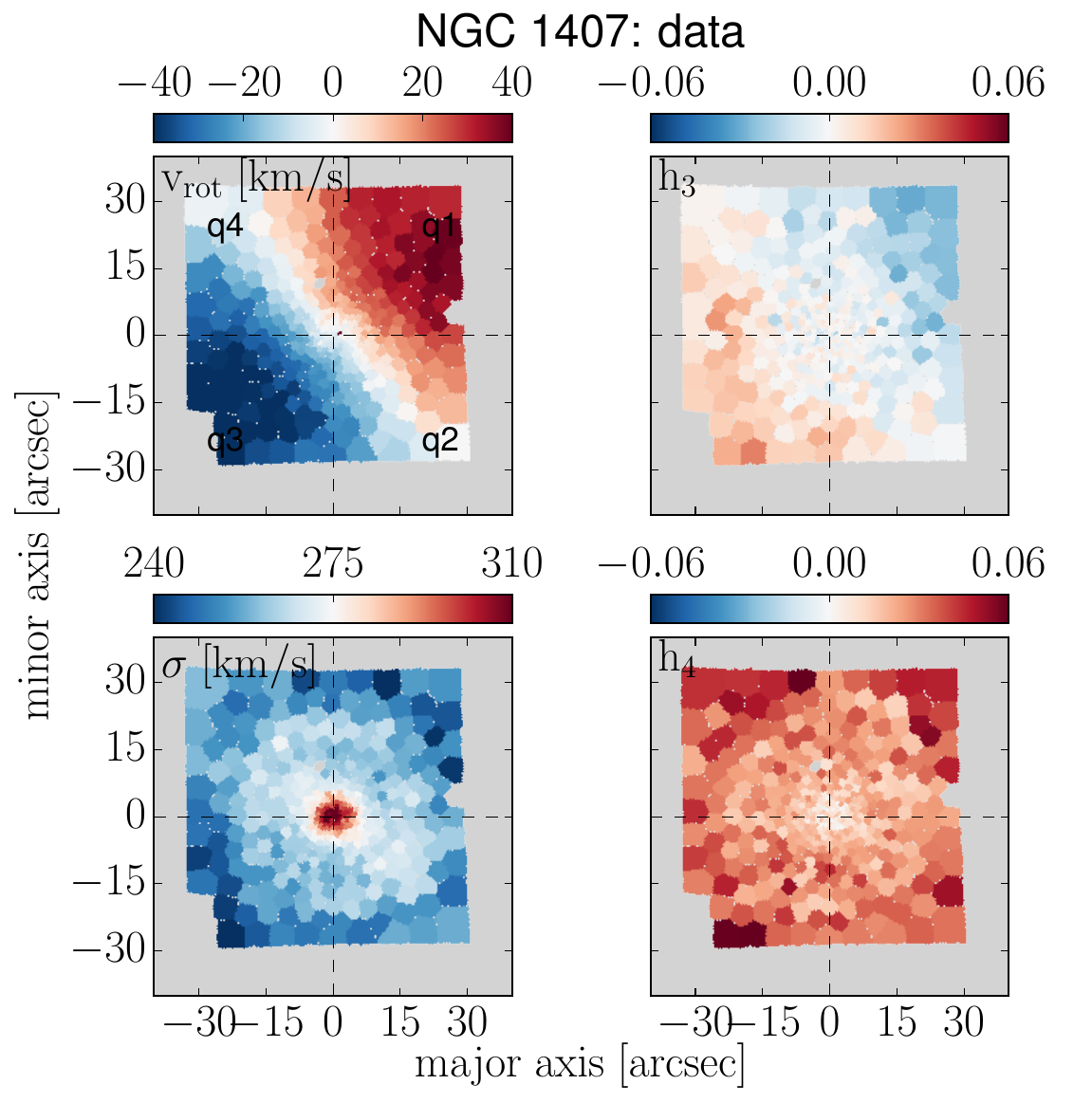}
\includegraphics[width = 0.49\columnwidth]{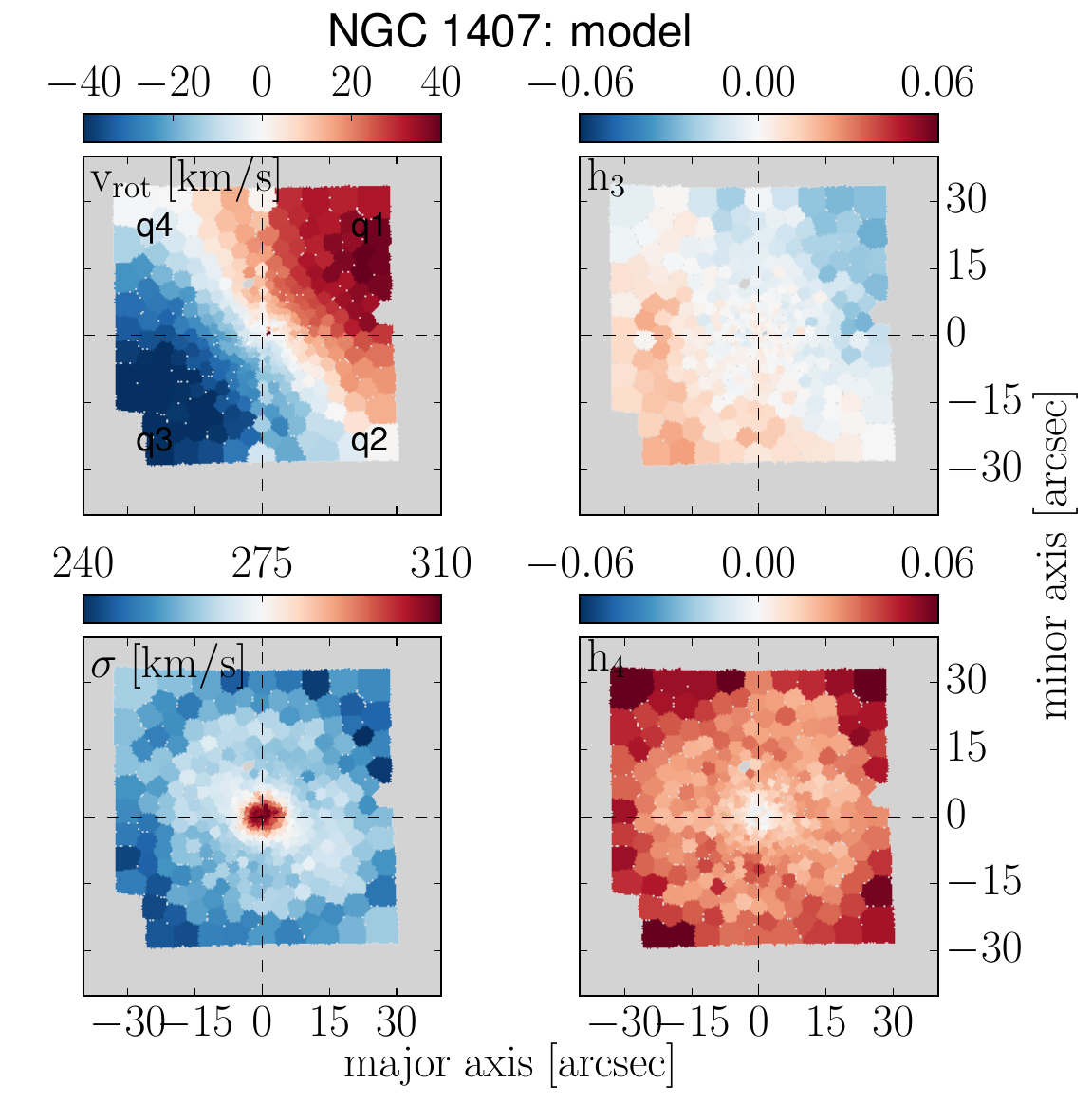}
 \caption{MUSE data-kinematics (left) and dynamical models (right) of NGC~1407 shown as 2D kinematic maps of Gauss-Hermite parameters. 
 X- and Y-axes are aligned with the major and minor axes of the galaxy. Gauss-Hermite parameters were derived from 8th order Gauss-Hermite polynomial fits to the non-parametric data and model LOSVDs -- though we only show the first four orders here. The model-map for $v_{\mathrm{rot}}$ shows a low-velocity artefact along the minor axis.}
   \label{fig:1407map}
\end{figure*}

\begin{figure*}
\centering
\includegraphics[width = 0.49\columnwidth]{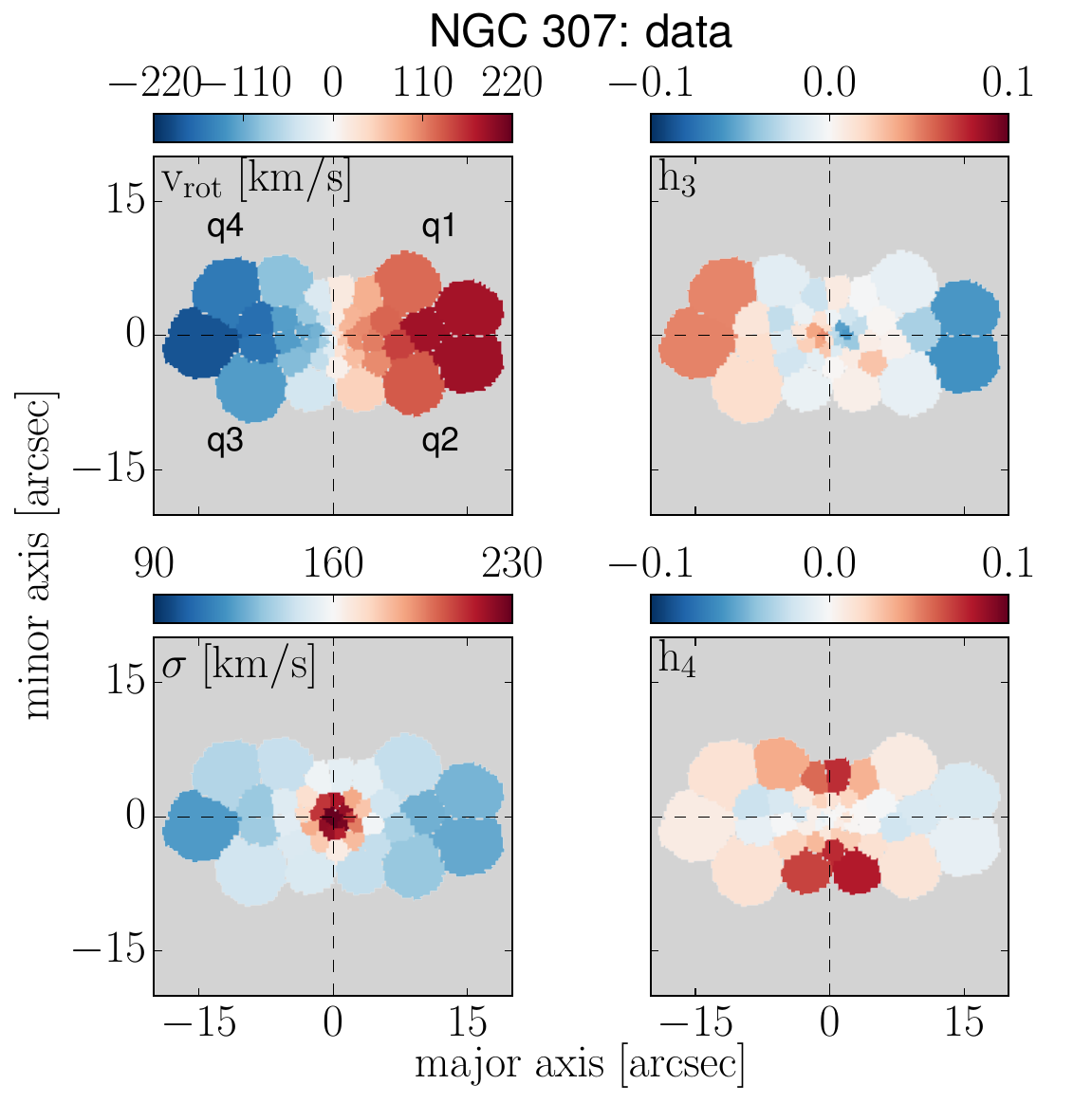}
\includegraphics[width = 0.49\columnwidth]{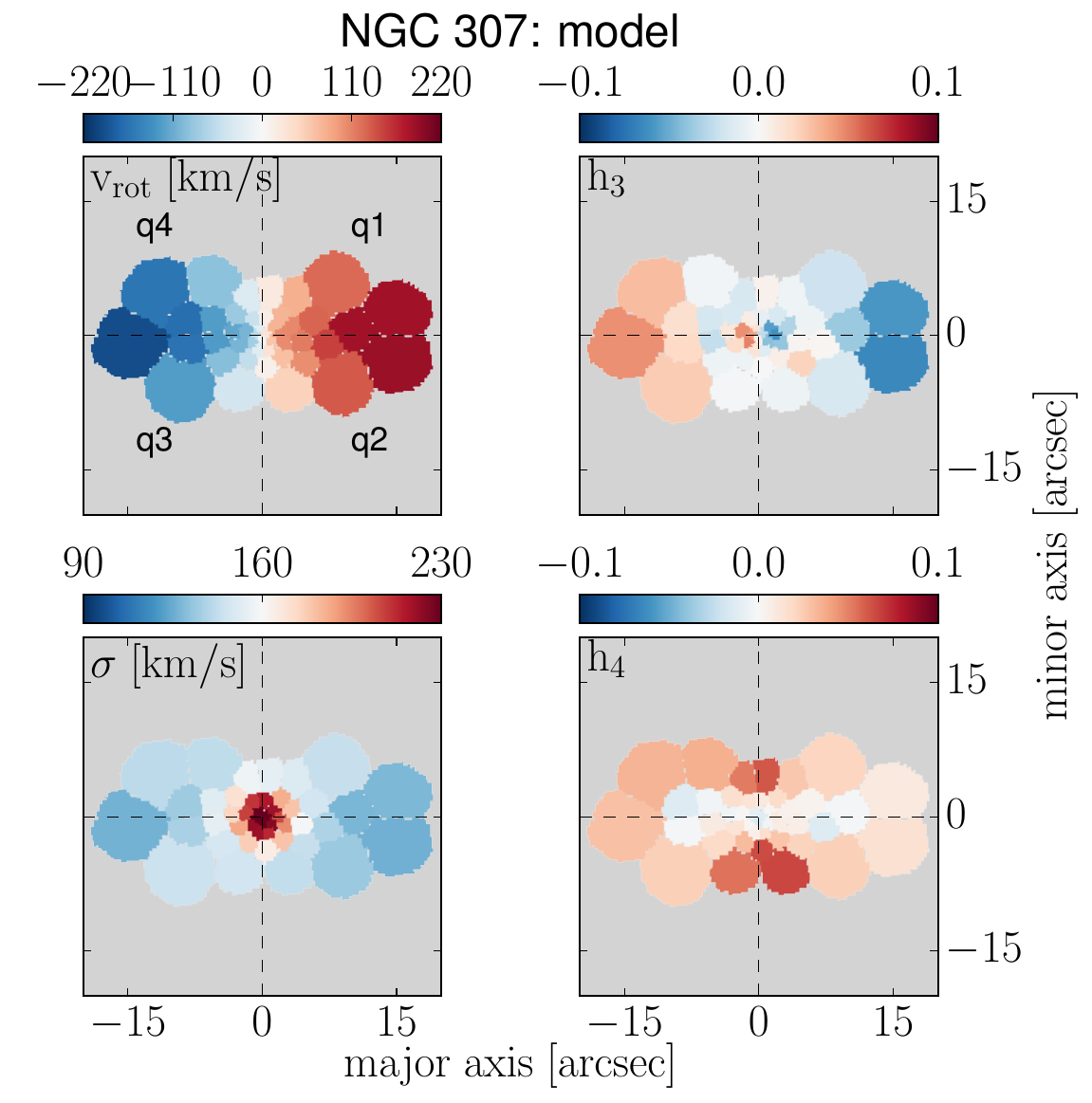}
 \caption{MUSE data-kinematics (left) and dynamical models (right) of NGC~307, same as Figure \ref{fig:1407map}.}
   \label{fig:0307map}
\end{figure*}
\FloatBarrier
\section{Constant-$\Upsilon$ models}
\label{ap:constant}

In Table \ref{tab:constantMLTab} we list the best-fit modeling parameters from our best-fit constant-$\Upsilon$ models. These models were \textit{fully} encompassed in the parameter space of our $\Upsilon$-gradient models. Best-fit $\Upsilon$-gradient models were in all cases better fits to the data than constant-$\Upsilon$ models, with an AIC$_p$-difference of around $10 -20$, slightly larger than the typical threshold for a black hole measurement, which is easily explained by the fact that the differences between the models primarily concern the central kiloparsec of the galaxies, which almost entirely accounts for the difference in AIC$_p$. This also indicates that outside this radius the slightly larger $\Upsilon$ of the constant-$\Upsilon$ models is taken out of the mass-buget of the DM component of the total dynamical mass profile.

\begin{table*}
\centering
 \begin{tabular}{l c c c c c}
 \hline
 \hline
Galaxy     &   Band &    $\Upsilon$ & $M_{\mathrm{BH}}$ & $\rho_{10}$ & $\alpha$\\
& & $\mathrm{[M_{\odot}/L_{\odot}]}$ & [$\mathrm{10^9 \ M_{\odot}}$] & [$\mathrm{10^8 \ M_{\odot}/kpc^3}$] & \\
\hline
NGC~1407 & B & $4.14 \pm 0.28$ & $8.50 \pm 0.87$ & $2.13 \pm 0.13$ & $0.75 \pm 0.32$\\
NGC~5328 & V & $5.81 \pm 0.31$ & $2.25 \pm 0.43$ & $0.85 \pm 0.15$ & $1.44 \pm 0.04$ \\
NGC~5516 & R &  $4.83 \pm 0.83$ & $2.88 \pm 1.13$ & $0.60 \pm 0.05$ & $1.61 \pm 0.18$\\
NGC~7619 & I &  $3.00 \pm 0.50$ & $4.38 \pm 0.38$ & $0.65 \pm 0.05$ & $1.61 \pm 0.20$\\

\hline
\end{tabular}
\caption{Results of Schwarzschild dynamical modeling using constant mass-to-light models. Photometric bands, as well as extinction corrections for $\Upsilon$-values for all galaxies were taken over from R+13, according to Table \ref{tab:generalTab}. Modeling parameters are listed as averages and standard deviations of values over all quadrants or sub-quadrants of each galaxy. In addition to the modeling parameters we also list the IMF mass normalization parameters relative to a Kroupa IMF.}

\label{tab:constantMLTab}
\end{table*}

\section{Testing our axisymmetric models with a triaxial N-body simulation}
\label{ap:stresstest}
\begin{figure*}
\centering
\includegraphics[width = 0.45\columnwidth]{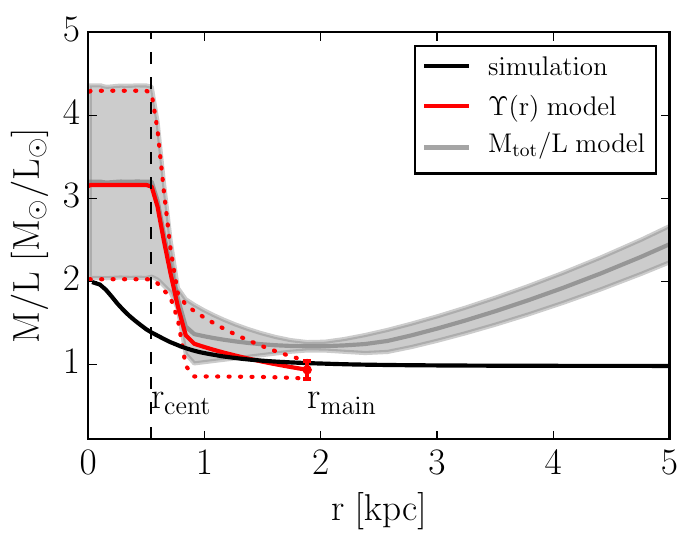}
 \caption{Mass-to-light ratio profile of our dynamical fits (solid red, the uncertainties are indicated by dotted red lines) to the triaxial $\Upsilon$-gradient galaxy simulation (black). Even though the simulation is triaxial, our axisymmetric models can correctly recover $\Upsilon_{\mathrm{main}}$. However, the stellar mass-to-light ratio inside $r_{\mathrm{cen}}$, $\Upsilon_{\mathrm{cen}}$ is overestimated by a factor of roughly $1.6$.}
   \label{fig:simulation}
\end{figure*}

As a stress test, we applied our axisymmetric models with $\Upsilon$ gradients to 
a numerical N-body  originally from \citet{Rantala2018}. It is the same simulation that we have used to test our triaxial Schwarzschild code SMART and details about how we extract mock LOSVDs and images can be found in the respective papers \citep{deNicola2022,Neureiter2023a}. We model the projection of the simulation along its intermediate axis. To match the simulation with the average galaxy in our sample we shrunk it in radius and mass by a factor of two, such that all particle velocities stay the same. 
Originally, the stellar particles all have the same mass. To introduce a gradient we have to assign a mass-to-light ratio to each particle. In a steady state system, a stable mass-to-light ratio gradient needs to be a function of the integrals of motion. Simply defining a $\Upsilon$ gradient as a function of radius is not a good option. Instead, we define the gradient as a function of energy. To do so we first fit a polynomial to the distribution E(r) of the particle energies.
Then we determine the average particle energies $E_{\mathrm{main}}$ at $\SI{2}{kpc}$ and $E_{\mathrm{cen}}$ at $\SI{0.5}{kpc}$. For all particles with $E<E_{\mathrm{cen}}$ we set the mass-to-light ratio equal to two and for all particles with $E>E_{\mathrm{main}}$ we set the mass-to-light ratio to one. In between, we interpolated the mass-to-light ratios log-linearly over $E$. With the mass-to-light ratio defined for each particle, we can assign a luminosity to each particle and derive LOSVDs and images respectively. The mock galaxy that we have constructed in this way has a stellar mass-to-light ratio gradient that is similar to our observed gradients, but somewhat steeper, a bit more extended, and without a central $\Upsilon$-plateau -- $\Upsilon$ increases to  $\Upsilon^{\mathrm{sim}}_{\mathrm{cen}}$ essentially in the very center. This can be seen in Figure \ref{fig:simulation}. 

To prepare the simulation for Schwarzschild dynamical modeling we set out to generate mock kinematic data in analogy to the data we used in this study (see Section \ref{subsec:kinematics}).
We adopt the simulated MUSE and SINFONI binning from \citet{Neureiter2023}, asuming a distance of $D = \SI{56.2}{Mpc}$ (about the largest in our sample). The LOSVDs were generated over $v_{\mathrm{los}} = \pm \SI{1500}{km/s}$ with $N_{vel} = 15$ for both mock-data sets, in analogy to the sample galaxies. Dividing the galaxy in the spatial quadrants along the major and minor axis (aligned with the $x$ and $y$ axis of the FOVs), we derive a total of $\sim 80$ mock SINFONI plus MUSE LOSVDs per quadrant. 

Finally we generated images, in a way that mimics our use of HST and ground-based imaging for the sample galaxies: One $\SI{30}{\arcsec} \times \SI{30}{\arcsec}$ image with a pixel size of $\SI{0.05}{arcsec}$, and one $\SI{300}{\arcsec} \times \SI{300}{\arcsec}$ image with a pixel size of $\SI{0.2}{\arcsec}$. For the photometric analysis and combination of the images we proceed as with the sample galaxies (see Section \ref{subsec:lightDens}).

The dynamical models of the simulated galaxy use exactly the same setup as was used for the other sample galaxies.

The best-fit models achieved a good $(\chi^2 + m_{\mathrm{eff}})/N \sim 0.96$. The models recovered the mass of the central SMBH within one sigma, $M_{\mathrm{BH}} = (7.38 \pm 2.68) \times 10^9 M_{\odot}$. As for the sample galaxies we used a cored NFW halo with just one parameter, $\rho_{10}$, which necessarily under-predicts the central DM density of the simulation, which has an inner logarithmic density slope of $\gamma ~\sim -0.7$. Nonetheless, when comparing the enclosed mass within $r_{\mathrm{cen}} = \mathrm{FWHM}_{\mathrm{PSF}} = \SI{1.5}{\arcsec}$, we find that our models recover the enclosed central DM mass within $8 \%$, $M_{DM}(r \leq r_{\mathrm{cen}}) = (4.02 \pm 1.12) \times 10^8 M_{\odot}/kpc^3$, versus $M^{\mathrm{sim}}_{DM}(r \leq r_{\mathrm{cen}}) \sim 5.56 \times 10^8 M_{\odot}/kpc^3$. 
We also correctly recover the main-body mass-to-light ratio of the stars within one sigma $\Upsilon_{\mathrm{main}} = 0.93 \pm 0.11$. This precision in the SMBH mass, DM recovery and main-body stellar mass is quite remarkably in view of the fact that the simulation is triaxial but our models assume axial symmetry.

The central mass-to-light ratio is more uncertain. On average, we overestimate its value by a factor of roughly $1.6$, $\Upsilon_{\mathrm{main}} = 3.16 \pm 1.13$, as shown in Figure \ref{fig:simulation}. This bias could have been caused by the fact that the simulation is triaxial. As triaxial effects are viewing angle dependant, with just one viewing angle tested it is difficult to draw a final conclusion at this point.

The test presented here should be considered a stress test for our approach. We have shown that even under difficult conditions (triaxial object, large sphere of influence) the main-body mass-to-light ratio and the spatial scale of the gradient are very robust. The central amplitude of the gradient -- if any -- could be shallower than inferred. 
We plan a more thorough and comprehensive investigation of how accurate stellar mass-to-light ratio gradients can be recovered dynamically in a future paper.

\bibliography{bibliography} 
\end{document}